\documentclass[aps,prx,twocolumn,superscriptaddress,showpacs]{revtex4-2}
\usepackage{amsmath}
\usepackage{amssymb,amsfonts}
\usepackage{amsthm,amsmath,amssymb}
\usepackage{dsfont}
\usepackage{mathrsfs}

\usepackage{graphicx}
\usepackage[colorlinks]{hyperref}

\newcommand{\be}{\begin{equation}}
\newcommand{\ee}{\end{equation}}

\def\bea{\begin{eqnarray}}
\def\eea{\end{eqnarray}}
\def\a{\alpha}
\def\b{\beta}

\def\d{\delta}
\def\D{\Delta}

\def\g{\gamma}

\def\la{\langle}
\def\ra{\rangle}

\def\s{\sigma}

\begin{document}

\title {CFT$_D$ from TQFT$_{D+1}$ via Holographic Tensor Network, and \\
Precision Discretisation of CFT$_2$ }

\author{Lin Chen}
\affiliation{State Key Laboratory of Surface Physics, Fudan University, 200433 Shanghai, China}
\affiliation{Department of Physics and Center for Field Theory and Particle Physics, Fudan University, Shanghai 200433, China}
\affiliation{School of Physics and Optoelectronics, South China University of Technology, 510641 Guangzhou, China}
\author{Haochen Zhang}
\affiliation{Department of Physics, Shandong University, Jinan 250100, China}
\affiliation{International Centre for Theoretical Physics Asia-Pacific, University of Chinese Academy of Sciences, 100190 Beijing, China}
\author{Kaixin Ji}
\affiliation{State Key Laboratory of Surface Physics, Fudan University, 200433 Shanghai, China}
\affiliation{Department of Physics and Center for Field Theory and Particle Physics, Fudan University, Shanghai 200433, China}
\author{Ce Shen}
\affiliation{State Key Laboratory of Surface Physics, Fudan University, 200433 Shanghai, China}
\affiliation{Department of Physics and Center for Field Theory and Particle Physics, Fudan University, Shanghai 200433, China}
\author{Ruoshui Wang}
\affiliation{Cornell University, Ithaca, New York 14853, USA}
\affiliation{Institute for Advanced Study, Tsinghua University, Beijing 100084, China}
\author{Xiangdong Zeng}
\affiliation{State Key Laboratory of Surface Physics, Fudan University, 200433 Shanghai, China}
\affiliation{Department of Physics and Center for Field Theory and Particle Physics, Fudan University, Shanghai 200433, China}
\author{Ling-Yan Hung}
\email{lyhung@fudan.edu.cn}
\affiliation{State Key Laboratory of Surface Physics, Fudan University, 200433 Shanghai, China}
\affiliation{Department of Physics and Center for Field Theory and Particle Physics, Fudan University, Shanghai 200433, China}
\affiliation{Institute for Nanoelectronic devices and Quantum computing, Fudan University, 200433 Shanghai, China}
\affiliation{Yau Mathematical Sciences Center, Tsinghua University, Beijing 100084, China}
\date{\today}


\begin{abstract}

We show that the path-integral of conformal field theories  in $D$ dimensions (CFT$_D$) can be constructed by solving for eigenstates of an RG operator following from the Turaev-Viro formulation of  a topological field theory in $D+1$ dimensions  (TQFT$_{D+1}$), explicitly realising the holographic sandwich relation between a symmetric theory and a TQFT. Generically, exact eigenstates corresponding to symmetric-TQFT$_D$  follow from Frobenius algebra in the TQFT$_{D+1}$.  For $D=2$, we constructed eigenstates that produce 2D rational CFT path-integral exactly, which, curiously connects a continuous field theoretic path-integral with the Turaev-Viro state sum. We also devise and illustrate numerical methods for $D=2,3$ to search for CFT$_D$ as phase transition points between symmetric TQFT$_D$. Finally since the RG operator is in fact an exact analytic holographic tensor network, we compute ``bulk-boundary'' correlator and compare with the AdS/CFT dictionary at $D=2$. Promisingly, they are numerically compatible given our accuracy, although further works will be needed to explore the precise connection to the AdS/CFT correspondence.
\end{abstract}
\maketitle

\section{Introduction}
In recent years, we have witnessed an explosion of activity in the study of generalization of the notion of global symmetries, loosely termed ``categorical symmetry'' \cite{Bhardwaj:2017xup, Chang:2018iay, Thorngren:2019iar,
Thorngren:2021yso, Ji:2019jhk, Kong:2020cie, Albert:2021vts, Chatterjee:2022kxb, Liu:2022cxc, Lootens:2021tet,Freed:2018cec, Levin:2019ifu},
which was preceded by an extensive and systematic study of defects in rational conformal field theories in 2D \cite{Verlinde:1988sn, Petkova:2000ip, Frohlich:2006ch, Frohlich:2004ef, Fuchs:2002cm, Fuchs:2003id, Fuchs:2004dz, Fuchs:2004xi,  Quella:2006de, Fuchs:2007vk,  Fuchs:2007tx,  Bachas:2007td, Davydov:2011kb, Davydov:2013lma, Kong:2019amd,  Petkova:2009pe, Carqueville:2012dk, Kitaev:2011dxc, Brunner:2013xna, Petkova:2013yoa, Aasen:2016dop, Williamson:2017uzx, Aasen:2020jwb, Frank1, Makabe:2017ygy, Thorngren:2018bhj, Ji:2019ugf, Lin:2019hks, Chang:2020imq, Huang:2021zvu, Burbano:2021loy} and topological field theories  \cite{Davydov:2011kb, Carqueville:2012dk, Brunner:2014lua, Carqueville:2016kdq, Carqueville:2017aoe}.
It has been applied to constrain theories and renormalization group flows \cite{Chang:2018iay, Thorngren:2019iar, Komargodski:2020mxz, Thorngren:2021yso,
Kikuchi:2021qxz, Rudelius:2020orz, Heidenreich:2021xpr, McNamara:2021cuo, Cordova:2022rer, Arias-Tamargo:2022nlf}.
Categorical symmetry usually refers to the algebra of topological defects in a quantum theory, and group symmetry is a special case. In \cite{Ji:2019jhk} categorical symmetry particularly refers to the enlarged collection of symmetries generated by the charges together with gauge fluxes which appear when the symmetry is gauged.
This consideration of enlarged symmetry leads to a very general holographic relation.  It has been suggested that to describe a $D$ dimensional system $S$ with topological defects associated to a tensor category $\mathcal{C}$, the system $S$ can be associated to a boundary condition of a topological theory in $D+1$ dimensions that is described by the center of $\mathcal{C}$ denoted $\mathcal{Z}(\mathcal{C})$, and whose topological excitations include the complete set of  charges and fluxes \cite{Freed:2018cec, Lootens:2021tet, Ji:2019jhk, Kong:2020cie, Albert:2021vts, Chatterjee:2022kxb, Chatterjee:2022tyg, Liu:2022cxc, Kong:2017etd, Kong:2019byq, Kong:2019cuu, Kong:2020iek, Kong:2020jne, Kong:2021equ, Xu:2022rtj, Aasen:2016dop, Frank1,Lootens:2019ghu, Gaiotto:2020iye, Bhardwaj:2020ymp, Apruzzi:2021nmk, Moradi:2022lqp}. Particularly, the symmetries of the field theory are made explicit making use of this holographic relation. Its path-integral can be realised as a sandwich in which the associated TQFT in one higher dimensions is sandwiched between a non-trivial boundary condition and a topological boundary condition \cite{Gaiotto:2020iye,Ji:2019jhk, Apruzzi:2021nmk, Freed:2022qnc}.

Separately, it is observed that infinite classes of well known $2$-dimensional critical partition functions of integrable statistical models can be expressed as the overlap between the ground state wave-functions $|\Psi\rangle$ of topological orders in 2+1 D and some carefully chosen state $\langle \Omega |$ \cite{Aasen:2020jwb,Frank1,Lootens:2019ghu, frank_haagerup}, often called a ``strange correlator''. The choice of the topological order associated to a modular higher category $\mathcal{C}$ captures the categorical symmetry of the critical model constructed, while $\la \Omega |$ is chosen by making comparison with known lattice models, or via some educated guesses. This method has been applied to construct the partition function of a novel CFT believed to exist \cite{frank_haagerup,Tachikawa_Haagerup}, whose categorical symmetry is related to the Haaegerup category.

This is an explicit realization of the holographic sandwich mentioned above that applies way more generally than $D=2$ CFTs and $D+1=3$ TQFTs  with $\langle \Omega | $ supplying the non-trivial boundary condition, and the topological boundary hidden inside $|\Psi\rangle $ (see  section 5 of \cite{Lootens:2020mso}), although apparently only in lattice models. It is part of our goal to show that this construction can be worked out for continuous field theory through the use of an explicit  renormalization group (RG) flow operator. This RG operator is going to play a central role connecting ideas in categorical symmetries,  renormalisation group  and holographic tensor network that leads to a general framework for explicitly constructing path-integrals of symmetric CFT$_D$ from TQFT$_{D+1}$. 

For a concise summary of the paper, we are presenting five main results. 
\begin{enumerate}
\item Construct an explicit discrete RG operator from Turaev-Viro formulation of TQFT$_{D+1}$. We argue that exact CFT$_D$ partition functions follow from eigenstates of these RG operators. Taking the overlap of these eigenstates with the ground state of TQFT$_{D+1}$ reproduces the path-integral of the CFT$_D$, generalising the strange correlators discussed above that work only for lattice models. (Section  \ref{sec:RG} )

\item To illustrate the idea we construct {\it topological} eigenstates of RG operators of TQFT$_{D+1}$ at $D=1,2,3$. 
They follow from Frobenius algebra in the fusion category associated to the TQFT$_{D+1}$. Strange correlators constructed from these eigenstates reproduce partition functions of symmetric TQFT$_D$.  
 ($D=1$ in section \ref{sec:2D}, $D=2$ in section \ref{sec:3D}, $D=3$ in section \ref{sec:4D})

\item We construct {\it analytic and exact} eigenstates of RG operators of TQFT$_3$ and show that their strange correlators recover path-integrals of 2D RCFT precisely. This provides a curious connection between continuous path-integrals and the discrete Turaev-Viro state sum.   (Section \ref{sec:exactcft})

\item CFT$_D$ can also be searched for numerically. We devise numerical algorithm to search for RG operator eigenstates corresponding to CFT$_D$. Our method makes use of the fact that \emph{CFTs are phase transition points between the topological eigenstates we found above}. We illustrate our method in $D=2,3$ taking specific examples. Our numerics can {\it generate numerically} the product map defining a Frobenius algebra. The $D=3$ algorithm we constructed and illustrated using the 3D Ising is, to our knowledge, a novel symmetry preserving tensor network renormalisation procedure.  ($D=2$ in section \ref{sec:num1} - \ref{sec:num2}, $D=3$ in section \ref{sec:3Dnum}.)

\item The RG operator and subsequent strange correlator we constructed is an exact holographic tensor network that can be understood as a discretisation of a (Euclidean) AdS$_{D+1}$ space while describing precise CFTs. We give numerical evidence at $D=2$ that the bulk-boundary propagator {\it agrees} with that in AdS$_3$/CFT$_2$. (Section \ref{sec:holo}.)

\end{enumerate}

\section{RG operator from a Topological Ground State Wave-Function as an Exact Holographic Network} \label{sec:RG}
Consider the path-integral of a $D+1$-dimensional topological order (TQFT) associated to a braided category $\mathcal{C}
$ on a $D+1$ ball $B_{D+1}$ with a $D$-spherical boundary $S_{D}$.
This produces a ground state wave-function $|\Psi\rangle$ of the topological order. Imposing boundary conditions on $S_{D}$ can
be interpreted as taking the overlap between some state $\langle \Omega|$ with the ground state wave-function :
\be \label{eq:partition_fn}
Z(\Omega, \mathcal{C}) = \langle \Omega | \Psi \rangle.
\ee
The ground state satisfies $\mathcal{O} |\Psi\rangle = |\Psi\rangle$, for any stringy/membrane operator $\mathcal{O} \in \mathcal{C}$.
This is inherited by $Z(\Omega,\mathcal{C})$ as topological symmetry. It is thus argued that any $D$-dimensional theory possessing  categorical symmetry $\mathcal{Z}(\mathcal{C})$ can be constructed by appropriate choice of $\langle \Omega |$ \cite{Gaiotto:2020iye}.
Since $|\Psi\rangle$ is the wave-function of a topological theory, it should be invariant under scale transformation, which is generated by some operator $\mathcal{H}_\mathcal{C}$.
Therefore
\begin{align}
Z(\Omega, \mathcal{C}) = \langle \Omega | \Psi \rangle = \langle \Omega| \exp( z \mathcal{H}_\mathcal{C}) |\Psi\rangle,
\end{align}
for some RG coordinate $z$. If $Z(\Omega, \mathcal{C})$ describes conformal/topological $D$-dimensional system, then
\begin{align}
\langle \Omega | \exp(z \mathcal{H}_\mathcal{C}) = \langle \Omega |,
\end{align}
and thus the construction of topological/conformal partition functions in $D$ dimensions with categorical symmetry $\mathcal{Z}(\mathcal{C})$ is reduced to a question of solving and classifying eigenstates of the RG operator $\mathcal{H}_\mathcal{C}$ \footnote{For $Z(\Omega, \mathcal{C})$ to be a partition function of a local theory in $D$-dimensions, we should require that  $ \Omega $ satisfies some local constraint, such as satisfying the area law in its entanglement. For a conformal theory one should probably require also rotation and translation invariance in $D$-dimensional plane. }.

These notions are useful if we could construct $\mathcal{H}_\mathcal{C}$ explicitly -- and this can readily be done, for example, in $D+1$ lattice topological models such as Dijkgraaf-Witten (DW) models in arbitrary dimensions, or Turaev-Viro type TQFT's in 2+1 dimensions, which would be discussed below.
In these models, the lattice spacing is a natural UV cutoff, and $\mathcal{H}_\mathcal{C}$ can be constructed out of $D+1$ simplices to connect the ground state wavefunction from one given triangulation of $S_D$ with lattice scale $\Lambda$ to another of lattice scale $\Lambda'$. The RG operator in these cases
would take the form of a holographic tensor network.

\section{Holographic Networks from 2D Dijkgraaf- Witten (DW) Theories}\label{sec:2D}
\begin{figure}
	\centering
	\includegraphics[width=0.4\linewidth]{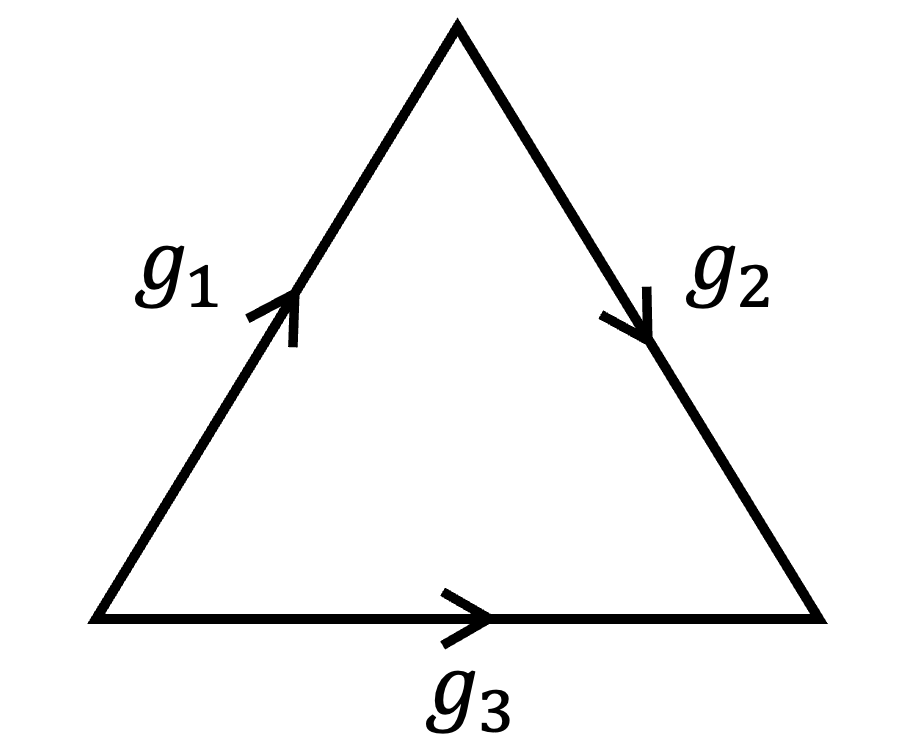}
	\caption{ A triangle with this type of orientation is assigned $\a_2(g_1,g_2)$ with $g_3=g_1\times g_2$.}
	\label{triorien}
\end{figure}

\begin{figure}
	\centering
	\includegraphics[width=0.8\linewidth]{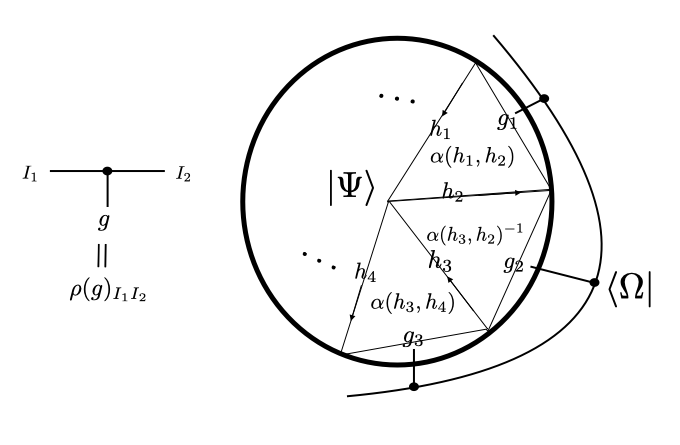}
	\caption{ A simple triangulation of a disk and a tensor network representation of the state $\langle \Omega|$.}
	\label{fig:2dDW}
\end{figure}

In the following, we will illustrate the idea above beginning with obtaining 1D theories from 2D topological models.
The first set of examples is 2D DW theories characterized by group $G$. To compute the path-integral
over a 2-manifold $M_2$, it is triangulated into triangles, where each edge is labeled by a group element $g_i\in G, \, i=\{1,2,3\}$
and the triangle is assigned a value $\alpha_2(g_1,g_2) \in H^2(G, U(1))$ \cite{Dijkgraaf:1989pz}, where $H^2(G, U(1))$ denotes the 2-cohomology, and $g_i \in G$
such that $g_1 \times g_2 = g_3$ for an orientation of the triangle chosen as shown in figure \ref{triorien}.
Consider the path-integral on a disk which produces the ground state wave-function of the 2D model on a circle. The simplest triangulation is given as in figure \ref{fig:2dDW}.
We note that this can be understood as a matrix product state
\begin{align}
&|\Psi\rangle_G =\sum_{\{g_i\}} \sum_{\{h_i\}} \cdots \tau_{h_i h_{i+1}}(g_i) \tau_{h_{i+1},h_{i+2}}(g_{i+1}) \cdots  \nonumber \\
&\qquad \qquad \qquad |\cdots g_i, g_{i+1}, \cdots\rangle, \\
& \tau_{h_ih_{i+1}}(g) = \alpha(h_i,h_{i+1})^{\varepsilon} \delta_{g, h_i h_{i+1}},
\end{align}
with $\varepsilon=\pm 1$ according to the orientations of the triangles.

\begin{figure}
	\centering
	\includegraphics[width=0.9\linewidth]{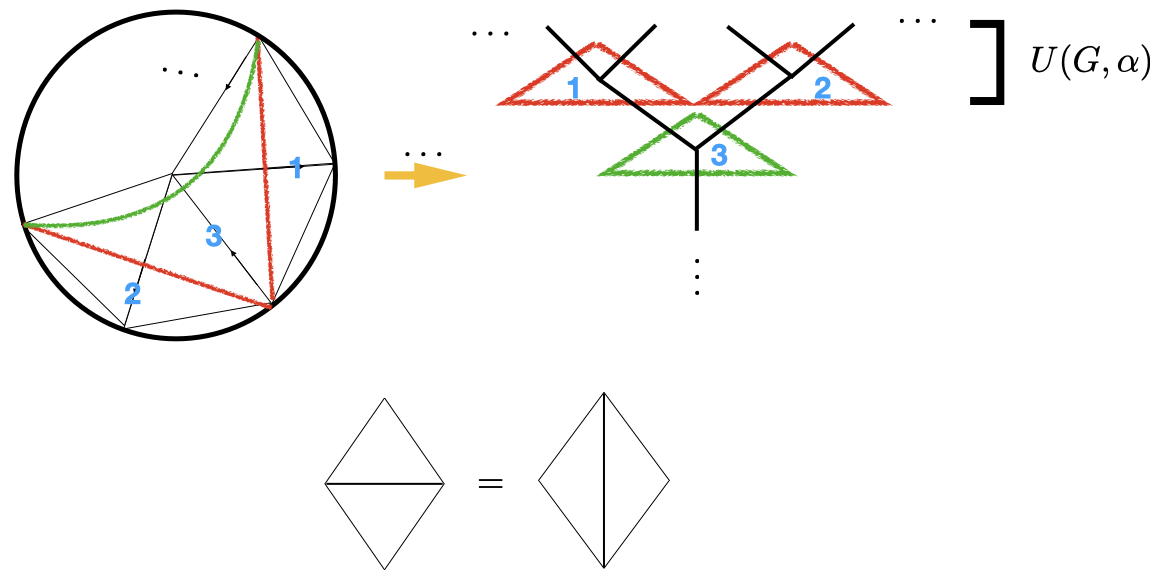}
	\caption{The triangulation of the disk can be converted to a tree-like tensor network by repeated use of the associativity condition (\ref{eq:alex1}) which is also illustrated pictorially in the lower half of the figure.  }
	\label{RG2d}
\end{figure}

These triangles satisfy the associativity constraint
\be \label{eq:alex1}
\frac{\alpha(g_1,g_2) \alpha(g_1g_2, g_3)}{\alpha(g_1, g_2g_3)\alpha(g_2,g_3)} = 1.
\ee
Using the relation (\ref{eq:alex1}), we can change the triangulation in the interior of the disk. Consider a change in triangulation such that red lines replace the original radial lines in figure \ref{fig:2dDW}. This is shown in figure \ref{RG2d}. We note that the red lines form the boundary of a smaller disk. The number of red edges is half the number of edges at the original boundary of the disk, i.e. if the original boundary circle of the disk has $2N$ edges, the red edges form a circle with $N$ edges as shown in figure \ref{RG2d}. The triangles that are between the original boundary edges and the red edges can be considered as a linear map between the ground state wave-function defined on the original boundary and the ground state wave-function defined on the red edges. Let us collect these triangles, and call them $U_N(G, \alpha)$, forming the first layer of the RG operator. 

We note that this process can be repeated indefinitely, as we change the triangulation of the disk bounded by the red lines. In the second re-triangulation, we would introduce green lines in figure \ref{RG2d} . Collecting the triangles between the red boundary and the green boundary  is thus $U_{N/2}(G, \alpha)$, forming the second layer of the RG operator. 

Consider the limit where $N$ approaches infinity. Then every layer of the RG operator would become identical, and RG process can be conducted indefinitely. 
We identify the ideal RG operator $\exp(z \mathcal{H}_\mathcal{C})$ with $ U(G, \alpha)$, the latter defined as
\be
\exp(z \mathcal{H}_\mathcal{C}) = U(G, \alpha) \equiv \lim_{N\to \infty} U_N(G,\alpha).
 \ee
Eigenstates $\langle \Omega|$ of $U(G,\a) $ would define scale invariant partition functions $Z(\Omega,G)$ with global symmetry $G$
through (\ref{eq:partition_fn}).

For simplicity, we will focus our discussion on the trivial element of $H^2$, such that $\alpha_2(g_1,g_2)= 1$.
There is a very simple class of eigenstates. The state is defined on a circle (approaching infinite size). Consider an $\langle \Omega|$ that can be represented by a matrix product state (MPS), meaning that the wave-function of which can be represented as the trace of products of matrices $\rho(g_i)_{a_ib_i}$ with $a_i,b_i$ internal auxilliary indices that have dimension $d$. Suppose
\begin{align}
&\langle \Omega | = \sum_{\{g\}}\textrm{tr}(\cdots \rho({g_i}) \rho({g_{i+1}}) \rho({g_{i+2}}) \cdots )  \langle \cdots g_i, g_{i+1}, g_{i+2} \cdots| , \,\, \nonumber \\
&(\rho(g_1)\rho(g_2))_{ac} \equiv \sum_{b}\rho(g_1)_{ab} \rho(g_2)_{bc} = \rho(g_1g_2)_{ac},  \label{eq:mps_gprep}
\end{align}
one can readily check that (\ref{eq:mps_gprep}) ensures that $\langle \Omega|$ is in fact an eigenstate of the RG operator $U(G,1)$.
Therefore every irreducible representation of $G$ constitutes an eigenstate.
One can also show that the most generic form of eigenstates to the RG operator can be decomposed as direct sum of representations
of the group $G$, i.e. $\rho(g) = \oplus \rho^\mu(g)$, where $\mu$ denotes different representations of the group $G$.
The details of the proof are relegated to the appendix \ref{2Dappendix}.

Therefore $Z(\Omega, G) = \langle \Omega | \Psi\rangle_G$ is a summation of finite dimensional traces of products of matrices $T_{I=(a,h), J=(b,k)}(g) = \rho(g)_{a,b} \alpha(h,k)^\varepsilon \delta_{g, hk}$.
All local correlation functions decay exponentially, proving that $Z(\Omega, G)$ is a topological 1$D$ theory. Physically, we do not expect critical models in 1$D$ and we believe
the construction gives a complete construction of 1$D$ models with symmetry $G$.

\section{Holographic Networks in 3D} \label{sec:3D}
\begin{figure}
	\centering
	\includegraphics[width=0.6\linewidth]{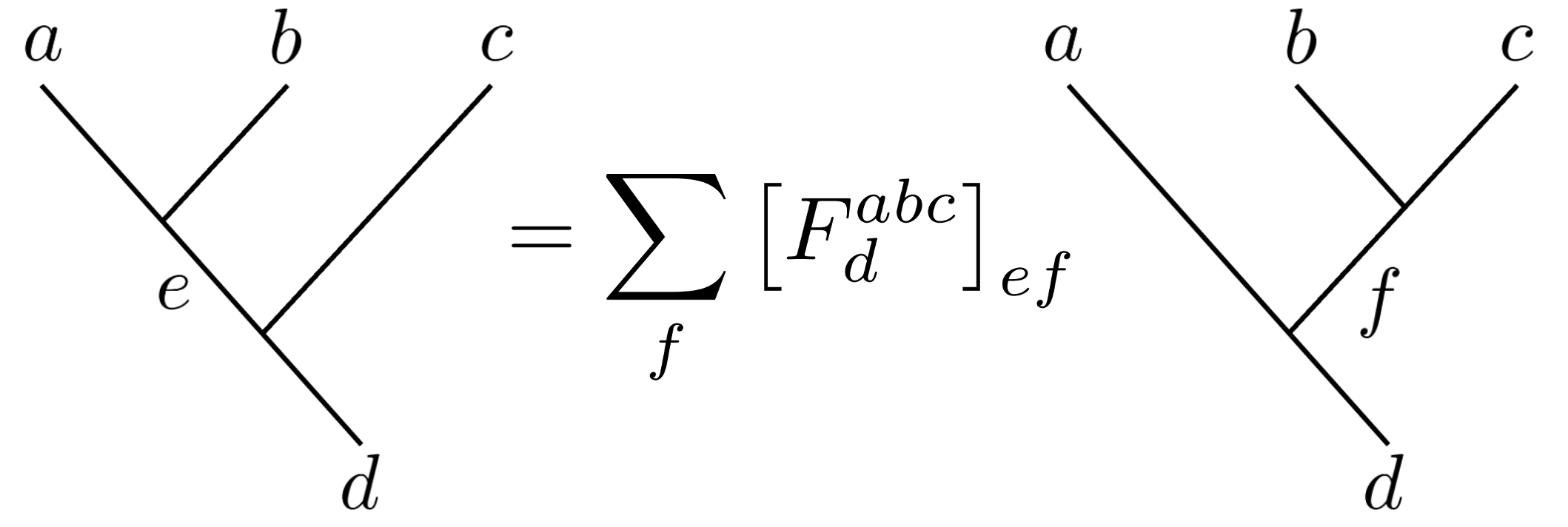}
	\caption{The associativity relation defines the F symbols $\left[F^{abc}_{d}\right]_{ef}$. And the corresponding 6j symbol is defined as $\left[\begin{smallmatrix}
		a & b & e  \\
		c & d & f
		\end{smallmatrix}\right]=\left[F^{abc}_{d}\right]_{ef}/\sqrt{d_ed_f}$. The details can be found in section 2 of \cite{Aasen:2020jwb}. This type of change of a diagram is typically referred to as an F-move.}
	\label{fig:6j}
\end{figure}

\begin{figure}
	\centering
	\includegraphics[width=0.4\linewidth]{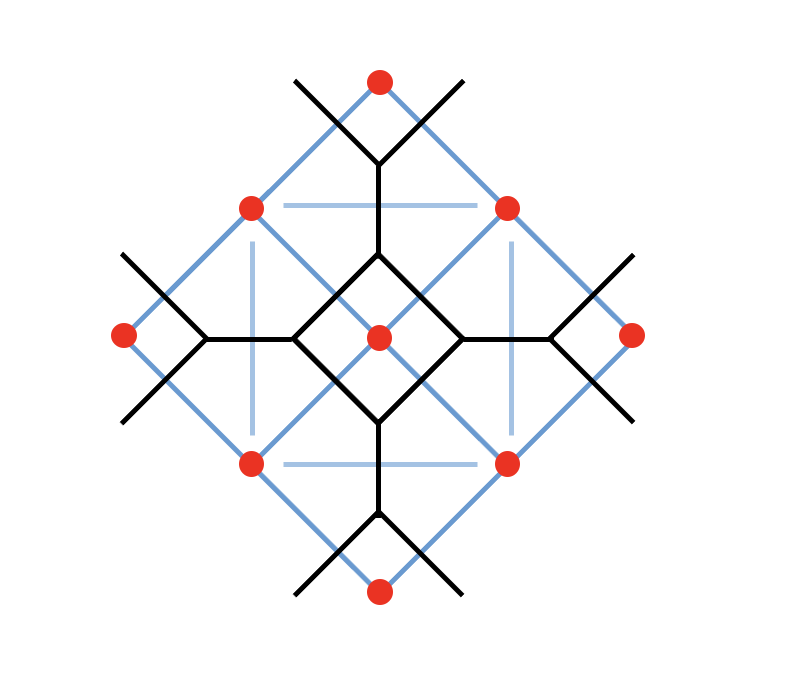}
	\caption{Tensor network representation of the ground state wave-function. The blue lines are ``physical sites'' of the topological order, and each colored by $i \in \mathcal{C}$.
	The red dots represent edges orthogonal to the $S_2$ surface which are weighted by a factor given in (\ref{eq:dweight}), and they are summed. The black lines are the dual graph of the blue lines.}
	\label{fig:strange_correlator}
\end{figure}

The story can be readily generalised to 3D. Consider a Turaev-Viro type topological theory associated to a fusion category $\mathcal{C}$ (section 2 of \cite{Aasen:2020jwb} provides a brief and clear introduction of fusion categories).
To generate the ground state wavefunction on a two-sphere $S_2$, we consider the path-integral on a three-ball which is triangulated into tetrahedra. A convenient choice  is chosen such that the two dimensional cross section would take the same form as in the 2D case shown in figure \ref{fig:2dDW} above, i.e. the $S_2$ is covered by  triangles whose edges are marked by blue lines in figure  \ref{fig:strange_correlator}, and each vertex of the triangle is attached an edge that extends into the ball and ends at the center of the ball. 
These radial edges are marked by red dots in figure  \ref{fig:strange_correlator}.  Therefore each surface triangle together with three red dots form a tetrahedron. Each edge of the triangulation is assigned an object of $\mathcal{C}$. Each tetrahedron is assigned the value of the 6j symbol. The 6j symbols of a fusion category $\mathcal{C}$ are defined through associativity relation in the fusion of objects of $\mathcal{C}$, as shown in figure \ref{fig:6j}. 
The object $a$ assigned to each radial edge (i.e. a red dot) has to be summed, weighted by a factor 
\be  \label{eq:dweight}
w_a=d_a S_{00}^{3/2},
\ee
where $d_a$ is the quantum dimension of individual object $a \in \mathcal{C}$,
and $S_{00}$ is a component of the  S- modular matrix, and the label $0$ labels the identity object of the category $\mathcal{C}$. 
It is well known that $S_{00} = 1/\mathcal{D}$, where $\mathcal{D}$ is the quantum dimension of $\mathcal{C}$ defined as
\be
\mathcal{D} \equiv \sqrt{\sum_{i\in \mathcal{C}} d_i^2}.
\ee

\begin{figure}
	\centering
	\includegraphics[width=1.0\linewidth]{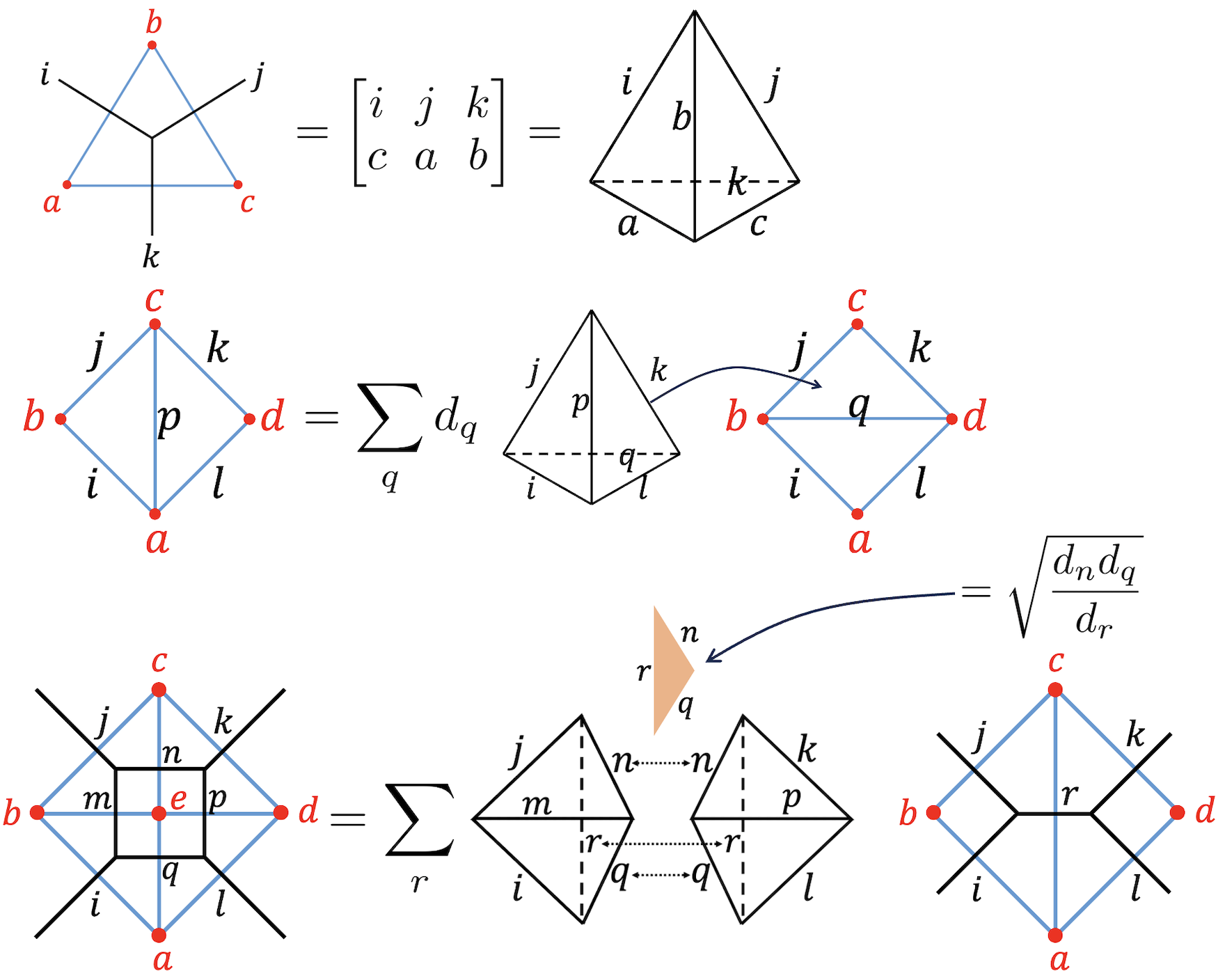}
	\caption{Values of the tensors are given by the 6j symbols (graphically we draw tetrahedron to represent it). The pentagon equations correspond to stacking an extra tetrahedron onto the surface of other tetrahedra, changing the surface triangulation. 
	Again the black lines are the dual graph of the blue lines.}
	\label{fig:Mong}
\end{figure}

This reproduces the tensor network representation of ground state wave-functions described for example in \cite{K_nig_2009,Luo:2016leh,  Williamson:2017uzx,  Aasen:2020jwb, Frank1}
and it is illustrated diagramatically as in figure \ref{fig:strange_correlator}. These triangles are tensors related to the 6j symbols. The precise relation is illustrated in figure \ref{fig:Mong}, with important relations they satisfy that follow from the pentagon equations.

Sumarising the above, the ground state wave-function for a given surface triangulation is given by
\be\label{Psistate}
|\Psi\rangle=\sum_{\{a_v\}}\sum_{\{i\}}\prod_e d^{1/2}_i\prod_v w_a \prod_{\triangle} 
\begin{bmatrix}
    i & j & k  \\
    c & a & b
    \end{bmatrix}|\{i\}\rangle.
\ee
The ket $|\{i\}\rangle$ are basis states living on the edges of triangles on the surface, with  $i\in \mathcal{C}$.
The factor of $d_i$ on each surface edge is the quantum dimension of the object $i$. They follow from normalisation of the surface edges commonly used in the literature.
The factor $w_a$ is the weight of the object $a$ living on each corner (red dot) already introduced in (\ref{eq:dweight}). 
The factors $\left[\begin{smallmatrix}
    i & j & k  \\
    c & a & b
    \end{smallmatrix}\right]$ are the 6j symbols that appeared in figure \ref{fig:Mong}.

\begin{figure}
	\centering
	\includegraphics[width=1.0\linewidth]{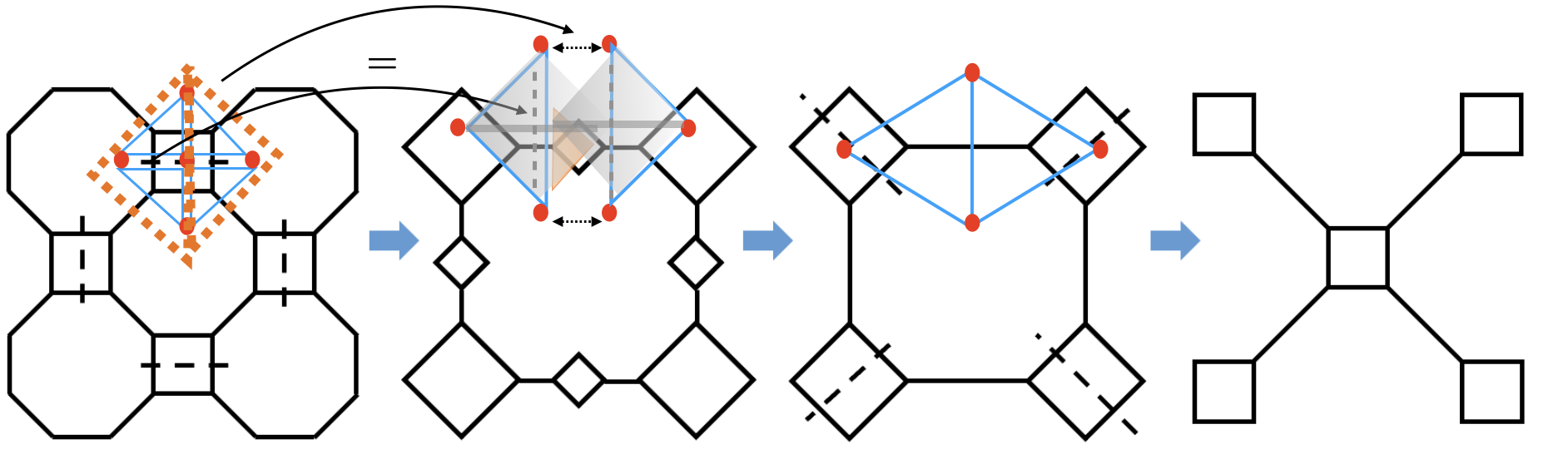}
	\caption{Coarse graining using the pentagone relations. The RG operator can be visualised as a collection of tetrahedra. Using the relations in figure \ref{fig:Mong}, we demonstrate the set of factors (grey pair of tetrahedra and a factor of quantum dimensions) needed to convert four triangles into two triangles.   }
	\label{fig:RG}
\end{figure}

Similar to the 2D case discussed above, the 3D RG operator is constructed by relating two boundaries with different lattice spacings.
 This can be done by considering a sequence of F-moves and making use of the relations in figure \ref{fig:Mong}. One choice of the blocking sequence is given in figure \ref{fig:RG} \cite{Frank1} which is represented by the dual graph.

One can express this sequence of blockings as a stack of tetrahedra. This is also illustrated in figure \ref{fig:RG}. The RG operator  $U(\mathcal{C})$ in this case, would be the collection of grey tetrahedra and yellow triangles collected in taking the wave-function from the $N$-th RG step to the $N+1$-th step.  One can see that $U(\mathcal{C})$ is determined purely by the topological data of the fusion category $\mathcal{C}$. Recursive application of the RG steps would result in a collection of tetrahedra that discretize a Euclidean AdS space. One can consider a ``vertical cross-section'' of this RG operator, cutting along a line from the UV fine-grained layer down towards the IR layer. The cross section would be triangulated like the tree as in figure \ref{RG2d} observed for the RG operator in a bulk 2D theory. 

\subsection{Frobenius Algebra gives Topological Fixed Points}
\begin{figure}
	\centering
	\includegraphics[width=0.6\linewidth]{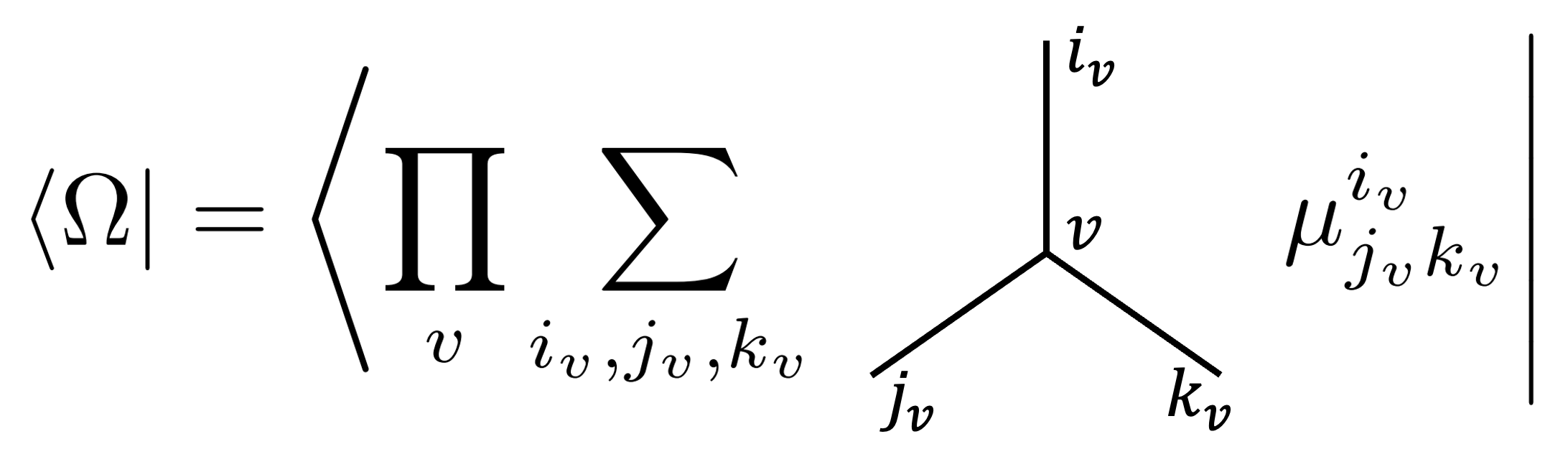}
	\caption{The bra-state $\langle \Omega \vert $ that follows from a Frobenius algebra is constructed such that every vertex on the 2D surface is weighted by the product coefficient $\mu$ defining the algebra. }
	\label{fig:Omega_gapped}
\end{figure}

There is a class of eigenstates of $U(\mathcal{C})$.
As it is well known, gapped boundaries of 2+1 dimensional topological orders described by say Levin-Wen models are classified by separable Frobenius algebra \cite{Fuchs:2002cm} of the {\it input} fusion categories \cite{Bullivant:2017qrv, Hu:2017faw, Hu:2017lrs}. It is thus very tempting to look for partition functions of these gapped boundaries making use of knowledge of the Frobenius algebra of the category $\mathcal{C}$.
The solution is given by
\bea\nonumber
\langle \Omega^\mu_\Lambda| &=&\sum_{\{i\}}\langle \{i\}|\prod_\triangle T^{ijk}_{\mu} \\
 &=& \prod_v  \prod_{l \sim v} \sum_{\{i_l\}} \langle i_l | T_{\mu}(v),
\eea
where $T_{\mu}(v)$ is a projector of every three-edged vertex which will be related to the (co-) product of a Frobenius algebra $A$. The sum over $\{i_l\}$ sums over the objects at edge $l$ connected to the vertex $v$, i.e. the partition function is such that we put an extra weight on each vertex. This is illustrated in figure \ref{fig:Omega_gapped}.

\begin{figure}
	\centering
	\includegraphics[width=0.6\linewidth]{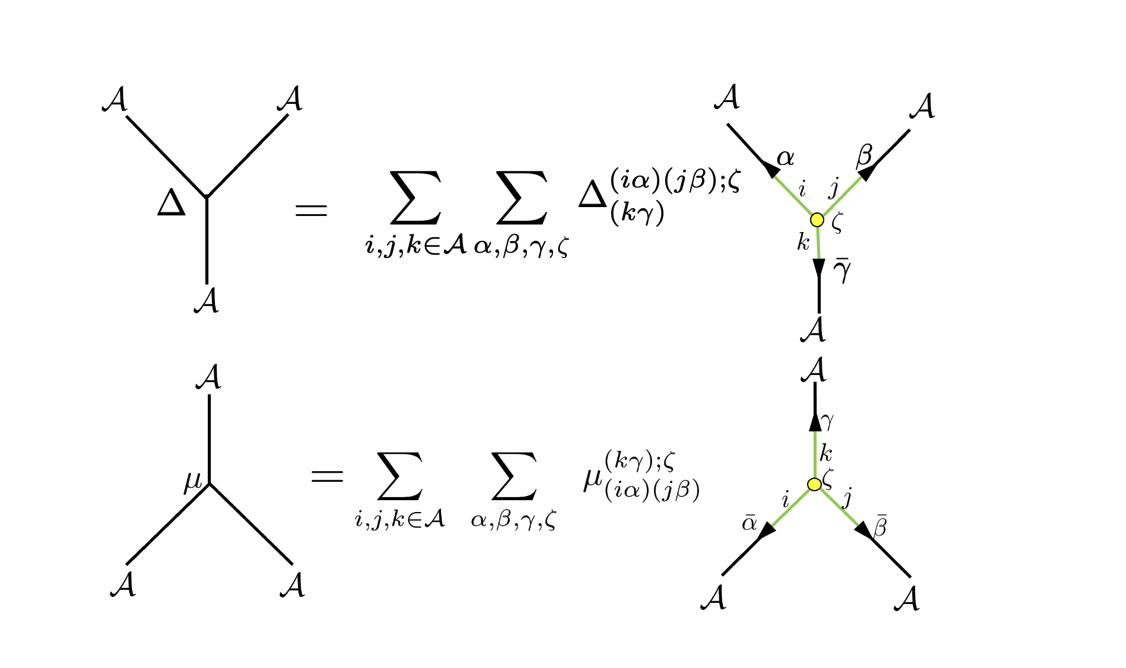}
	\caption{The product $\mu$ and co-product $\Delta$ of a Frobenius algebra expressed  in basis form. }
	\label{fig:Frobenius_basis}
\end{figure}

The product and co-product  of a Frobenius algebra is expressed as in figure \ref{fig:Frobenius_basis}.

For simplicity, where each object is its own dual, it is possible to have the co-product equal to flipping the product, and in such a case, we do not have to distinguish them. In this simple case, we can then require that the value assigned to each triangle is equated with the product, i.e.
\be T_{\mu} = \mu.
\ee
In the general case we need to specify the orientation on every edge and then distinguish the product from the co-product.

\begin{figure}
	\centering
	\includegraphics[width=0.6\linewidth]{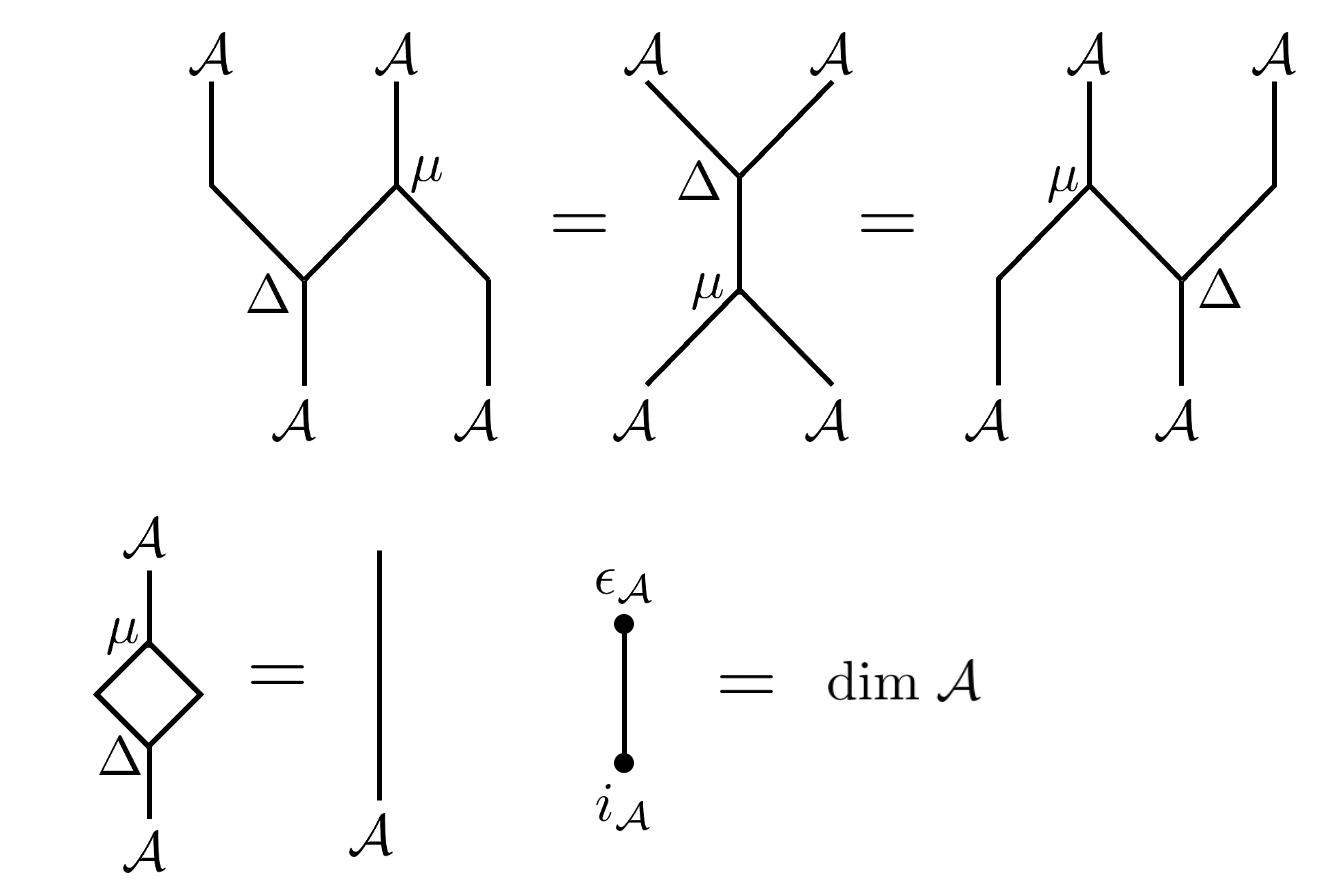}
	\caption{The upper half depicts the associativity of a Frobenius algebra. The lower half depicts the condition of  separability. }
	\label{fig:frobenius_separable}
\end{figure}

Since the algebra is Frobenius and separable -- as depicted in figure \ref{fig:frobenius_separable}, one can readily see that locally, going through the steps depicted in figure \ref{fig:RG}, $\langle \Omega^\mu|$  and $ \langle \Omega^\mu| U(\mathcal{C})$,  look exactly the same -- every vertex is still weighted by the same weight characterized by the product map $\mu$ of the Frobenius algebra $A$.
Every Frobenius algebra of $\mathcal{C}$ gives a topological 1+1 D TQFT with symmetry characterized by categorical symmetry $\mathcal{Z}(\mathcal{C})$.
We note that $\mathcal{Z}(\mathcal{C})$ is spontaneousely broken in each of these TQFT's. The order parameter in terms of defect operators can be constructed and will be presented in a different paper.

We note that every separable Frobenius algebra in the input category corresponds to a {\it Lagrangian algebra} in the output category associated to the topological order. Each Lagrangian algebra describes a spatial gapped boundary condition of the topological order which is physically prescribing a collection of anyons condensing at the boundary. We note however, that different Frobenius algebra of the input category might be mapped to the same or isomorphic Lagrangian algebra in the output category which is often considered as physically equivalent in the literature \cite{Hu:2017lrs}. As we are going to see below however, they lead to different 2D TQFT through the strange correlator, and there are CFT's describing phase transitions between them.

\subsection{CFTs from phase transitions between fixed points} \label{sec:num1}
\begin{figure}
	\centering
	\includegraphics[width=1.0\linewidth]{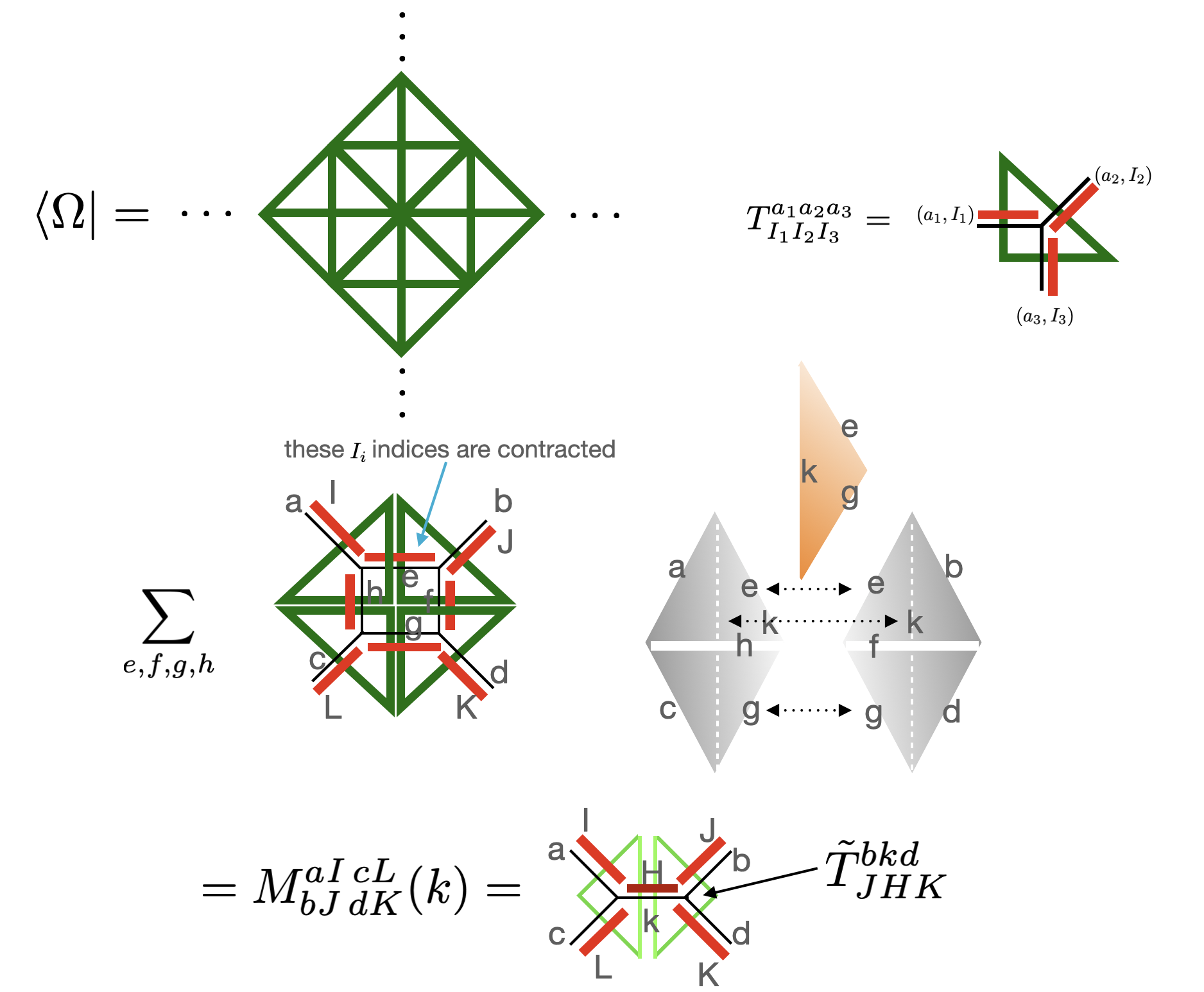}
	\caption{The top figure illustrates the PEP construction of the boundary state $\langle \Omega |$. The indices lives on the edges of the triangles (or edges on the dual graph). Each edge carries a pair of indices, $a_i \in \mathcal{C}$  and $I_i$ are auxiliary indices with bond dimension $\eta$. In the bottom figure, we break down the tensor $M(k)$ into the product of two $\tilde T$'s (light green triangles). This breaking down can be achieved by SVD. The new auxiliary index $H$ (colored dark red) is generically of dimension $\eta^2$, but we would cut it off down to dimension $\eta$.  }
	\label{fig:3dblocking}
\end{figure}

The most interesting question is to construct partition functions of critical points. A critical point is reached when non-commuting topological symmetries are preserved at the same time \cite{Levin:2019ifu, Ji:2019jhk, Chatterjee:2022tyg, Chatterjee:2022kxb}. Since $Z(\mathcal{C})$ is supposedly a local partition function, we expect that $\langle \Omega |$ should admit a PEPS tensor network construction. Inspired by the form of the topological solution above where $\langle \Omega|$ is essentially a direct product of states defined on each triangle, we construct an ansatz where each tensor covers a triangle with 6 legs i.e. $T^{a_1a_2a_3}_{I_1I_2I_3}$, i.e. each edge of the triangle carries a pair of indices, where $a_i\in \mathcal{C}$ and $I_i$ are auxilliary indices with bond dimension $\eta$.
In principle one could constrain the form of $T^{a_1a_2a_3}_{I_1I_2I_3}$ by imposing that these open topological string operators die off at most with a power-law as a function of distance between the end points. In practice, checking for power law decay  is highly non-trivial -- it suggests that $\eta$ should scale with system size and approach infinity for an infinite system. We will adopt a different approach to extract critical point data. Consider a single RG step $\langle \Omega(T) | U(\mathcal{C})$. As demonstrated in figure \ref{fig:RG}, four triangles are mapped to two triangles with entangled boundary conditions. By applying for example, singular value decomposition (SVD) and keeping a fixed bond dimension for auxiliary indices $I_i$, one can rewrite the products of four $T$'s as a contraction of two new tensors $\tilde T$, as shown in the figure \ref{fig:3dblocking}.
For finite $\eta$, the critical point is unstable and the sequence of $T$'s obtained from the RG process would eventually flow to one of the topological fixed points. By appropriate parametrisation of $T$, we would be able to find out critical points where the RG process begins to be confused about which topological fixed point to go to. For every fixed $\eta$ it is thus possible to obtain $T$'s that are best approximating the critical state.

\subsection{Some Explicit Examples} \label{sec:num2}

\subsubsection{The A-series minimal CFTs}
\begin{figure}
	\centering
	\includegraphics[width=0.8\linewidth]{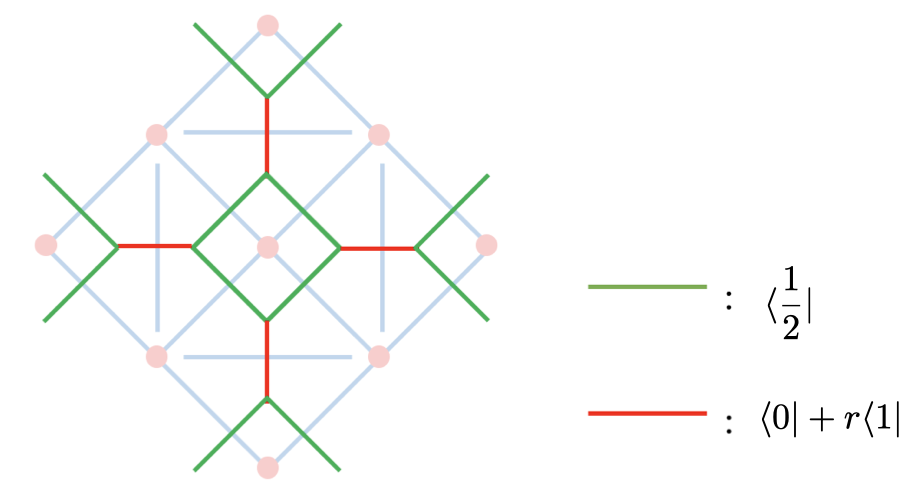}
	\caption{A family of boundary conditions for $SU(2)_k$ models with one parameter $r$. At $k=2$ one recovers the 2D Ising model at temperature $e^{-2\beta}=r$. }
	\label{fig:Isingbc}
\end{figure}

We can demonstrate in a class of examples constructed from the fusion category formed by representations of the quantum group $SU(2)_k$.
At level $k$, there are $k+1$ objects $j=0,1/2,\cdots, k/2$. They satisfy the fusion rule

\be
j_1 \otimes j_2 = |j_1- j_2| \oplus (|j_1-j_2| + 1) \oplus \cdots \oplus \textrm{min}\{j_1+j_2, k - j_1 - j_2\}.
\ee
In particular, one sees that $1/2 \otimes 1/2 = 0 \oplus 1 $.  Consider a family of boundary state $\la \Omega|$ taking the form
\be
\la\Omega|=\sum_{\{a\}}\la \{a\}|\prod_{\D_2}T_{\D_2},
\ee
where the summation is over all the configurations, and each triangle on the boundary contributes $T_{\D_2}$.
The boundary condition is such that each $T_{\D_2}$ attached to a triangle $\D_2$  is given by the three index tensor $T^{a_1a_2a_3}$ chosen to be
\be  \label{eq:diag_cft}
T^{a_1a_2a_3}=\left\{
 \begin{aligned}
 1&,& (a_1=a_3=\frac{1}{2},a_2=0),\\
 r&,& (a_1=a_3=\frac{1}{2},a_2=1),\\
 0&,& (\textrm{Otherwise}).
 \end{aligned}
 \right.
\ee
This is depicted in figure \ref{fig:Isingbc}.  Comparing with the top of figure \ref{fig:3dblocking}, this corresponds to the case where the extra index $I_i$ attached to an edge is trivial ( i.e. bond dimension 1).  

For every $k$,  there is a critical coupling $r= r_c(k)$ which is known to recover in the thermodynamic limit, the A-series minimal CFT. We would like to recover the critical coupling using the method we discussed above i.e. allowing the tensor $T$ to flow under repeated use of the RG operator.
Starting with this boundary state, we carry out the RG process described in figure \ref{fig:3dblocking} with bond dimension for the auxiliary indices kept, as a very crude approximation, at $\eta=1$. And we find that for small $r$, the boundary state flows to a fixed point where $T^{a_1a_2a_3}$ only has one non-vanishing component $T^{000}$, i.e. all the boundary legs are projected to the identity element, denoted 0 here. This is of course a separable Frobenius algebra for any fusion category. We call that $\mathcal{A}_0$. For bigger $r$, the boundary state flows to a fixed point where $T^{a_1a_2a_3}$ has the following non-vanishing components $T^{000}=T^{110}=T^{101}=T^{011}$ and
$T^{111}$ (this component appears when $k>2$). The ratio $T^{000}/T^{111}$ is a function of $k$. When $k=3$, $T^{000}/T^{111}=1.43463$. When $k=4$, $T^{000}/T^{111}=1.18921$.  We note that this coincides precisely with a family of separable Frobenius algebra of $SU(2)_k$ category. The objects in the algebra is given by $\mathcal{A}_1=\{0,1\}$, and
the ratio of the product of the algebra $T^{000}_{\mu(k)}/T^{111}_{\mu(k)}$ for arbitrary $k>2$ is given by
\be
\frac{T^{000}_{\mu(k)}}{T^{111}_{\mu(k)}} =  \frac{\sqrt[4]{2 \cos \left(\frac{2 \pi }{k+2}\right)+1}}{\sqrt{2\cos \left(\frac{2 \pi }{k+2}\right)}}.
\ee
The separable Frobenius algebra is generated by the RG algorithm!

\begin{figure}
	\centering
	\includegraphics[width=1\linewidth]{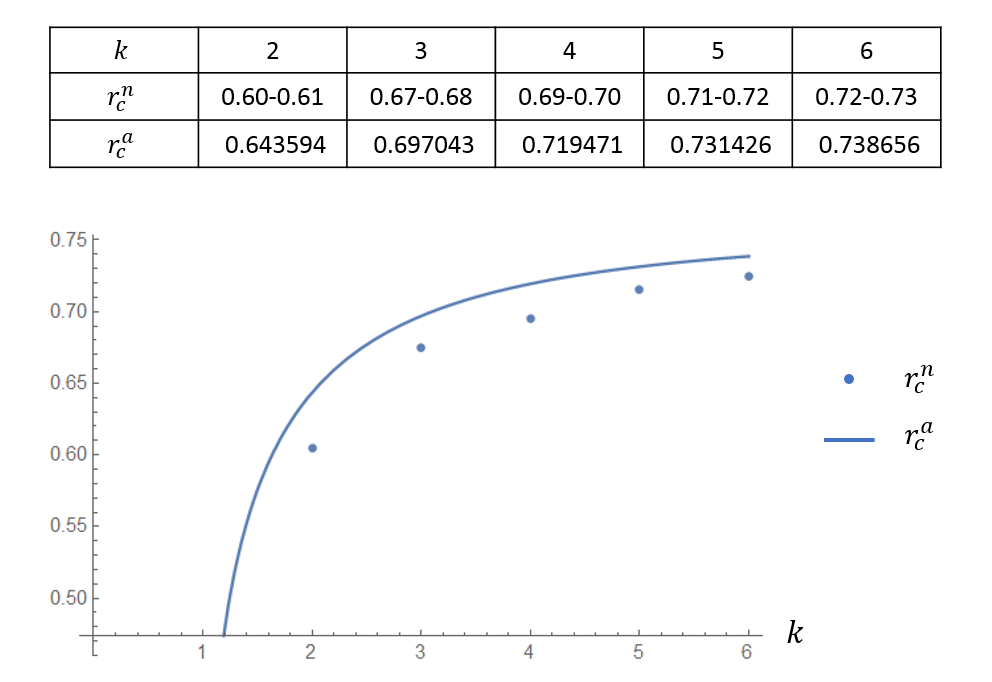}
	\caption{Here $r^n_c$ is the numerical result for the phase transition point, and $r^a_c$ is the analytical result which is given by
$r^a_c=\sqrt{\frac{\sqrt{2 \cos \left(\frac{2 \pi }{k+2}\right)+1}}{2 \cos \left(\frac{\pi }{k+2}\right)+1}}.$
}
	\label{Ak}
\end{figure}

The critical coupling $r_c(k)$ determining the phase transitions between the two fixed points characterized by the algebra $\mathcal{A}_0$ and $\mathcal{A}_1$ are determined numerically. We compare our numerical results with the analytical one in figure \ref{Ak}. At the critical point, Kramers-Wannar duality is preserved in this series of models \footnote{This is known for a long time. For a recent discussion see for example \cite{Aasen:2020jwb}.} and the critical couplings can thus be readily obtained as benchmarks.  The numerical results recover the analytical result to about 1 significant figure. This is quite remarkable given that we adopted such a brutal truncation with auxiliary bond dimension 1. The method can be readily adopted with larger bond dimensions and is currently work underway.

Here we would like to note that the CFT is a phase transition point between two topological fixed points of the RG operator described by two Frobenius algebras in the input category $\mathcal{C}$.
For example in the case of $k=2$, these two algebras $\mathcal{A}_0$ and $\mathcal{A}_1$ correspond to the same set of anyon condensations in the output category (the topological order the lattice model describes) \cite{Hu:2017lrs}.  Specifically, the output category is a doubled Ising topological order where there are 9 topological sectors that can be labeled by the pair $(j_1, \bar{j}_2)$, where $j_{1,2} \in \{0,1/2,1\}$. Here, both Frobenius algebra in the input category, $\mathcal{A}_0$ and $\mathcal{A}_1$, correspond to the same condensable Lagrangian algebra in the output category, namely $(0,\bar 0) \oplus (1/2, \bar{1/2}) \oplus (1, \bar 1)$. However,  we can see here that physically they lead to different 2D topological fixed points and a physical phase transition between them is exactly the 2D Ising CFT connects them, which is a point observed also in \cite{Ji:2019jhk}. The two topological fixed points are interpreted as the electric and magnetic condensate respectively if we do not include the Kramers-Wannier duality in the categorical symmetry, leaving the bulk reduced to the 2+1 D toric code. (The doubled Ising bulk and the toric code are related by anyon condensation too -- condensing $(0,\bar{0}) \oplus (1,\bar 1)$\cite{Bais:2008ni}.) In terms of categorical symmetry, the toric code version breaks spontaneousely the Kramers-Wannier duality that is kept explicit in the doubled Ising bulk. One should thus take extra care when seeking phase transition points between topological fixed points. The condensable algebra of the output category might undercount the number of fixed points, unless we take into account isomorphic algebras with the same collection of condensing anyons as physically distinct.

\subsubsection{A curious example}
\begin{figure}
	\centering
	\includegraphics[width=1\linewidth]{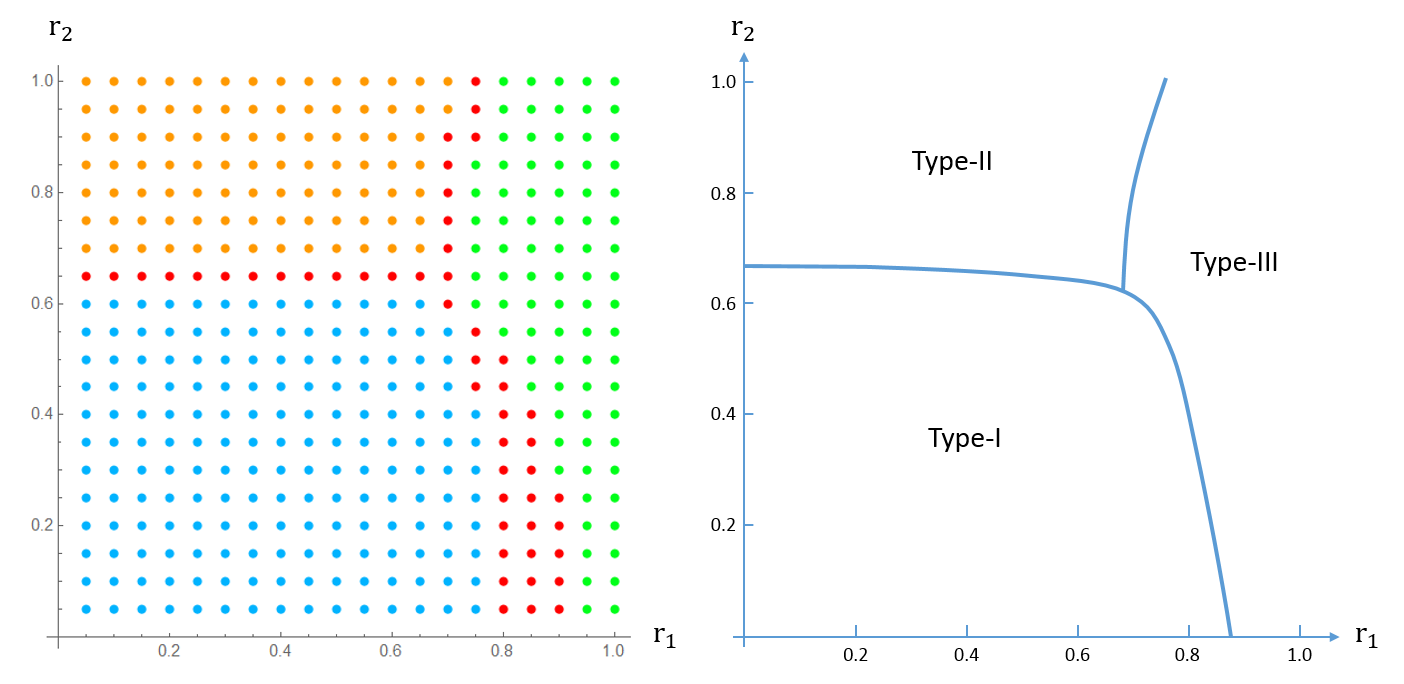}
	\caption{The left hand side is our numerical result. The blue/orange/green dots correspond to type I/II/III fixed points. The red dots are those we can't tell its fixed point type within 10 RG steps. According to the numerical result, we can obtain a phase diagram shown on the right hand side.
}
	\label{phasediagram}
\end{figure}

When $k=4$, we also consider the case that $T^{a_1a_2a_3}$ is given by
\be \label{eq:curiousbc}
T^{a_1a_2a_3}=\left\{
 \begin{aligned}
 1&,& (a_1=a_3=1,a_2=0),\\
 r_1&,& (a_1=a_3=1,a_2=1),\\
 r_2&,& (a_1=a_3=1,a_2=2),\\
 0&,& (\textrm{Otherwise}).
 \end{aligned}
 \right.
\ee
 We carry out the RG process described in figure \ref{fig:3dblocking} with bond dimension $\eta=1$, and find that the boundary state can flow to 3 different fixed points when we change $r_1,r_2$. In the type I fixed point, $T^{a_1a_2a_3}$ only has one non-vanishing component $T^{000}$, i.e. all the boundary legs are projected to 0. In the type II fixed point, $T^{a_1a_2a_3}$ has following non-vanishing components $T^{000}=T^{022}=T^{202}=T^{220}$. In the type III fixed point, $T^{a_1a_2a_3}$ has following non-vanishing components $T^{000}=T^{011}=T^{101}=T^{110}=T^{022}=T^{202}=T^{220}=T^{112}=T^{121}=T^{211}$ (the component $T^{111}$ is allowed by the fusion rules but is 0). And the phase diagram is shown in figure \ref{phasediagram}. The triple point is around $(r_1,r_2)=(0.7,0.6)$.

This family of boundary conditions parametrised as in (\ref{eq:curiousbc}) is mentioned in \cite{Aasen:2020jwb}, belonging to the integrable family of "19-vertex models".
The three topological fixed points correspond to three different Frobenius algebra in the input category. Each of them correspond to a Lagrangian algebra of the topological order which determines the collection of anyons (or topological defects) that condenses at the 2d boundary. From \cite{ostrik, Fuchs:2002cm} one can see that type I fixed point corresponds to condensation of $\mathcal{A}_I = \oplus_{j =\{0,1/2,1,3/2,2\}}  j \boxtimes \bar j$.
The type II fixed point corresponds to  $\mathcal{A}_{II} =   (0\oplus 2) \boxtimes (\bar 0\oplus  \bar 2) \oplus 2 (1\boxtimes \bar{1})$.  The type III fixed point involves the condensation of the same set of defects as $\mathcal{A}_I$. However, it is a different Lagrangian algebra. This is the analogue of the $k=2$ case discussed in the previous example where there are two isomorphic algebras characterising topological boundary conditions. While both correspond to a diagonal condensation, physically they describe two different condensation -- namely condensing "electric" and "magnetic" defects.  and they are related by an electromagnetic duality.

We note that in the three topological condensates, it is clear that they definitely share the subset of $\mathcal{A}_{sub}= (0 \boxtimes \bar 0) \oplus (2 \boxtimes \bar 2)$. All the three boundary conditions can be considered as a sequential condensation, first condensing $\mathcal{A}_{sub}$. Therefore we conclude that the tri-critical point should also naturally have $\mathcal{A}_{sub}$ condensed.  It has a reduced categorical symmetry compared to the critical point considered in (\ref{eq:diag_cft}).

We will leave working out further details of the CFT in future works.

\subsection{Exact Eigenstates corresponding to RCFTs } \label{sec:exactcft}
In the previous section, we searched for CFTs by interpolating between topological eigenstates of the RG operator. 
In this section we will demonstrate that exact eigenstates of the RG operator can be constructed, leading to an exact tensor network representation of path-integrals of RCFTs.  For each RCFT, their primaries admit holomorphic- anti-holomorphic factorisation, so that the Hilbert space can be decomposed as
\be
H = \oplus_{a,\bar a} \Lambda_{a, \bar a}  V_a \otimes V_{\bar a},
\ee
where $\Lambda_{a,\bar a}$ are positive integers, and $V_a$ and $V_{\bar a} $ label the vector spaces of the holomorphic and anti-holomorphic conformal primaries, with labels $a $ and $ \bar a$ given by objects in a (modular) fusion category $\mathcal{C}$. 
Our discussion is applicable to general RCFTs that admit such a factorisation, but for concreteness, we will focus on the case where $\Lambda_{a, \bar a} = \delta_{a, \bar a}$. These are called diagonal RCFTs. 
%
We will demonstrate that the RCFT path integral/ partition function can indeed be expressed as a strange correlator, 
\be \label{eq:rstrange}
Z_{\textrm{RCFT}}=\langle\Omega|\Psi\rangle,
\ee 
where $|\Psi\rangle$ is the ground state of the Levin-Wen model
corresponding to the fusion category $\mathcal{C}$ characterising the CFT, and whose explicit expression was given in (\ref{Psistate}),
and 
\be\label{Omegastate}
\langle\Omega|=\sum_{\{(i,I)\}}\langle\{i\}|\prod_{\triangle}(\gamma^{ijk}_{IJK}S_{00}^{1/4})
\ee 
which is an exact eigenstate of the RG operator defined in $\textrm{TQFT}_3$.  
Here $\gamma^{ijk}_{IJK}$ are constructed from three-point conformal blocks and play the role of fixed point tensor
with $i, j, k$ representing the primaries, and $I, J, K$  representing the descendants of their respective primaries.

\begin{figure}
	\centering
	\includegraphics[width=0.35\linewidth]{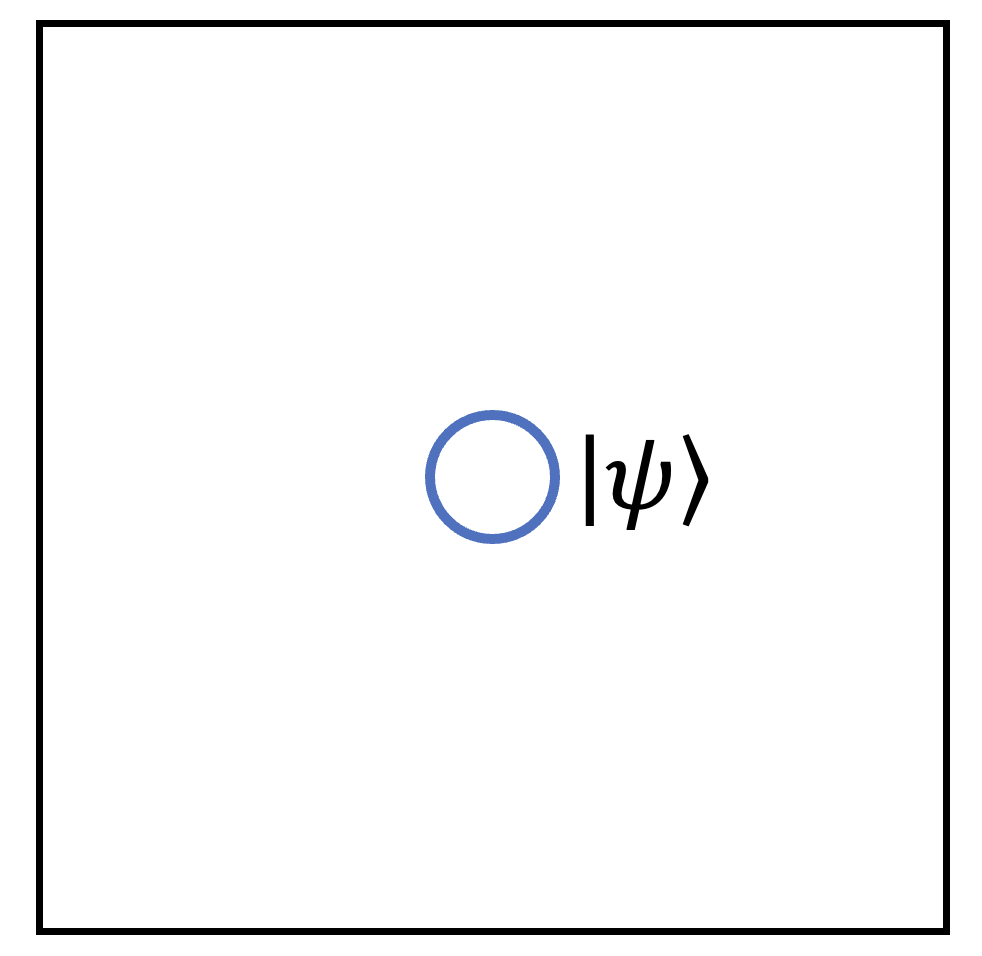}
	\caption{Open a hole on the plane with boundary condition $|\psi\rangle$.}
	\label{onehole}
\end{figure}

The construction is based on the use of contractible conformal boundary conditions first discussed in \cite{Hung:2019bnq}, 
which was introduced as appropriate boundary condition when computing entanglement entropy. These will be introduced below. 
Let us start by considering a rational conformal field theory on the manifold $\mathcal{M}$ whose partition function is denoted 
by $Z_{\mathcal{M}}$. We can open a hole with radius $r$ on this manifold, and set the 
boundary condition by choosing a state $|\psi\rangle$ as shown in figure \ref{onehole}. One notes that when 
$|\psi\rangle=|0\rangle \;\textrm{and}\; r\to 0 $, we can recover the partition function 
$Z_{\mathcal{M}}$. However the vacuum state is not a conformal boundary condition.  The closest alternative is the Ishibashi state $|0\rangle\rangle$, which includes 
$|0\rangle$, and it is a superposition of Cardy states. Choosing $|\psi\rangle=|0\rangle\rangle$,
the descendants of the vacuum state contained in the Ishibashi state would contribute to irrelevant perturbations in the CFT, which would approach 0 as $r \to 0$.  This partition function defined on a lattice of holes thus recovers the exact partition on a smooth plane  $Z_{\mathcal{M}}$. The relation between Cardy states and Ishibashi states is 
given by:
\be
|a\rangle=\sum_{i\in \mathcal{C}}\frac{S_{ai}}{\sqrt{S_{0i}}}|i\rangle\rangle,
\ee 
where $S_{ab}$ is the modular S matrix relating characters evaluated on a torus with modulus $\tau$ and $-1/\tau$ : 
\be
\chi_a(e^{-2\pi i/\tau})=\sum_{b\in \mathcal{C}} S_{ab}\chi_b(e^{2\pi i\tau}).
\ee

\begin{figure}
	\centering
	\includegraphics[width=0.8\linewidth]{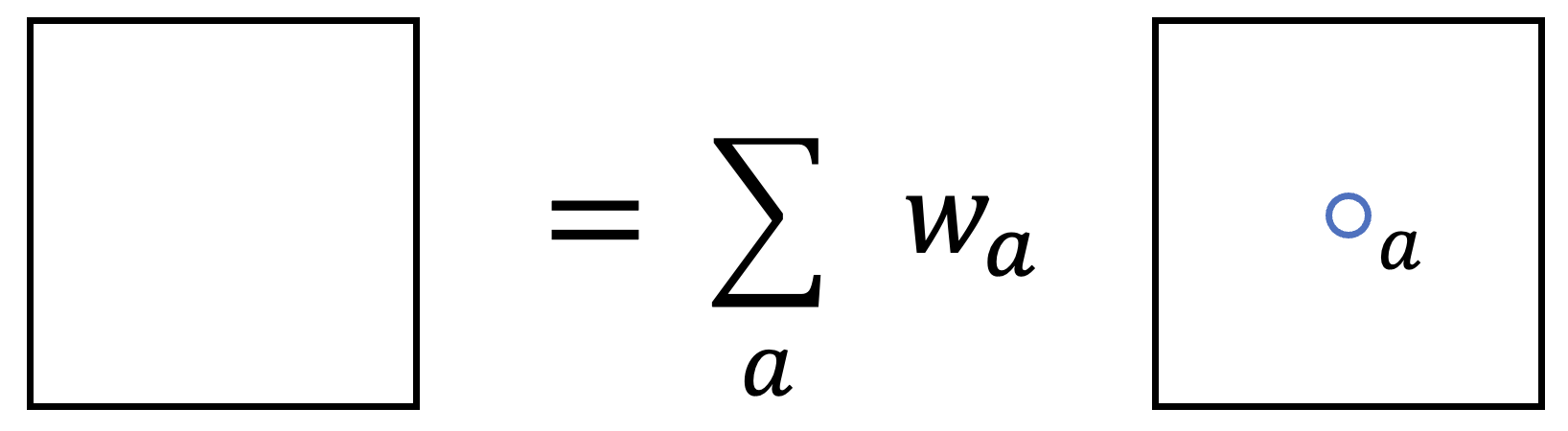}
	\caption{Sum over $a$ to remove the infinitesimal hole.}
	\label{Suma}
\end{figure}

After some calculations, we can get:
\be
|0\rangle\rangle=\sum_{a\in \mathcal{C}} S_{a0}\sqrt{S_{00}}|a\rangle,
\ee
which is exactly the entanglement brane boundary conditions discussed in \cite{Hung:2019bnq}.
Let us write it as
\be \label{eq:weight1}
|0\rangle\rangle=\sum_{a\in \mathcal{C}} w_a|a\rangle,\;\;w_a\equiv S_{a0}\sqrt{S_{00}}=d_a S_{00}^{3/2},
\ee
where $d_a=S_{a0}/S_{00}$ has been used.
Now we can rewrite $Z_{\mathcal{M}}$ as 
\be
Z_{\mathcal{M}}=\sum_a w_a Z_a,
\ee
with $Z_a$ denoting the partition function on $\mathcal{M}$ with an infinitesimal hole whose 
boundary condition is the physical boundary condition $|a\rangle$ as shown in figure \ref{Suma}.

\begin{figure}
	\centering
	\includegraphics[width=0.85\linewidth]{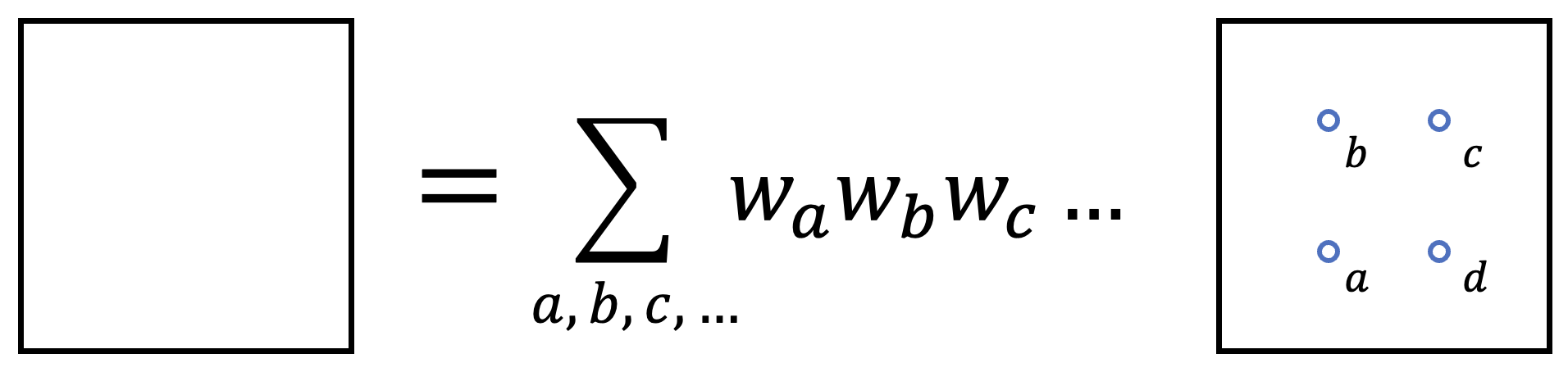}
	\caption{Sum over $a,b,c,\dots$ to remove the infinitesimal holes.}
	\label{Many_holes}
\end{figure}

Having this, we can insert more infinitesimal holes and get
\be
Z_{\mathcal{M}}=\sum_{a,b,c,\dots} w_a w_b w_c \dots Z_{a,b,c,\dots},
\ee
where $Z_{a,b,c,\dots}$ denotes the partition function on $\mathcal{M}$ with many infinitesimal holes whose 
boundary conditions are labelled by $a,b,c,\dots$. See figure \ref{Many_holes}. 

\begin{figure}
	\centering
	\includegraphics[width=0.35\linewidth]{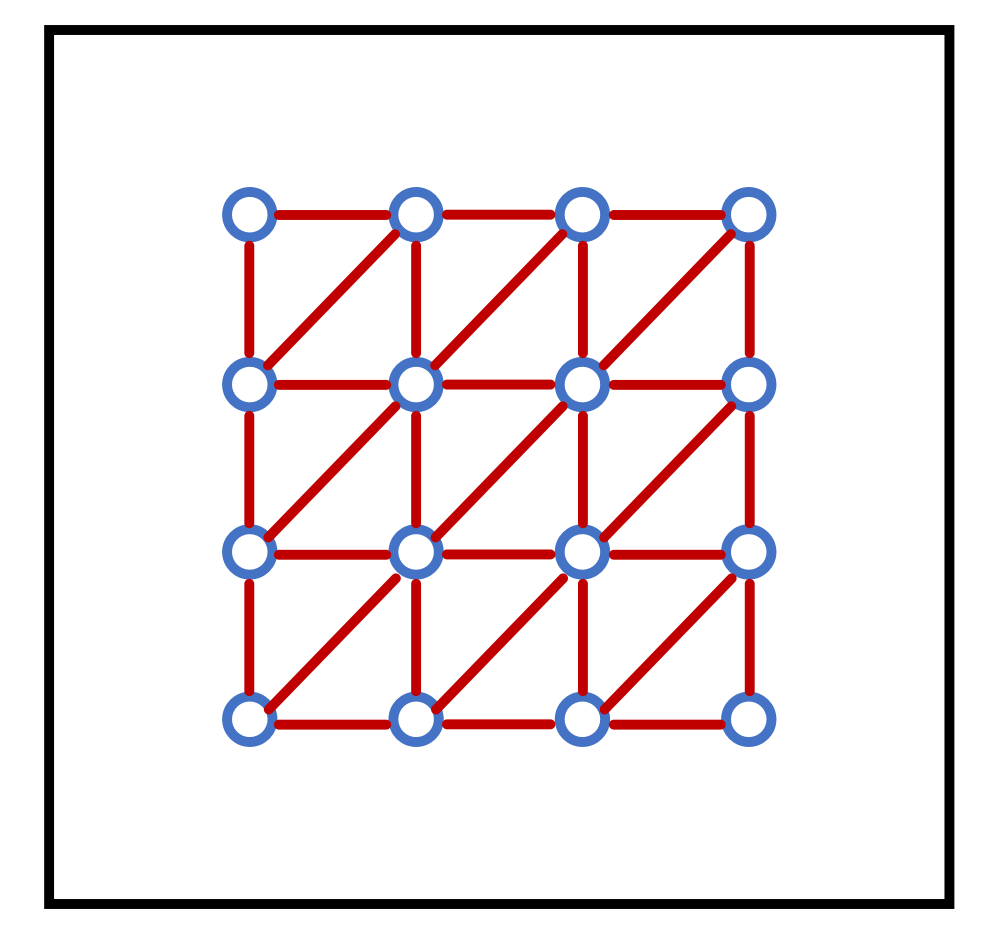}
	\caption{Triangulation of the plane by considering the holes as vertices.}
	\label{Triangulation}
\end{figure}

\begin{figure}
	\centering
	\includegraphics[width=0.8\linewidth]{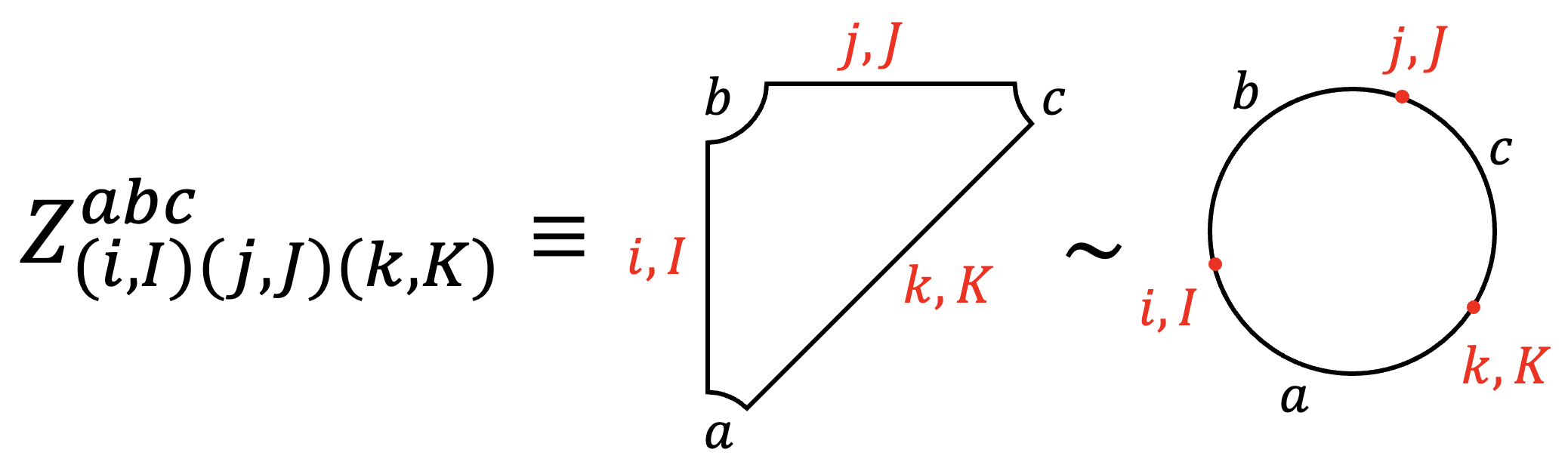}
	\caption{The partition function on the triangle.}
	\label{Onetri}
\end{figure}

Now we can get the triangulation of the manifold $\mathcal{M}$ by considering the holes as vertices as shown in figure \ref{Triangulation}. 
On the edges of the 
triangles, we insert a complete set of basis:
\be
1=\sum_{(i,I)}|i,I\rangle\langle i,I|,
\ee
where $i$ labels the primaries and $I$ labels the descendants in the conformal family of $i$. 
There is an implicit normalization condition here:
\be 
\label{norm1}
\langle i,I|j,J\rangle=\delta_{ij}\delta_{IJ}.
\ee
Then $Z_{a,b,c,\dots}$ can be written as
\be
Z_{a,b,c,\dots}=\sum_{\{(i,I)\}}\prod_{\triangle}Z^{abc}_{(i,I)(j,J)(k,K)},
\ee
where $Z^{abc}_{(i,I)(j,J)(k,K)}$ is the partition function on each triangle whose vertices are $a,b,c$ 
and edges are labelled by $|i,I\rangle,|j,J\rangle,|k,K\rangle$ as shown in figure \ref{Onetri}. 
To avoid redundancy, we have omitted the dependence on the coordinates of three vertices here. 
\footnote{We discovered \cite{Brehm:2021wev} during the revision and expansion of the current paper. It was observed there that a 2D RCFT partition function can be decomposed into triangles using the entanglement brane boundary condition we described above and one of us proposed back in \cite{Hung:2019bnq}.  We independently arrived at this decomposition as triangles while constructing exact eigenstates of the RG operator. The connection to strange correlators thus follows naturally in our construction, and is to the best of our knowledge novel in the literature.  } 
In the following, we will express $Z^{abc}_{(i,I)(j,J)(k,K)}$ in two different ``gauges" of three point conformal blocks.

\subsubsection{Computing  $Z^{abc}_{(i,I)(j,J)(k,K)}$ in a conventional {\it block gauge} }
$Z^{abc}_{(i,I)(j,J)(k,K)}$ can be conformally mapped to a boundary three point function on the upper-half plane. 
Our conventions for primary boundary operators are:
\bea 
\langle \mathds{1}\rangle_a&=&\alpha^a=\frac{S_{a0}}{\sqrt{S_{00}}}=d_a\sqrt{S_{00}},\\
\label{norm2}
\langle\phi^{ab}_i(0)\phi^{cd}_{j}(x)\rangle&=&\delta_{ad}\delta_{bc}\delta_{ij}x^{-2\D_i}\alpha^{ab}_i,\\
\phi^{ab}_i(0)\phi^{bc}_j(x)&\sim& \sum_k C^{abc}_{ijk}x^{\D_k-\D_i-\D_j}\phi^{ac}_k(x)+\dots.
\eea 
Then 
\be
\langle\phi^{ab}_{(i,I)}(x_1)\phi^{bc}_{(j,J)}(x_2)\phi^{ca}_{(k,K)}(x_3)\rangle=C^{abc}_{ijk} \alpha^{ac}_k \beta^{ijk}_{IJK}(x_1,x_2,x_3).
\ee
In particular,
\bea
&&\beta^{ijk}_{000}(x_1,x_2,x_3)=\\
\nonumber
&&\frac{1}{|x_1-x_2|^{\D_1+\D_2-\D_3}|x_1-x_3|^{\D_1+\D_3-\D_2}|x_3-x_2|^{\D_3+\D_2-\D_1}},
\eea
where $I = J = K = 0$ denotes the fact that the inserted operators are all primaries.
Such conventional normalisation for three point block is often called {\it block gauge}  \cite{Kojita:2016jwe}. 

Taking into account the normalization condition (\ref{norm1}) and (\ref{norm2}),
$Z^{abc}_{(i,I)(j,J)(k,K)}$ is conformally related with 
\bea
\nonumber 
&&\langle\phi^{ab}_{(i,I)}(x_1)\phi^{bc}_{(j,J)}(x_2)\phi^{ca}_{(k,K)}(x_3)\rangle/\sqrt{\a^{ab}_i\a^{bc}_j\a^{ca}_k}\\
&&=\frac{C^{abc}_{ijk} \alpha^{ac}_k }{\sqrt{\a^{ab}_i\a^{bc}_j\a^{ca}_k}}\beta^{ijk}_{IJK}(x_1,x_2,x_3).
\eea  
So $Z^{abc}_{(i,I)(j,J)(k,K)}$ takes the form
\be\label{Zabc1}
Z^{abc}_{(i,I)(j,J)(k,K)}=\frac{C^{abc}_{ijk} \alpha^{ac}_k }{\sqrt{\a^{ab}_i\a^{bc}_j\a^{ca}_k}}\alpha^{ijk}_{IJK},
\ee 
where $\alpha^{ijk}_{IJK}$ is fully determined by $\beta^{ijk}_{IJK}$ via the conformal map from the upper-half plane to a triangle. The details of this conformal map is given in \cite{Cheng:2023kxh}, where we focus on numerical illustration of the construction here.  
Note that $\alpha^{ijk}_{IJK},\beta^{ijk}_{IJK}$ are three-point conformal blocks which do not depend on any normalization of boundary operators.
And they satisfy
\be
\label{crossingblock}
\sum_P \a^{ijp}_{IJP}\a^{klp}_{KLP}=\sum_{q,Q} F^{\textrm{blocks}}_{pq}\left[\begin{smallmatrix}
	l& i\\
	k& j\\
	\end{smallmatrix}\right]
	\a^{liq}_{LIQ}\a^{jkq}_{JKQ},
\ee 
where $F^{\textrm{blocks}}_{pq}\left[\begin{smallmatrix}
	l& i\\
	k& j\\
	\end{smallmatrix}\right]$ is the F matrix relating two sets of four-point conformal blocks.
(Here we take the notation from \cite{Kojita:2016jwe}.)

The crossing symmetry of boundary RCFT requires that 
\be \label{Ccross}
\sum_q F^{\textrm{blocks}}_{qp}\left[\begin{smallmatrix}
	l& i\\
	k& j\\
	\end{smallmatrix}\right] C^{abc}_{ijq}C^{cda}_{klq}\a^{ac}_q=
C^{dab}_{lip}C^{bcd}_{jkp}\a^{db}_p.
\ee 
Since the two point functions can be obtained from the three point functions, we have 
\be \label{2from3}
\a^{ab}_i=C^{aba}_{ii0}\a^{a},
\ee 
where $0$ labels the identity operator.
Considering the two- and three-point functions, it is easy to note that
\be \label{2and3}
\a^{ab}_{i}=\a^{ba}_i,\;\;C^{abc}_{ijk}\a^{ac}_k=C^{bca}_{jki}\a^{ba}_i=C^{cab}_{kij}\a^{cb}_j.
\ee 
Using (\ref{2from3}) and (\ref{2and3}), eqation (\ref{Ccross}) can be simplified to
\be
C^{abd}_{ipl}C^{bcd}_{jkp}=\sum_q F^{\textrm{blocks}}_{qp}\left[\begin{smallmatrix}
	l& i\\
	k& j\\
	\end{smallmatrix}\right] C^{abc}_{ijq}C^{acd}_{qkl}.
\ee
One solution for this equation is
\be
C^{abc}_{ijk}=F^{\textrm{blocks}}_{bk}\left[\begin{smallmatrix}
	a& c\\
	i& j\\
	\end{smallmatrix}\right],
\ee 
thanks to the pentagon equation of F matrices.

Following \cite{Kojita:2016jwe}, we introduce
\bea
g^\prime_a&=&\frac{1}{F^{\textrm{blocks}}_{00}\left[\begin{smallmatrix}
	a& a\\
	a& a\\
	\end{smallmatrix}\right]},\\ 
\theta(a,b,c)&=&\frac{1}{F^{\textrm{blocks}}_{00}\left[\begin{smallmatrix}
	a& a\\
	a& a\\
	\end{smallmatrix}\right]
	F^{\textrm{blocks}}_{0a}\left[\begin{smallmatrix}
		b& c\\
		b& c\\
		\end{smallmatrix}\right]},\\
\tilde{\theta}(a,b,c)&=&\frac{1}{F^{\textrm{blocks}}_{00}\left[\begin{smallmatrix}
			a& a\\
			a& a\\
			\end{smallmatrix}\right]
			F^{\textrm{blocks}}_{a0}\left[\begin{smallmatrix}
				b& b\\
				c& c\\
				\end{smallmatrix}\right]}.
\eea
Here $\theta(a,b,c)$ and $\tilde{\theta}(a,b,c)$ both possess the $S_3$ permutation symmetry.
And we can write $C^{abc}_{ijk}$ as
\be
\label{Cabcijk}
C^{abc}_{ijk}=F^{\textrm{blocks}}_{bk}\left[\begin{smallmatrix}
	a& c\\
	i& j\\
	\end{smallmatrix}\right]
=g^\prime_k\sqrt\frac{\theta(c,j,b)\theta(i,a,b)}{\theta(i,j,k)\theta(c,a,k)}
\begin{bmatrix}
    i & j & k  \\
    c & a & b
    \end{bmatrix},
\ee 
where $\left[\begin{smallmatrix}
    i & j & k  \\
    c & a & b
    \end{smallmatrix}\right]$ are the 6j symbols (tetrahedral symbols).

And $\a^{ab}_i$ becomes
\bea
\nonumber
\a^{ab}_i&=&C^{aba}_{ii0}\a^a=
F^{\textrm{blocks}}_{b0}\left[\begin{smallmatrix}
	a& a\\
	i& i \\
	\end{smallmatrix}\right]\a^a=
	\frac{g^\prime_b\a^a}{\tilde{\theta}(a,b,i)},\\
 &=&C^{bab}_{ii0}\a^b=
 F^{\textrm{blocks}}_{a0}\left[\begin{smallmatrix}
	b& b\\
	i& i\\
	\end{smallmatrix}\right]\a^b
=\frac{g^\prime_a\a^b}{\tilde{\theta}(a,b,i)}.
\eea 

Setting $i=0$, we obtain
\be
\a^a=\a^{aa}_{0}=F^{\textrm{blocks}}_{a0}\left[\begin{smallmatrix}
	a& a\\
	0& 0\\
	\end{smallmatrix}\right]\a^a \;\; \Rightarrow \;\; F^{\textrm{blocks}}_{a0}\left[\begin{smallmatrix}
		a& a\\
		0& 0\\
		\end{smallmatrix}\right]=1.
\ee
In particular, $g^\prime_0=1/F^{\textrm{blocks}}_{00}\left[\begin{smallmatrix}
	0& 0\\
	0& 0\\
	\end{smallmatrix}\right]=1$.

Setting $b=0$, we obtain
\be 
g^\prime_0 \a^a=g^\prime_a \a^0,
\ee 
which implies
\be 
g^\prime_a=\frac{\a^a}{\a^0}=\frac{d_a\sqrt{S_{00}}}{d_0\sqrt{S_{00}}}=d_a.
\ee 
Then
\be
\label{aabi}
\a^{ab}_i=\frac{d_ad_b\sqrt{S_{00}}}{\tilde{\theta}(a,b,i)}.
\ee 
Plugging (\ref{Cabcijk}) and (\ref{aabi}) into (\ref{Zabc1}), using the property 
\be 
d_ad_bd_c=\theta(a,b,c)\tilde{\theta}(a,b,c),
\ee
the expression of $Z^{abc}_{(i,I)(j,J)(k,K)}$ can be simplified into
\be \label{Zabcdirect}
Z^{abc}_{(i,I)(j,J)(k,K)}=S_{00}^{-\frac{1}{4}}\sqrt{\frac{d_id_jd_k}{\theta(i,j,k)}}\begin{bmatrix}
    i & j & k  \\
    c & a & b
    \end{bmatrix} \a^{ijk}_{IJK}.
\ee

The computation in this section has shown that a CFT partition function can be decomposed into products of 
open three point correlation functions. The conformal blocks are normalised in the {\it block gauge}. 
To express the CFT partition as a strange correlator as in (\ref{eq:rstrange}) so that $|\Psi\rangle$ is precisely given by (\ref{Psistate}), 
we need to make a change of normalisation of the conformal blocks to the so called {\it Racah gauge} \cite{Kojita:2016jwe}. 

\subsubsection{The expression of  $Z^{abc}_{(i,I)(j,J)(k,K)}$ in {\it Racah gauge} and strange correlator}
Previously, we fix normalization of boundary operators and three point conformal blocks in a conventional way,  which is often called {\it block gauge}.
In this gauge the connection with the expression in (\ref{Psistate}) is not completely obvious. To make it explicit, we wish to change the normalisation of three point blocks, so that
the structure coefficients of the open three point function is given simply by the 6j symbol \cite{Fuchs:2004xi}.
We will denote the corresponding three point block in this gauge as $\gamma^{ijk}_{IJK}$ (closely related with $\zeta^{ijk}_{IJK}$ below). As it turns out, this conformal block is in the so called {\it Racah gauge}, as we will explain below. 
For readers not so much interested in the derivation, the conversion between the conventionally normalised conformal blocks (\ref{Zabc1}) and $\gamma^{ijk}_{IJK}$ are given in (\ref{eq:blockrelate}). The Racah gauge has been discussed for example also in \cite{Kojita:2016jwe}.

In this gauge, 

\be
\label{Ztri}
Z^{abc}_{(i,I)(j,J)(k,K)}=\begin{bmatrix}
    i & j & k  \\
    c & a & b
    \end{bmatrix}
    \zeta ^{ijk}_{IJK},
\ee
where $\left[\begin{smallmatrix}
    i & j & k  \\
    c & a & b
    \end{smallmatrix}\right]$ are the 6j symbols encoding the information about OPE coefficients, 
and $\zeta^{ijk}_{IJK}$ encodes the information about three-point conformal blocks. 
The 6j symbols possess all the symmetries of a tetrahedron:
\be
\label{tsym}
\begin{bmatrix}
    a & b & c  \\
    d & e&  f
    \end{bmatrix}=
    \begin{bmatrix}
        a & c & b  \\
        d & f & e
        \end{bmatrix}=
        \begin{bmatrix}
            c & b & a  \\
            f & e & d
            \end{bmatrix}=
            \begin{bmatrix}
                a & e & f  \\
                d & b & c
                \end{bmatrix}=\dots,
\ee
and satisfy the unitarity condition:
\bea
\label{unitarity}
\sum_y d_y\begin{bmatrix}
    a & b & x  \\
    c & d & y
    \end{bmatrix}
    \begin{bmatrix}
        a & b & x^\prime  \\
        c & d & y
        \end{bmatrix}
        =\frac{\delta_{xx^\prime} N^x_{ab}N^x_{cd}}{d_x},
\eea
together with the pentagon equation:
\be
\label{pentagon}
\sum_x d_x \begin{bmatrix}
    \rho & \nu & g  \\
    \beta & \gamma & x
    \end{bmatrix}
    \begin{bmatrix}
        \mu & \rho & h \\
        \gamma & \alpha & x
        \end{bmatrix}
        \begin{bmatrix}
            \nu & \mu & j  \\
            \alpha & \beta & x
            \end{bmatrix}
            =\begin{bmatrix}
                g & h & j  \\
                \alpha & \beta & \gamma
                \end{bmatrix}
                \begin{bmatrix}
                    g & h & j  \\
                    \mu & \nu & \rho
                    \end{bmatrix}.
\ee

\begin{figure}
	\centering
	\includegraphics[width=0.9\linewidth]{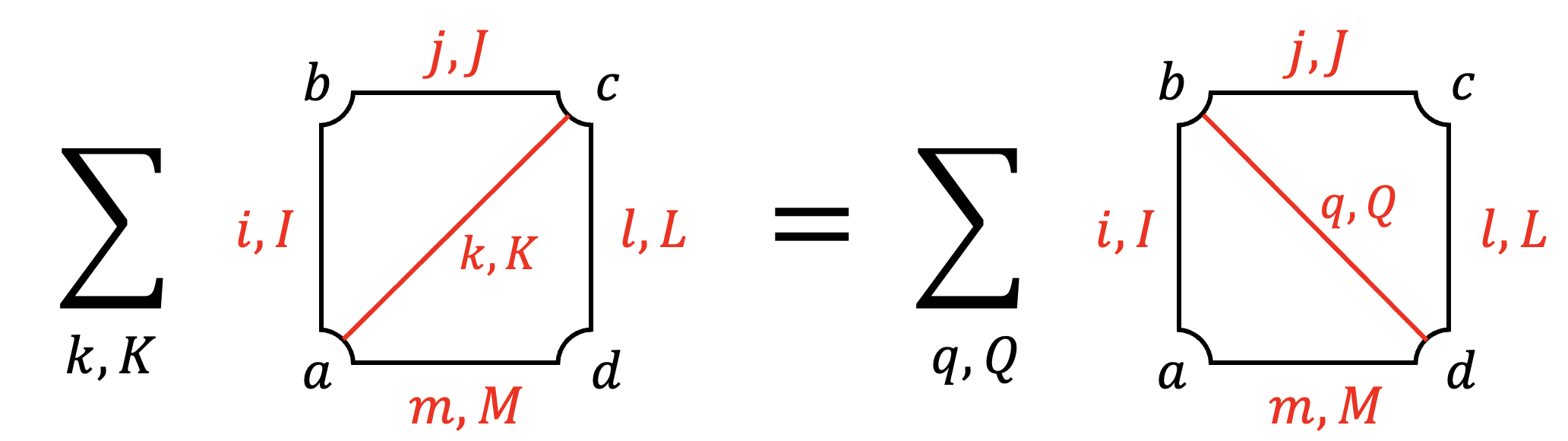}
	\caption{Two ways of dividing the quadrilateral into two triangles.}
	\label{Twodivide}
\end{figure}

To have a better understanding of $\zeta^{ijk}_{IJK}$, let's consider a quadrilateral 
which can be divided into two triangles in two ways as shown in figure \ref{Twodivide}. 
And we get the equation:  
\bea
\nonumber
&&\sum_{k,K}Z^{abc}_{(i,I)(j,J)(k,K)}Z^{cda}_{(l,L)(m,M)(k,K)}\\
&&=\sum_{q,Q}Z^{dab}_{(m,M)(i,I)(q,Q)}Z^{bcd}_{(j,J)(l,L)(q,Q)}.
\eea

Plugging (\ref{Ztri}) into it, we find

\bea
\label{crossing}
\nonumber
&&\sum_{k,K}\begin{bmatrix} i & j & k  \\  c & a & b  \end{bmatrix}
        \begin{bmatrix} l & m & k  \\  a & c & d  \end{bmatrix}
        \zeta_{IJK}^{ijk}\zeta_{LMK}^{lmk}\\
&&=
\sum_{q,Q}\begin{bmatrix} m & i & q  \\  b & d &  a \end{bmatrix}
       \begin{bmatrix} j & l & q  \\  d& b & c  \end{bmatrix}
       \zeta_{MIQ}^{miq}\zeta_{JLQ}^{jlq}.
\eea

\begin{figure}
	\centering
	\includegraphics[width=0.9\linewidth]{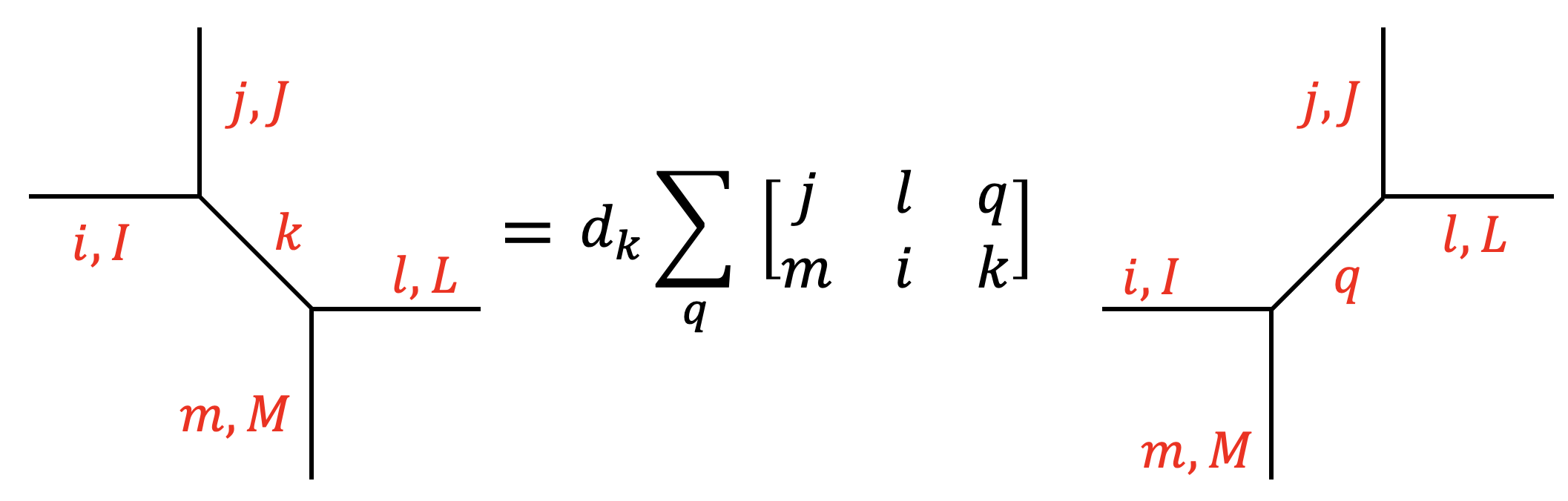}
	\caption{On the internal legs, the capital letters $K,Q$ corresponding to descendants have be summed over, 
    leaving $k,q$ representing conformal families.}
	\label{Fmove}
\end{figure}

Using the symmetries (\ref{tsym}) of tetrahedral symbols, we have 
\be
       \begin{bmatrix} m & i & q  \\  b & d & a \end{bmatrix}
       \begin{bmatrix} j & l & q  \\  d & b & c  \end{bmatrix}=
       \begin{bmatrix} b & d & q  \\  m & i & a \end{bmatrix}
       \begin{bmatrix} b & d & q  \\  l & j & c  \end{bmatrix}.
\ee
Then the pentagon equation (\ref{pentagon}) tells us that
\be
\begin{bmatrix} b & d & q  \\  m & i & a \end{bmatrix}
       \begin{bmatrix} b & d & q  \\  l & j & c  \end{bmatrix}
=\sum_k d_k \begin{bmatrix}
    c & j & b  \\
    i & a & k
    \end{bmatrix}
    \begin{bmatrix}
        l & c & d \\
        a & m & k
        \end{bmatrix}
        \begin{bmatrix}
            j & l & q  \\
            m & i & k
            \end{bmatrix} 
\ee
Plugging this into (\ref{crossing}), we find that
\be
\label{crossing1}
\sum_{K} \zeta_{IJK}^{ijk}\zeta_{LMK}^{lmk}=
d_k\sum_{q}\begin{bmatrix} j & l & q  \\  m & i &  k \end{bmatrix}
\sum_{Q}\zeta_{MIQ}^{miq}\zeta_{JLQ}^{jlq},
\ee
which is actually the transformation rule for four-point conformal blocks resulting from crossing symmetry
(differing in normalization from Equation (\ref{crossingblock})).
Graphically, this relation is shown in figure \ref{Fmove}.

\begin{figure}
	\centering
	\includegraphics[width=1\linewidth]{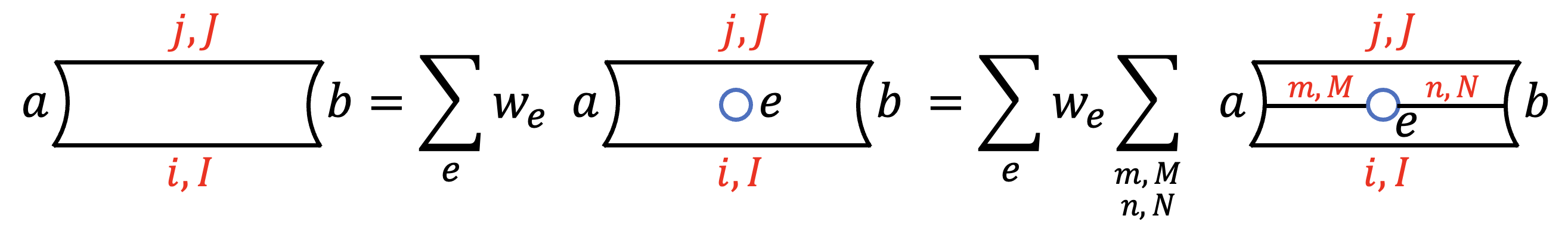}
	\caption{The equality for an infinitely narrow strip.}
	\label{Strip}
\end{figure}

\begin{figure}
	\centering
	\includegraphics[width=0.65\linewidth]{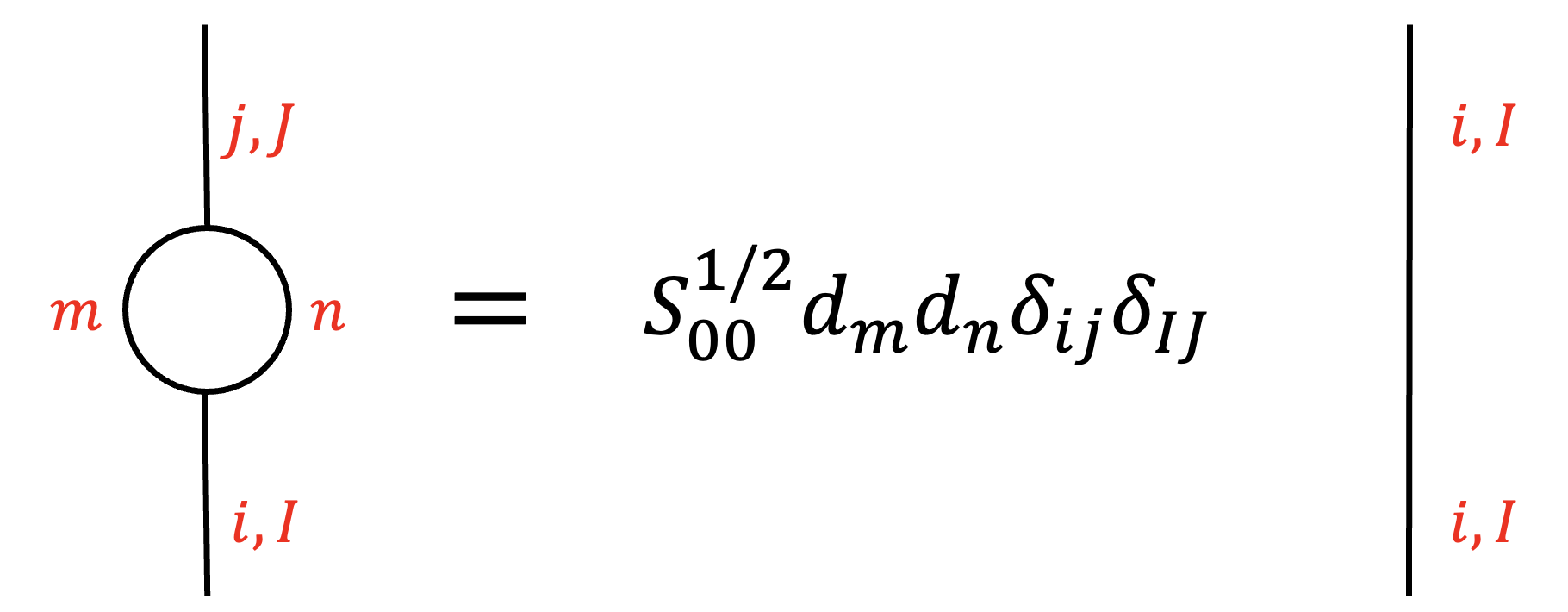}
	\caption{As before, on the internal legs, the capital letters $M,N$ corresponding to descendants have be summed over.}
	\label{Bubble}
\end{figure}

Now let us consider an infinitely narrow strip as shown in figure \ref{Strip}. This time we get
\be
\label{eqnstrip}
\begin{aligned}
    \delta_{ij}\delta_{IJ}N^i_{ab}&=\sum_e w_e \sum_{m,M}\sum_{n,N}Z^{abe}_{(m,M)(n,N)(i,I)} Z^{abe}_{(m,M)(n,N)(j,J)}\\
    &=\sum_e w_e \sum_{m,M}\sum_{n,N}
    \begin{bmatrix} m & n& i  \\  b & a&  e \end{bmatrix}
    \begin{bmatrix} m & n& j  \\  b & a&  e \end{bmatrix}
    \zeta_{MNI}^{mni}\zeta_{MNJ}^{mnj}.
    \end{aligned}
\ee
The appearance of $\delta_{ij}\delta_{IJ}$ is due to the strip being infinitely narrow.
Recall $w_a=d_a S_{00}^{3/2}$, and using the unitarity condition (\ref{unitarity}), we get
\be
\sum_e w_e  \begin{bmatrix} m & n& i  \\  b & a&  e \end{bmatrix}
\begin{bmatrix} m & n& j  \\  b & a&  e \end{bmatrix}
=\frac{S_{00}^{3/2}\delta_{ij}N^i_{mn}N^i_{ba}}{d_i}.
\ee
Plugging it into (\ref{eqnstrip}), we obtain
\be
\begin{aligned}
\delta_{ij}\delta_{IJ}N^i_{ab}&=\sum_{m,M}\sum_{n,N}\frac{S_{00}^{3/2}\delta_{ij}N^i_{mn}N^i_{ba}}{d_i}\zeta_{MNI}^{mni}\zeta_{MNJ}^{mnj}\\
&=\frac{S_{00}^{3/2}\delta_{ij}N^i_{ba}}{d_i}\sum_{m,n}N^i_{mn}\sum_{M,N}\zeta_{MNI}^{mni}\zeta_{MNJ}^{mnj}
\end{aligned}
\ee
Note that there is an identity for the quantum dimensions:
\be
\sum_{m,n} N_{mn}^i d_m d_n =d_i \mathcal{D}^2 ,\;\;\;\textrm{with}\;\;\; \mathcal{D}^2 \equiv \sum_{k\in\mathcal{C}} d_k^2=S_{00}^{-2}.
\ee
It is natural to guess that
\bea
\sum_{M,N}\zeta_{MNI}^{mni}\zeta_{MNJ}^{mnj} 
=  S_{00}^{1/2}d_m d_n \delta_{ij}\delta_{IJ}.
\label{eqnbubble}
\eea
Graphically, this equation is shown in figure \ref{Bubble}.

\begin{figure}
	\centering
	\includegraphics[width=0.95\linewidth]{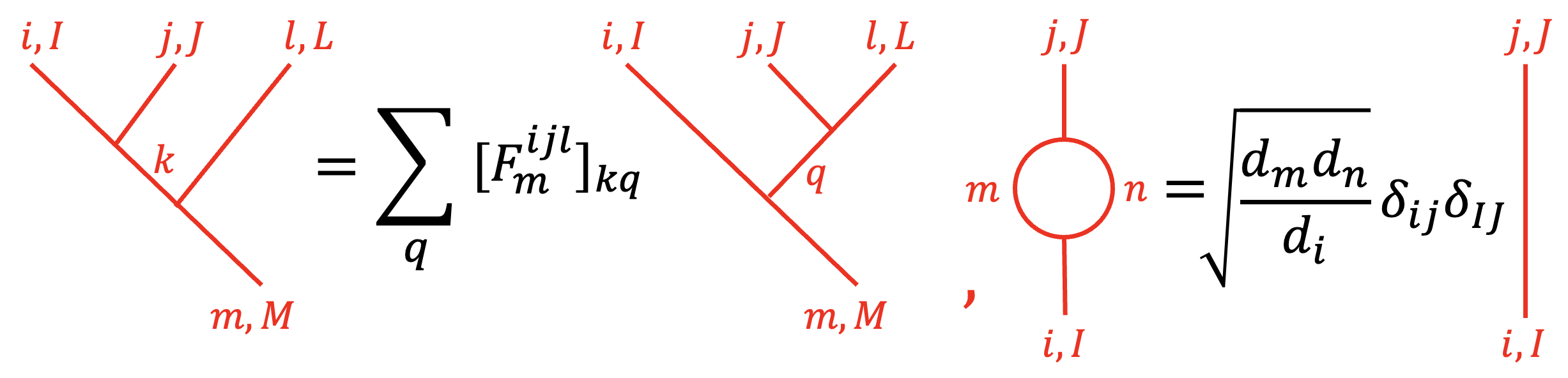}
	\caption{The properties of $\gamma_{IJK}^{ijk}$. These graphs are the same as 
    those commonly used in the fusion categories. }
	\label{Fusionrelation}
\end{figure}

Figure \ref{Fmove} and \ref{Bubble} shows the basic properties of $\zeta_{IJK}^{ijk}$.
To obtain the graphs commonly used in the fusion categories, let's define:
\be
\gamma_{IJK}^{ijk}=\left(S_{00}d_id_jd_k\right)^{-\frac{1}{4}}\zeta_{IJK}^{ijk}.
\ee
Then $\gamma_{IJK}^{ijk}$ satisfies:
\be
\label{crossing2}
\sum_{K} \gamma_{IJK}^{ijk}\gamma_{LMK}^{lmk}=
\sum_{q}\sqrt{d_k d_q}\begin{bmatrix} j & l & q  \\  m & i &  k \end{bmatrix}
\sum_{Q}\gamma_{MIQ}^{miq}\gamma_{JLQ}^{jlq},
\ee
and
\be
\sum_{M,N}\gamma_{MNI}^{mni}\gamma_{MNJ}^{mnj} = \sqrt{\frac{d_m d_n}{d_i}} \delta_{ij}\delta_{IJ}.
\ee
Using the commonly used F-symbols
\be
\left[F^{abc}_d\right]_{xy}\equiv \sqrt{d_xd_y}\begin{bmatrix} a & b & x  \\  c & d &  y \end{bmatrix},
\ee
the graphical representation of the properties of $\gamma_{IJK}^{ijk}$ is shown in figure \ref{Fusionrelation}. 

In terms of $\gamma_{IJK}^{ijk}$, finally $Z^{abc}_{(i,I)(j,J)(k,K)}$ takes the form:
\be
Z^{abc}_{(i,I)(j,J)(k,K)}=\left(S_{00}d_id_jd_k\right)^{\frac{1}{4}}
\begin{bmatrix}
    i & j & k  \\
    c & a & b
    \end{bmatrix}
	\gamma_{IJK}^{ijk}.
\ee 

Comparing with (\ref{Zabcdirect}), we observe that
\be \label{eq:blockrelate}
\gamma^{ijk}_{IJK}=S_{00}^{-\frac{1}{2}}\sqrt{\frac{\sqrt{d_id_jd_k}}{\theta(i,j,k)}} \a^{ijk}_{IJK},
\ee 
which in turn indicates that 
\be
[F^{ijk}_l]_{mn}=[F^{ijk}_l]^{\textrm{blocks}}_{mn}\frac{\mathcal{N}_{jkn} \mathcal{N}_{inl} }{\mathcal{N}_{ijm}\mathcal{N}_{mkl}},
\ee 
with
\be 
[F^{ijk}_l]^{\textrm{blocks}}_{mn} \equiv F^{\textrm{blocks}}_{mn}\left[\begin{smallmatrix}
	l& i\\
	k&j \\
	\end{smallmatrix}\right], \;\;\;\;
\mathcal{N}_{ijk} \equiv \sqrt{\frac{\theta(i,j,k)}{\sqrt{d_id_jd_k}}}.
\ee
In other words, $[F^{ijk}_l]_{mn}$ and $[F^{ijk}_l]^{\textrm{blocks}}_{mn}$ are related by a gauge transformation.

~\\

\begin{figure}
	\centering
	\includegraphics[width=0.9\linewidth]{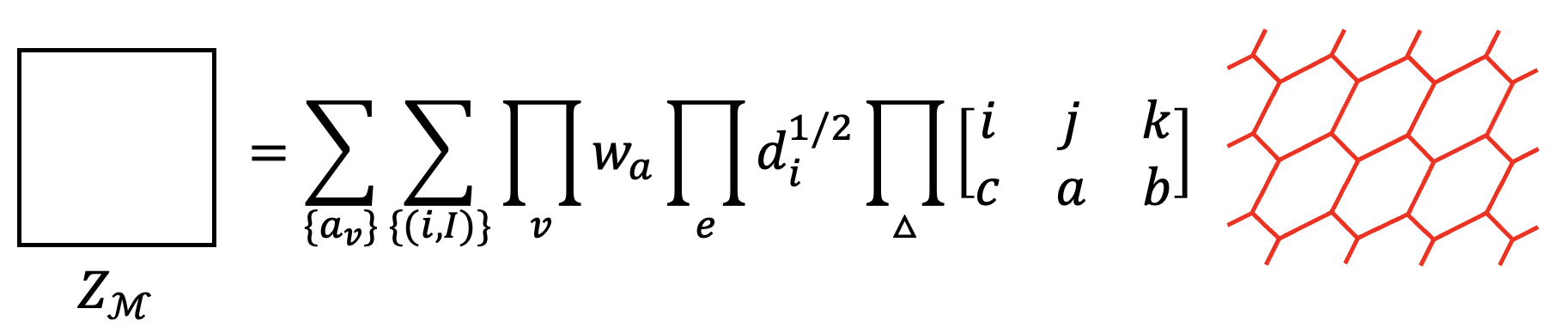}
	\caption{The network representation of the partition function. 
	Here we have absorbed the factor $S_{00}^{1/4}$ into the network.}
	\label{Partition}
\end{figure}

Now we can write down the partition function on $\mathcal{M}$ in terms of $\gamma_{IJK}^{ijk}$ explicitly:

\be
Z_{\mathcal{M}}=\sum_{\{a_v\}} \sum_{\{(i,I)\}} \prod_{v} w_a \prod_{e}d_i^{1/2} \prod_{\triangle }\left(\begin{bmatrix}
    i & j & k  \\
    c & a & b
    \end{bmatrix} \gamma_{IJK}^{ijk}S_{00}^{1/4}\right).
\ee
Note that each vertex (if labelled by $a$) contributes $w_a$, 
each edge (if labelled by $i$) contributes $d_i^{1/2}$,
each triangle contributes $[\dots]\gamma S_{00}^{1/4}$. 
Using the graphical representation for $\gamma_{IJK}^{ijk}$, we finally obtain the network representation of the CFT partition function 
as shown in figure \ref{Partition}. 

Now it is explicit that $Z_{\mathcal{M}}$ can be written as $Z_{\mathcal{M}}=\langle\Omega|\Psi\rangle$ with $|\Psi\rangle$ and $\langle\Omega|$ given by 
(\ref{Psistate}) and (\ref{Omegastate}) respectively.
The fact that $\langle\Omega|$ is an eigenstate of the RG operator is guaranteed by the properties of 
$\gamma^{ijk}_{IJK}$ as shown in figure \ref{Fusionrelation}.

In the above, we have focussed on the explicit reconstruction of the path-integral of diagonal RCFTs. For general non-diagonal full RCFTs, each of these theories is characterised by a different module category $\mathcal{M}_\mathcal{C}$ of the (modular) fusion category $\mathcal{C}$ characterising the conformal primaries $\{i,j,k,\cdots\}$ of the theory.  The objects $\{A,B,C, \cdots\} \in  \mathcal{M}_\mathcal{C}$ label different conformal boundary conditions of the given non-diagonal RCFT. 
The strange correlator reconstruction here works again in the same manner by cutting the theory into open three point correlation functions. The entanglement brane boundary condition in this case is obtained by replacing the factor $d_a$ in (\ref{eq:weight1}) by $d_A$, where $d_A$ is the quantum dimension of the object $A$ in  $\mathcal{M}_\mathcal{C}$ .
The end result is that the strange correlator representation of the path-integral is given by the same $\langle \Omega |$ constructed from three point blocks, but with $|\Psi\rangle$ in (\ref{Psistate}) replaced by
\be\label{offdiagonal}
|\Psi\rangle=\sum_{\{A_v\}}\sum_{\{i\}}\prod_e d^{1/2}_i\prod_v d_A \prod_{\triangle} 
\begin{bmatrix}
    i & j & k  \\
    C & A & B
    \end{bmatrix}|\{i\}\rangle,
\ee
where $\begin{bmatrix}
    i & j & k  \\
    C & A & B
    \end{bmatrix}$ is the open structure coefficients of the said RCFT, with boundaries $A,B,C \in \mathcal{M}_\mathcal{C} $ .
By an alternative tensor network representation described in \cite{Lootens:2020mso}, this  expression (\ref{offdiagonal}) would agree precisely with the alternative tensor network representation of the ground state $|\Psi\rangle$ now with a different topological boundary condition characterised by module category $\mathcal{M}_\mathcal{C}$, i.e. we obtain a different RCFT by changing the topological boundary of the sandwich in \cite{Gaiotto:2020iye,Ji:2019jhk, Apruzzi:2021nmk, Freed:2022qnc}.  It has been observed that strange correlators of lattice models built out of (\ref{offdiagonal}) can produce off-diagonal RCFT in the thermodynamic limit \cite{Vanhove:2021nav}. Here we are basically producing the RG fixed point that would have been reached by these lattice models at critical points in the thermodynamic limit, thus explaining field theoretically how the RCFT emerges from these previous lattice constructions.

\section{Holographic Networks and RG operator in 4d} \label{sec:4D}

One can construct 3D partition functions from strange correlators with ground state wave-functions of 4D TQFT.
Generically, the topological input data of the 4D TQFT involves a 2-fusion category \cite{Lan_2019, Johnson_Freyd_2022, fuse2, fuse22}.  The geometrical building block of 4D TQFT is a 4-simplex.
One assigns objects of the 2-category to the edges, fusion 1-morphism to each surface and  2-morphism to the 3-sub-simplex.
Here, we would focus on the 4D Dijkgraaf-Witten model. In that case, each model is characterized by a group $G$, and an element of $H^4(G,U(1))$. Group elements of $G$ play the role of objects and are again assigned to the edges, and the 4-simplex is a function  $\alpha(\{g_i\}) \in H^4(G, U(1))$ of the 10 edges.  Fusion between objects is reduced to the group product.

The 3+1 D TQFT ground state $|\Psi\rangle$ admits tensor network representations that are natural generalization of the 3D case. The ground state wave-function can be obtained from a path-integral of a 4-ball with a 3-sphere boundary. The path-integral is obtained by considering a triangulation of boundary 3-space into 3-simplices. Each of these 3-simplices belongs to a 4-simplex. They all share an index in the center of the 4-ball, which is a higher dimensional generalization of figure \ref{fig:2dDW}. As a tensor network, the building block can be written as a tetrahedron with indices both on the edges and on the vertices. The 4D ground state wave-function defined on a 3D surface is obtained by putting these tetrahedra together filling the 3D space, and summing over the indices on the shared vertices between tetrahedra.
A version of this tensor network is discussed in \cite{Delcamp:2020rds} for the 3+1 D toric code model. Here we follow a direct generalization of the picture in the 2+1 D and 1+1 D case with a notation adapting to that in \cite{Aasen:2020jwb}. The discussion should apply quite generally to any 3+1 D topological models based on 2-fusion categories.

\begin{figure}
	\centering
	\includegraphics[width=0.35\linewidth]{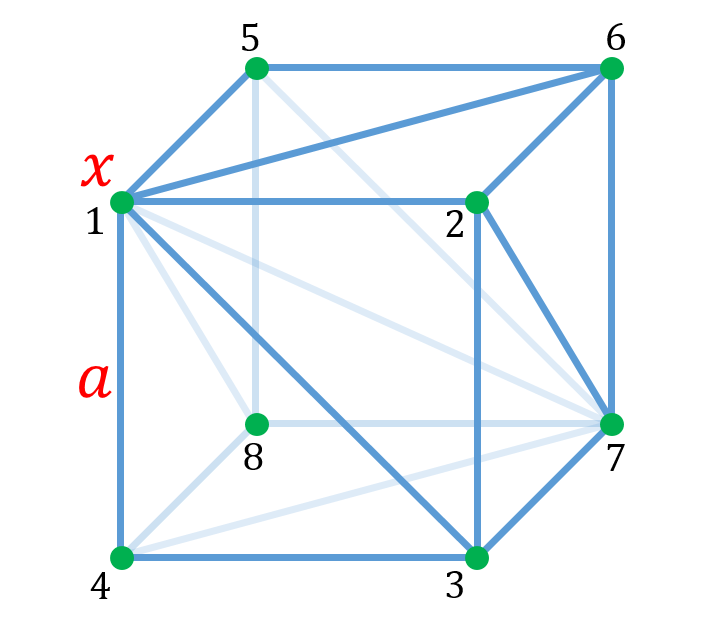}
	\caption{Each cube is divided into 6 tetrahedra. They are $1237$, $1267$, $1567$, $1587$, $1487$, $1437$. We remark that they are one to one correspondence to the paths from vertex 1 to vertex 7. Vertex 1 and vertex 7 are diagonal on the cube, and they determine the triangulation of the cube. Green dots represent the bulk edges. The bulk edges are labeled by $x$. The boundary edges are labeled by $a$.
}
	\label{cube}
\end{figure}

For concreteness, we pick a specific triangulation of the 3D surface. First fill the 3D space with cubes. Each vertex (colored green in figure \ref{cube}) is shared between 8 cubes and the group element attached to it would be summed over.
Then each cube is divided into 6 tetrahedra, as shown in figure \ref{cube}. In this triangulation of the cube, two vertices (1 and 7) play a special role.

\begin{figure}
	\centering
	\includegraphics[width=0.8\linewidth]{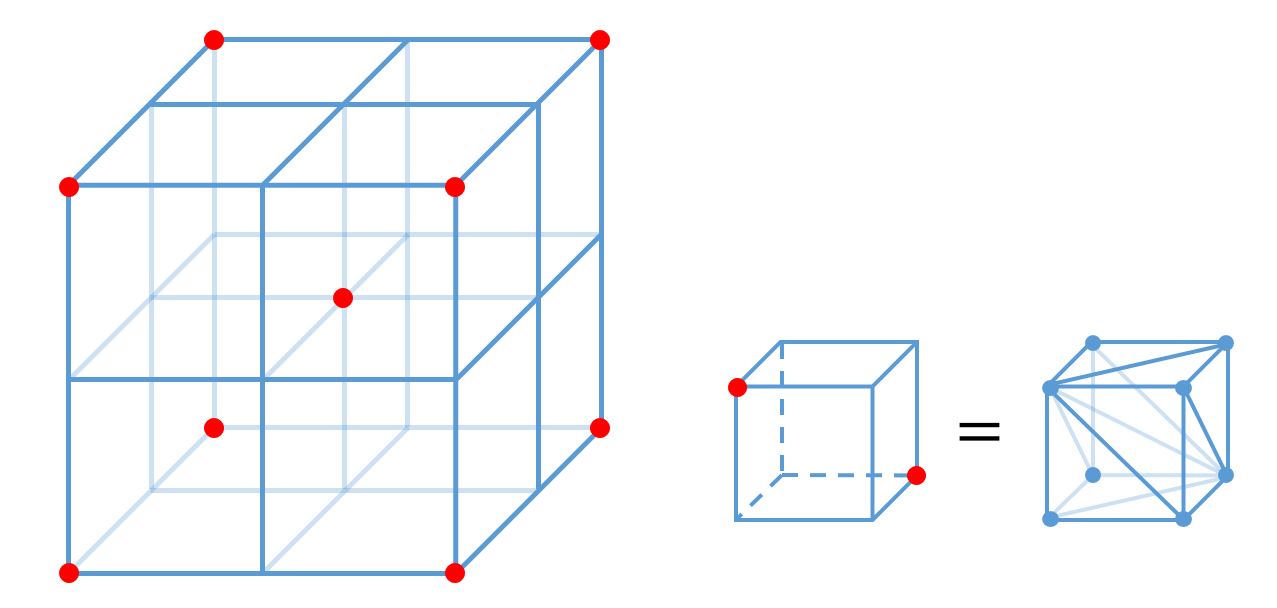}
	\caption{Each $2\times2\times2$ cube contains 8 small cubes. Each small cube has two red dots on its diagonal vertices. And the triangulation of small cube is determined by these two red dots. Previously we use green dots to represent bulk edges. Here the blue dots are just the vertices of the small cubes.
}
	\label{bigcube}
\end{figure}

A complete specification of the triangulation therefore involves marking the cubic lattice vertices that play the special role used in the above triangulation which are marked red in figure \ref{bigcube}. Including this structure, the smallest self-repeating unit is a $2\times 2 \times 2$ collection of cubes.

\begin{figure}
	\centering
	\includegraphics[width=0.8\linewidth]{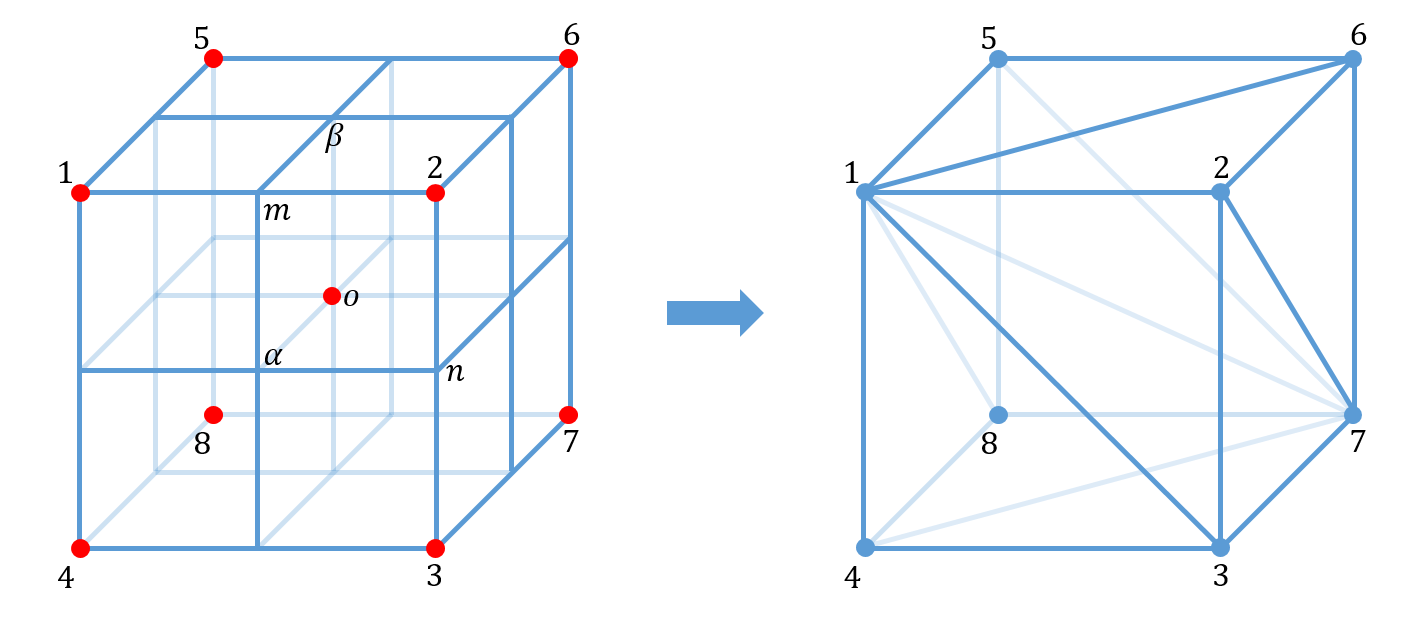}
	\caption{Coarse grain the $2\times2\times2$ cube into $1\times1\times1$ cube.
}
	\label{RG0}
\end{figure}

The RG operator we will construct below involves mapping the $2\times2\times2$ cube into 1 cube as shown in figure \ref{RG0}.

This is achieved in three steps. But to that end, we need to use a generalization of the pentagon relation leading to figure \ref{fig:Mong}.
In the case of the Dijkgraaf-Witten models, this is just the 4-cocycle relation \cite{Dijkgraaf:1989pz}, which is explained in the beginning of Appendix \ref{DWF}.

\begin{figure}
	\centering
	\includegraphics[width=0.7\linewidth]{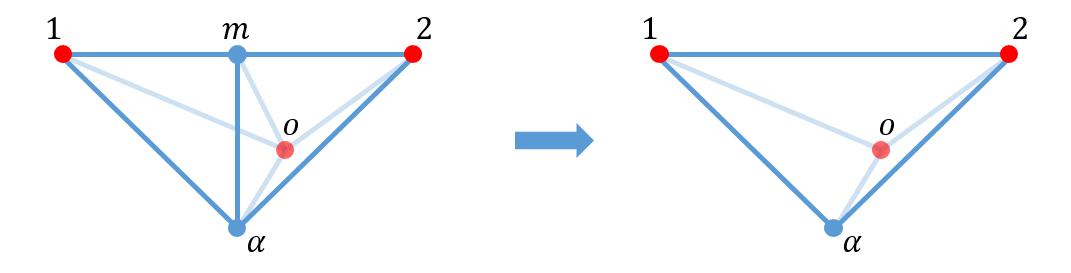}
	\caption{Combine $1m\a o$ and $2m\a o$ to get a bigger tetrahedron $12\a o$.
}
	\label{combine1}
\end{figure}

Using the 4-cocycle relation, we can thus map two tetrahedra into one, as illustrated in figure \ref{combine1}.
Through this map, one vertex is eliminated as a result. We will thus in the following discuss the RG process as a sequence of steps that eliminate designated vertices through repeated use of this relation in figure \ref{combine1}, keeping the collection of 4-cocycles, or often called 10-j symbols accumulating implicitly.

The first step is to eliminate the vertices $m,n,\dots$ in the middle of the edges of the $2\times2\times2$ cube and obtain the edges $12,23,\dots$ of the target $1\times1\times1$ cube. Let's consider the vertex $m$ as an example. It is shared by 4 tetrahedra: $1m\a o,1m\b o, 2m\a o,2m\b o$. Now we combine $1m\a o$ and $2m\a o$ to get a bigger tetrahedron $12\a o$ as shown in figure \ref{combine1}. Similarly combining $1m\b o$ and $2m\b o$ will give us tetrahedron $12\b o$. And then the vertex $m$ disappears and the edge $12$ is obtained.
In the 4D perspective, we have a 4-simplex $12m\a o$ whose boundary consists of 5 tetrahedra: $1m\a o,2m\a o, 12\a o,12mo,12m\a$. One can note that $1m\a o,2m\a o,12\a o$ appear in the combining procedure. While $12mo$ is shared by a neighboring 4-simplex $12m\b o$, and $12m\a$ is similarly shared by another neighboring 4-simplex.
 So the combining procedure is actually achieved by inserting these 4-simplices.
They form a 4D body which has two boundaries. One boundary consists of small tetrahedra $1m\a o,1m\b o, 2m\a o,2m\b o$, and the other boundary consists of bigger tetrahedra $12\a o,12\b o$.

\begin{figure}
	\centering
	\includegraphics[width=0.6\linewidth]{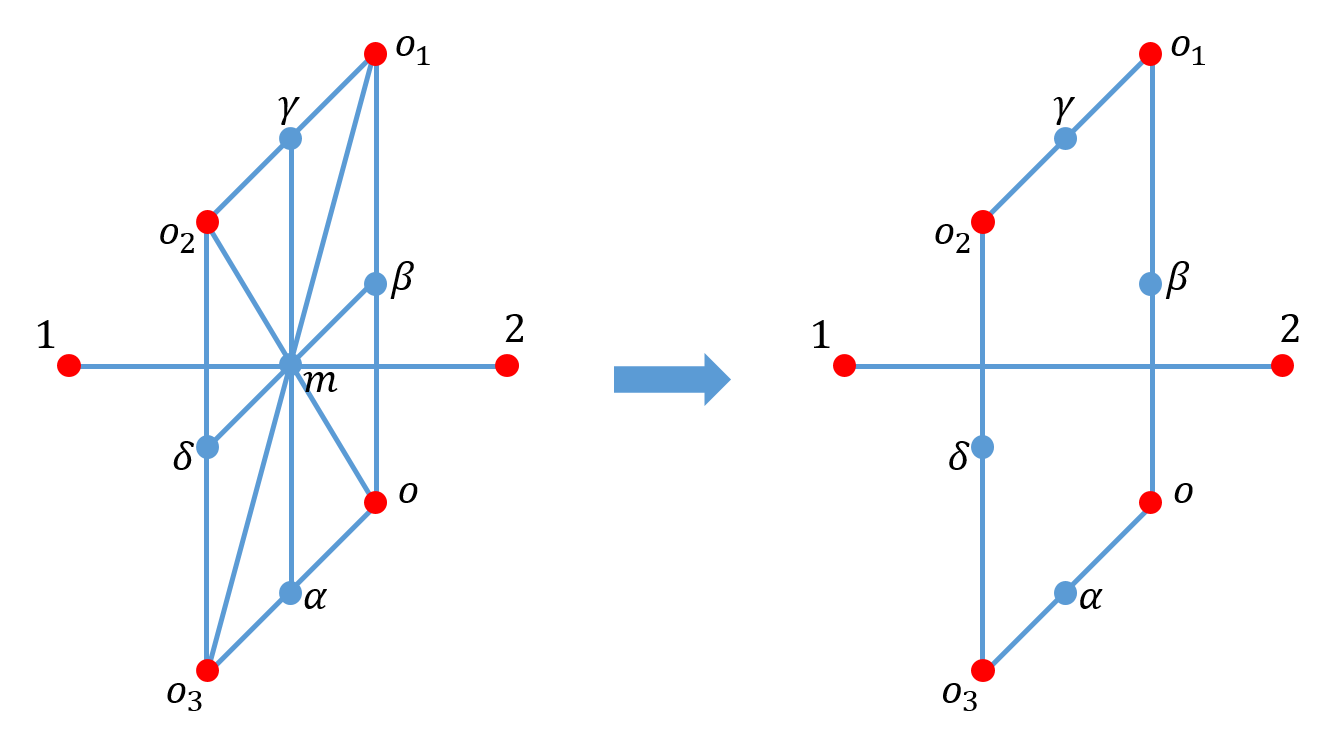}
	\caption{The first step is to eliminate the vertices like $m$ and to obtain edges like $12$. To avoid clutter, we omit the edges connecting $1,2$ with $\a,\b,\g,\d,o,o_1,o_2,o_3$. On the left hand side, there is a vertex $m$, and there are 16 small tetrahedron $1m\a o,1m\b o,1m\b o_1, \dots,2m\a o,2m\b o,2m\b o_1\dots$. On the right hand side, there is no $m$, and there are 8 bigger tetrahedron $12\a o,12\b o,12\b o_1,\dots$. They are on the two boundaries of a 4D body which consists of eight 4-simplices $12m\a o$, $12m\b o$, $ 12m\b o_1$, $ 12m\g o_1$, $ 12m\g o_2$, $ 12m\d o_2$, $ 12m\d o_3$, $ 12m\a o_3$.
}
	\label{step1}
\end{figure}

Notice that the vertex $m$ is shared by four $2\times2\times2$ cubes. In the above we only consider one of them. The whole picture is given by figure \ref{step1}, where $o_1,o_2,o_3$ are the centers of other three $2\times2\times2$ cubes. We have eight 4-simplices $12m\a o,12m\b o$, $12m\b o_1, 12m\g o_1$, $ 12m\g o_2, 12m\d o_2$, $ 12m\d o_3, 12m\a o_3$. The adjacent simplices share the same tetrahedron: for example, $12m\a o,12m\b o$ share the tetrahedron $12mo$. Finally these 4-simplices form a 4D body which has two boundaries. One boundary consists of 16 small tetrahedra $1m\a o,1m\b o,1m\b o_1, \dots,2m\a o,2m\b o,2m\b o_1\dots$, and the other boundary consists of 8 bigger tetrahedra $12\a o,12\b o,12\b o_1,\dots$. The 4D body is a map between these two boundaries. After applying this type of maps, we remove the vertices $m,n,\dots$ in the middle of the edges of the $2\times2\times2$ cube, and obtain edges $12,23,\dots$ of the target $1\times1\times1$ cube.

\begin{figure}
	\centering
	\includegraphics[width=0.7\linewidth]{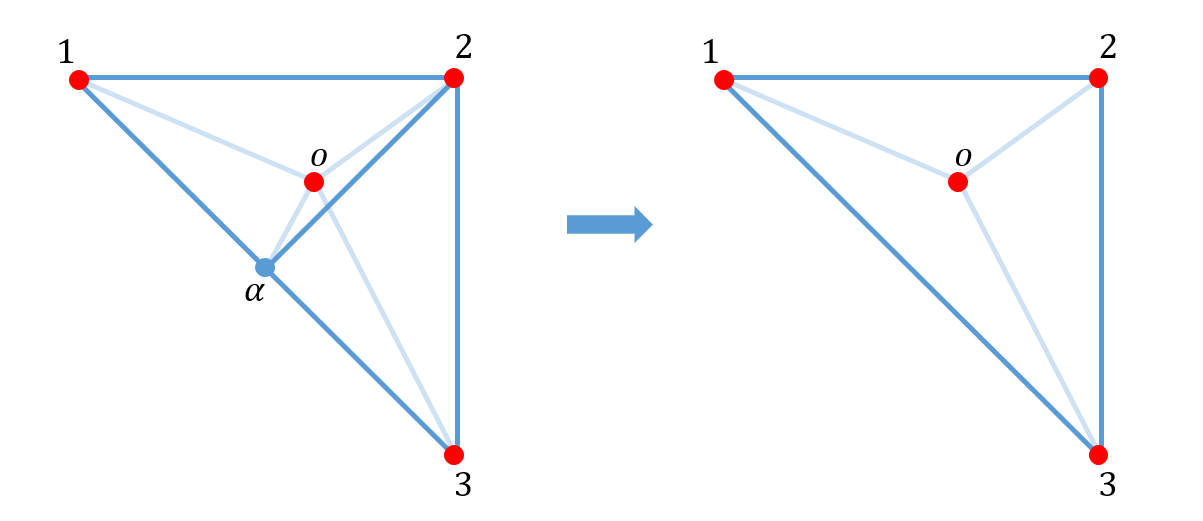}
	\caption{Combine $12\a o$ and $23\a o$ to get a bigger tetrahedron $123 o$.
}
	\label{combine2}
\end{figure}

\begin{figure}
	\centering
	\includegraphics[width=0.6\linewidth]{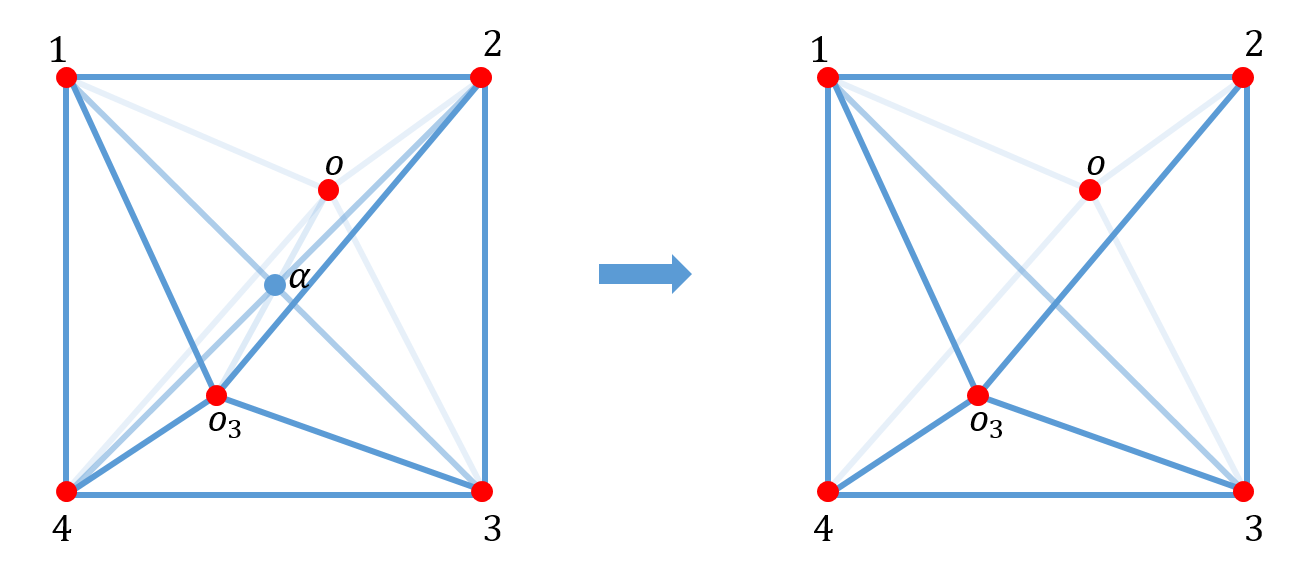}
	\caption{The second step is to eliminate the vertices like $\a$. Here we choose to connect vertices $1,3$ since in the target coarse grained cubic there is a $13$ edge as shown in figure \ref{RG0}.
On the left hand side, there is a vertex $\a$, and there are 8 tetrahedra $12\a o,23\a o,34\a o,41\a o,12\a o_3,23\a o_3,34\a o_3,41\a o_3$. On the right hand side, there is no $\a$, and there are 4 tetrahedra $123o,341o,123o_3,341o_3$. They are on the two boundaries of a 4D body which consists of four 4-simplices $123\a o,123\a o_3,341\a o_3,341\a o$.
}
	\label{step2}
\end{figure}

The second step is to eliminate the vertices $\a,\b,\dots$ in the center of faces of the $2\times2\times2$ cube and obtain the edges $13,16,\dots$ in the target $1\times1\times1$ cube. Let's consider the vertex $\a$ as an example. After the first step, it is shared by 4 tetrahedra: $12\a o,23\a o,34\a o,41\a o$. Now we combine $12\a o$ and $23\a o$ to get a bigger tetrahedron $123 o$ as shown in figure \ref{combine2}. Similarly combining $34\a o$ and $41\a o$ will give us tetrahedron $341o$. And then the vertex $\a$ disappears, and the edge $13$ is obtained. The vertex $\a$ is actually shared by two $2\times2\times2$ cubes. The whole picture is given by figure \ref{step2}, where $o_3$ is the center of the other $2\times2\times2$ cube. We have four 4-simplices $123\a o,123\a o_3,341\a o_3,341\a o$. And these 4-simplices form a 4D body which has two boundaries. One boundary consists of 8 tetrahedra
$12\a o,23\a o,34\a o,41\a o,12\a o_3,23\a o_3,34\a o_3,41\a o_3$, and the other boundary consists of 4 tetrahedra $123o,341o,123o_3,341o_3$. The 4D body is a map between these two boundaries. After applying this type of maps properly, we remove the vertices $\a,\b,\dots$ in the center of faces of the $2\times2\times2$ cube, and obtain edges $13,16,18,72,74,75$ in the target $1\times1\times1$ cube.

\begin{figure}
	\centering
	\includegraphics[width=0.8\linewidth]{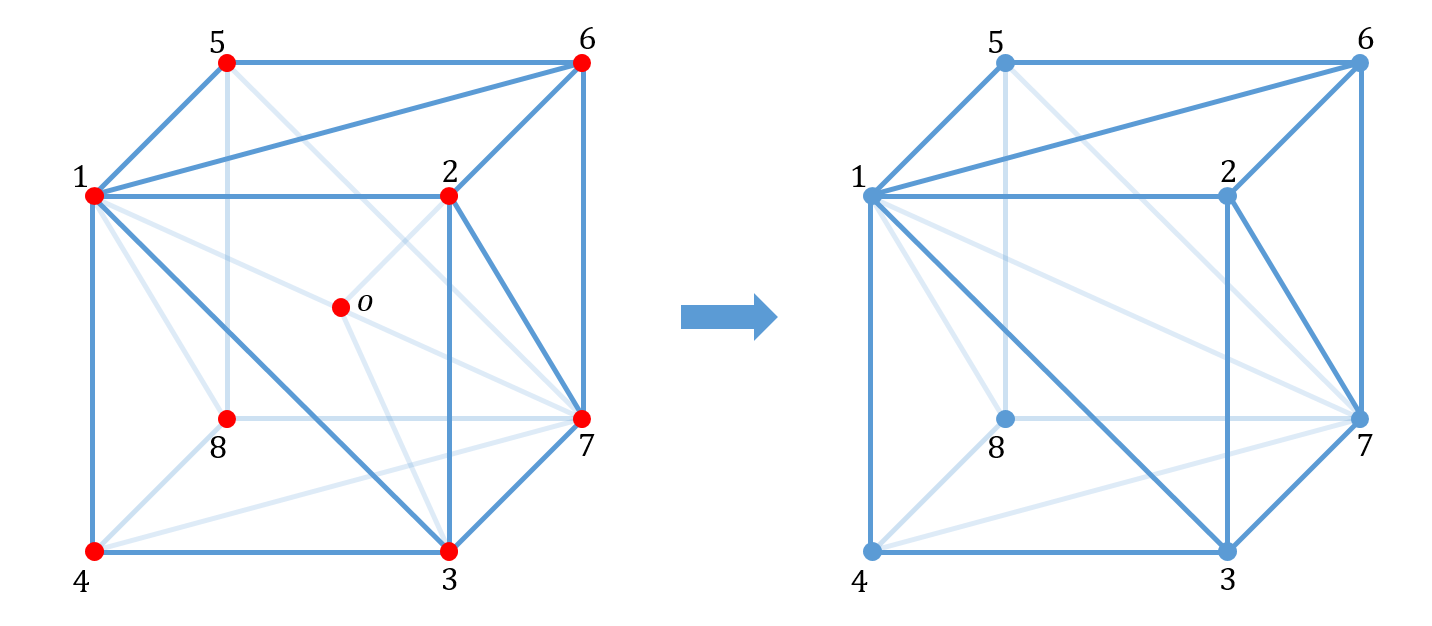}
	\caption{The third step is to eliminate the vertex $o$ and to obtain the edge $17$. To avoid clutter, we only show some of the edges connecting $o$ with $1,2,3,4,5,6,7,8$. On the left hand side, there is a vertex $o$, and there are 12 tetrahedra $123o,143o$, $237o,267o$, $126o,156o$, $148o,158o$, $487o,437o$, $567o,587o$. On the right hand side, there is no $o$, and there are 6 tetrahedra $1237$, $1267$, $1567$, $1587$, $1487$, $1437$. They are on the two boundaries of a 4D body which consists of six 4-simplices $1237o,1267o,1567o,1587o,1487o,1437o$. Combining $123o$ and $237o$ to get the tetrahedron $1237$ can be read off from this figure.
}
	\label{step3}
\end{figure}

The third step is to eliminate the vertex $o$ in the center of the $2\times2\times2$ cube and obtain the edge $17$. After the second step, the vertex $o$ is shared by 12 tetrahedra $123o,143o,237o,267o,126o,156o$, $148o,158o,487o,437o,567o,587o$ as shown on the left hand side of figure \ref{step3}. Now we combine $123o$ and $237o$ to get a bigger tetrahedron $1237$. Similarly the tetrahedra $1267,1567,1587,1487,1437$ in the target cube can be obtained by combining two of these 12 tetrahedra. And then the vertex $o$ disappears, and the edge $17$ is obtained. The whole picture is shown by figure \ref{step3}. We have six 4-simplices $1237o,1267o,1567o,1587o,1487o,1437o$. And these 4-simplices form a 4D body which has two boundaries. One boundary consists of 12 tetrahedra
$123o,143o$, $237o,267o$, $126o,156o$, $148o,158o$, $487o,437o$, $567o,587o$, and the other boundary consists of 6 tetrahedra $1237$, $1267$, $1567$, $1587$, $1487$, $1437$. The 4D body is a map between these two boundaries. After applying this map, we remove the vertex $o$ in the center of the $2\times2\times2$ cube, and obtain the edge $17$ in the target $1\times1\times1$ cube.

After the 3 steps described above, we coarse grain the $2\times2\times2$ cubes into exactly 1 cube. Eight  such cubes can be arranged into  a structure of $2\times2\times2$ cubes shown in figure \ref{bigcube} . This is the same as the original structure. And these 3 steps form one RG step.

\subsection{Higher Frobenius Algebra gives Topological Fixed Points}

Like in the case of RG operators from 3D topological theories where topological fixed points are given by Frobenius algebra,
the RG operator constructed from 4D topological order presented above also admits fixed points based on higher dimensional generalization of Frobenius algebra.
In the case where the bulk is given by DW theory, the Frobenius algebra characterizing topological boundary conditions has been studied \cite{Wang:2018qvd, Zhao:2022yaw}. Together with a higher generalization of ``separability'' that we will discuss below, we obtain fixed points of the 4D topological RG operators.

These topological eigenstates are constructed as follows. Previously in figure \ref{fig:Omega_gapped}, the ground state wave-function is constructed by assigning a weight to every triangle on the 2 dimensional surface, and these weights define the product of a Frobenius algebra of the input fusion category.
In the current case the wave-function is defined in 3 dimensions, and we instead assign a value $\tilde\beta(g,h,k)$ to every tetrahedron, which has three independent edge degrees of freedom after taking into account of the face constraints.
The wavefunction is thus given by
\be
\la \Omega(\tilde\beta) | = \sum_{\{g_l\}}   \prod_{\Delta} \tilde \beta(g_{\Delta_1},g_{\Delta_2},g_{\Delta_3}) \la  \{g_l\} |,  \label{eq:productstate}
\ee
where $\Delta$ runs over all tetrahedra that triangulate the three dimensional space, $\Delta_{1,2,3}$ denote 3 independent edges of a tetrahedron, and $\{g_l\}$ denotes the collection of all edge degrees of freedom.

\begin{figure}
	\centering
	\includegraphics[width=0.9\linewidth]{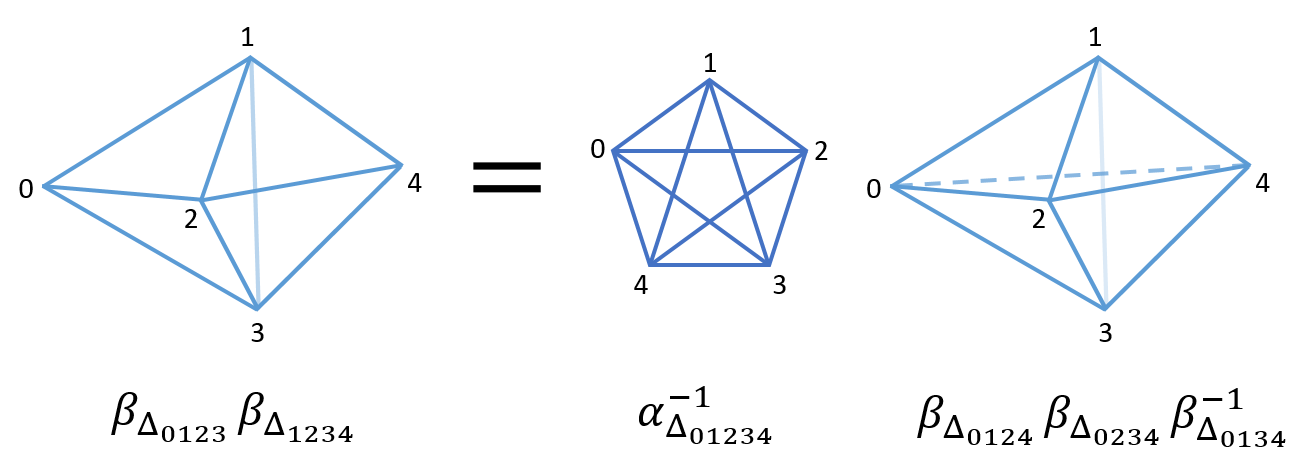}
	\caption{There are 2 tetrahedra on the left and 3 tetrahedra on the right corresponding to two different triangulations of the boundary. We have $\b_{\D_{0123}}\b_{\D_{1234}}=
\a_{\D_{01234}}^{-1}\b_{\D_{0124}}\b_{\D_{0234}}\b_{\D_{0134}}^{-1}$. The powers of $-1$ are related to the orientations.
}
	\label{F3D}
\end{figure}

Frobenius algebras associated to 4D DW models have been studied and each is associated with a 2+1 D topological boundary condition \cite{Wang:2018qvd, Zhao:2022yaw}.
The data that goes into defining a higher Frobenius algebra involves the collection of objects $\mathcal{A} $ in the fusion category, which is a subgroup in the current case.
We also need to define the product between objects and also the associativity map  \cite{Wang:2018qvd, Zhao:2022yaw}. The data can be encapsulated in assigning a function $\beta(g_{\Delta_1},g_{\Delta_2},g_{\Delta_3})$ to every tetrahedron. This function satisfies the following relation:
\begin{align} \label{eq:frobenius3d}
&\frac{\alpha(h_0,h_1, h_2, h_3) \beta(h_0, h_1.h_2, h_3)\beta(h_0, h_1,h_2)\beta(h_1,h_2,h_3)}{\beta(h_0, h_1, h_2. h_3)\beta(h_0. h_1, h_2, h_3)} \nonumber \\
&=1,
\end{align}
where $\alpha$ is the 4-cocycle in $H^4(G,U(1))$ that defines the DW theory. The 4-cocycle $\alpha$ involved above is precisely a 4-simplex, that maps between two sets of tetrahedra.
The geometric meaning of (\ref{eq:frobenius3d}) is demonstrated in figure \ref{F3D}.

\begin{figure}
	\centering
	\includegraphics[width=0.9\linewidth]{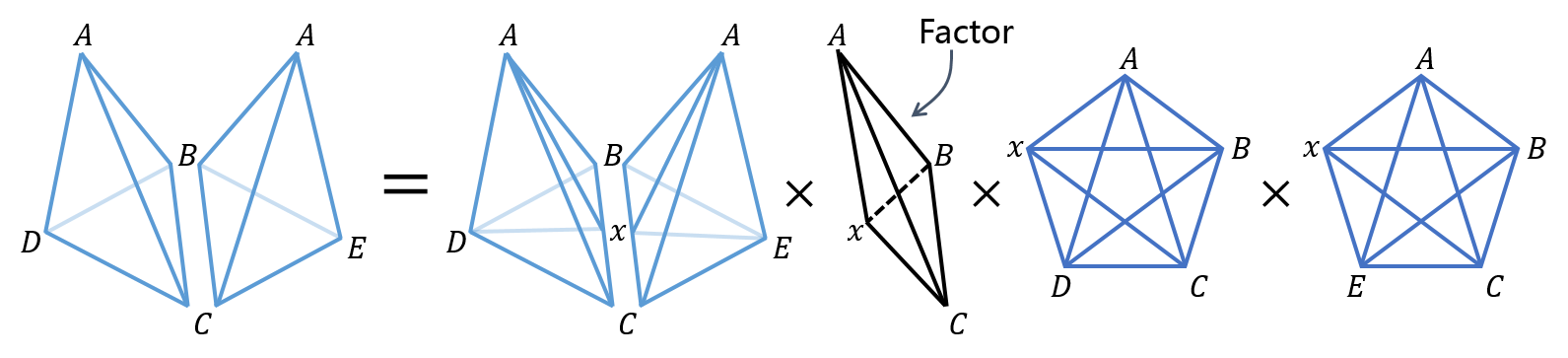}
	\caption{The blue tetraheron corresponds to the boundary factors $\beta$. The pair of 4-simplices on the right hand side corresponds to the 4-cocycles of the DW theory. The black tetrahedron referred to as a ``factor" is the analogue of a bubble that is contracted. The equality is based on absorbing this black tetrahedron and is thus the analogue of separability in 2+1 dimensional topological order. In the current model however the factor is equal to unity.}.
	\label{fig:3dseparable}
\end{figure}

The above condition is the higher dimensional analogue of associativity in a higher Frobenius algebra. (The condition at least for Frobenius algebra in a n-dimensional DW theory is reviewed in the appendix.) In addition to that, we require also a condition analogous to the separability in 2+1 dimensional topological order.
This condition is illustrated in figure \ref{fig:3dseparable}.

We note that algebra satisfying (\ref{eq:frobenius3d}) were considered as spatial boundary conditions of the DW models. Here however, when inserted into (\ref{eq:productstate}) and then subsequently constructing the strange correlator $\langle \Omega(\beta) | \Psi\rangle$, each such boundary condition $\beta$ is in 1-1 correspondence with an SPT phase in 2+1 D with global symmetry $G$ when the 4-cocycle $\alpha$ is trivial. This generalizes explicit realization of the holographic relation discussed in the previous sections to the case of 3+1 D topological order / 2+1 D gapped theories with symmetries $G$. The construction should work for a generic categorical symmetry describable by 3+1 D lattice models with a  2-fusion category as input category.

\subsection{CFTs from phase transitions between fixed points -- 2+1 D Ising as an example} \label{sec:3Dnum}

Much like the situation of constructing partition functions in 2D from 3D topological order, we can obtain 3D CFT partition functions by looking for phase transitions between the topological fixed points to the RG operator we have defined above.

A judicious parametrization of the boundary state $\langle \Omega |$ is important towards obtaining a critical point numerically.

We will work with an explicit example of the 3+1 D toric code \cite{Hamma:2004ud}, corresponding to the $G= \mathbb{Z}_2$ DW theory in 4D.
In this case $H^4(\mathbb{Z}_2, U(1))$ is trivial and the 4-cocycle satisfies $\alpha =1$ when the edge group elements $h_i \in \mathbb{Z}_2$ satisfy $h_1 h_2 = h_3$ wherever the edges concerned form a closed triangle, and zero otherwise.

\begin{figure}
	\centering
	\includegraphics[width=0.4\linewidth]{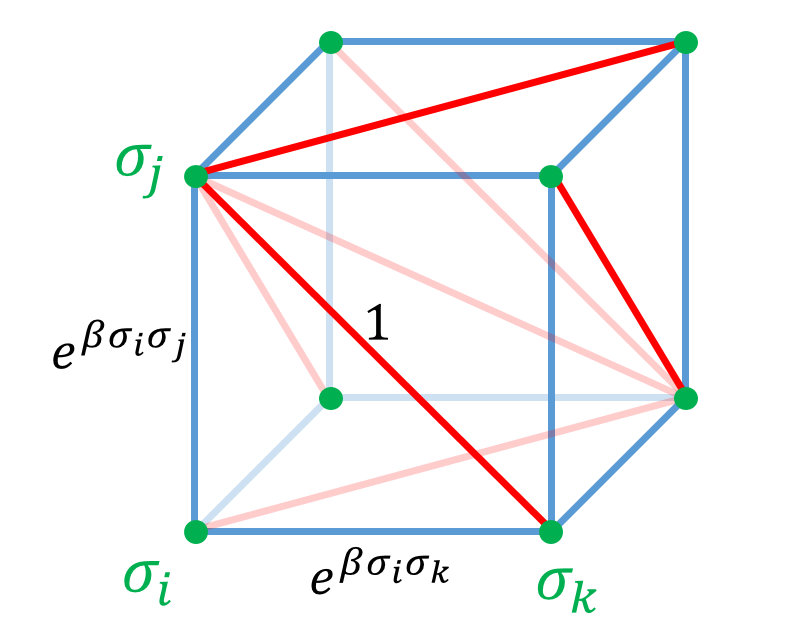}
	\caption{There are two types of boundary edges: the blue ones and the red ones. The boundary edges are determined by the bulk configuration $\{\sigma_i\}$, and we can read off their contribution to $\bigotimes_e\la w_e|\{a\}_{\{x\}}\ra $.
}
	\label{bluered}
\end{figure}

First, we note that the partition function of the well-known 2+1 D lattice Ising model is recovered as a strange correlator between a direct product state $\langle \Omega|$ and  the 3+1D toric code ground state wavefunction $|\Psi\rangle$. We will take the triangulation as shown in figure \ref{bigcube} and make use of the tensor network construction described there to recover its wave-function.
The boundary state is chosen as follows:
\be\label{Omega1}
\la \Omega|=\sum_{\{a\}}\Omega_{\{a\}}\la \{a\}|=\bigotimes_e\la w_e|,
\ee
where $\la w_e|$ is a state on the boundary edge $e$. We divide the boundary edges into two types: the blue ones and the red ones as shown in figure \ref{bluered}. And $\la w_e|$ is chosen as
\be\label{Omega2}
\la w_e|=\left\{
 \begin{aligned}
 \la 1|+\la -1|&,& \textrm{for red edges},\\
 e^\beta\la 1|+e^{-\beta}\la -1|&,&\textrm{for blue edges}.
 \end{aligned}
 \right.
\ee

Each group element $\sigma \in \mathbb{Z}_2$ assigned to an edge is denoted $\sigma \in \{1,-1\}$.
Then the bulk configuration $\{x\}$ becomes $\{\sigma_i\}$, where $\sigma_i$ labels the bulk edges. Given a specific bulk configuration $\{\sigma_i\}$, each boundary edge is determined by the bulk edges living on its two ends. If these two bulk edges are $\sigma_i$ and $\sigma_j$, then the boundary edge is $\sigma_i\sigma_j$ according to the fusion rules. When we consider $\bigotimes_e\la w_e|\{a\}_{\{x\}}\ra $ inside which $\{a\}_{\{x\}}$ is the boundary configuration fully determined by $\{x\}$, this boundary edge will contribute a factor
\be
\left\{
 \begin{aligned}
 1&,& \textrm{for red boundary edges},\\
 e^{\beta \sigma_i\sigma_j} &,&\textrm{for blue boundary edges},
 \end{aligned}
 \right.  \label{eq:ising2}
\ee
as shown in Fig\ref{bluered}. Finally we obtain
\be  \label{eq:state_ising}
\la \Omega|\Psi\ra_{\mathcal{M}} = \sum_{\{\sigma_i\}}\prod_{\la ij\ra}e^{\b \s_i\s_j},
\ee
where $\la ij\ra$ means the nearest neighbor on the cubic lattice. One can note that it is exactly the partition function of 3D Ising model.

\subsection{Boundary state under RG}
\begin{figure}
	\centering
	\includegraphics[width=0.3\linewidth]{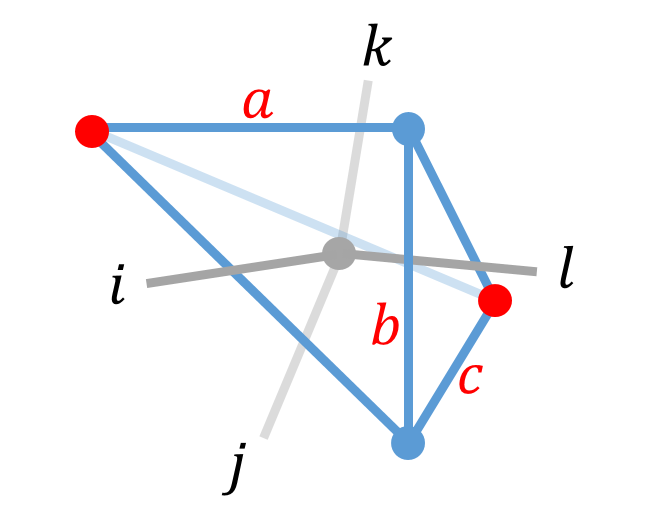}
	\caption{The labels on a tetrahedron. Here $a,b,c$ are the objects on the edges, and $i,j,k,l$ are labels on the faces. The grey lines here are for clear indication of the faces.
}
	\label{block}
\end{figure}

The 3D boundary is divided into similar tetrahedra.
Focusing on one tetrahedron, each face of it is shared by another tetrahedron. This inspires us to introduce face labels which will be contracted. And then we can construct a tensor network type wave function. Its building blocks are these tetrahedra. Given a tetrahedron labeled as shown in figure \ref{block}, it will contribute a tensor $u^{ijkl}(a,b,c)$, where $i,j,k,l$ are labels with dimension $N$ on the faces and $a,b,c$ are the objects (1 or -1) on the edges. One tetrahedron has 6 edges, but the objects on them are related by fusion rules, and where the bulk is a Dijkgraaf-Witten type lattice gauge theory, there are only 3 independent labels. So the tensor $u$ only has 3 arguments $a,b,c$.

\begin{figure}
	\centering
	\includegraphics[width=0.8\linewidth]{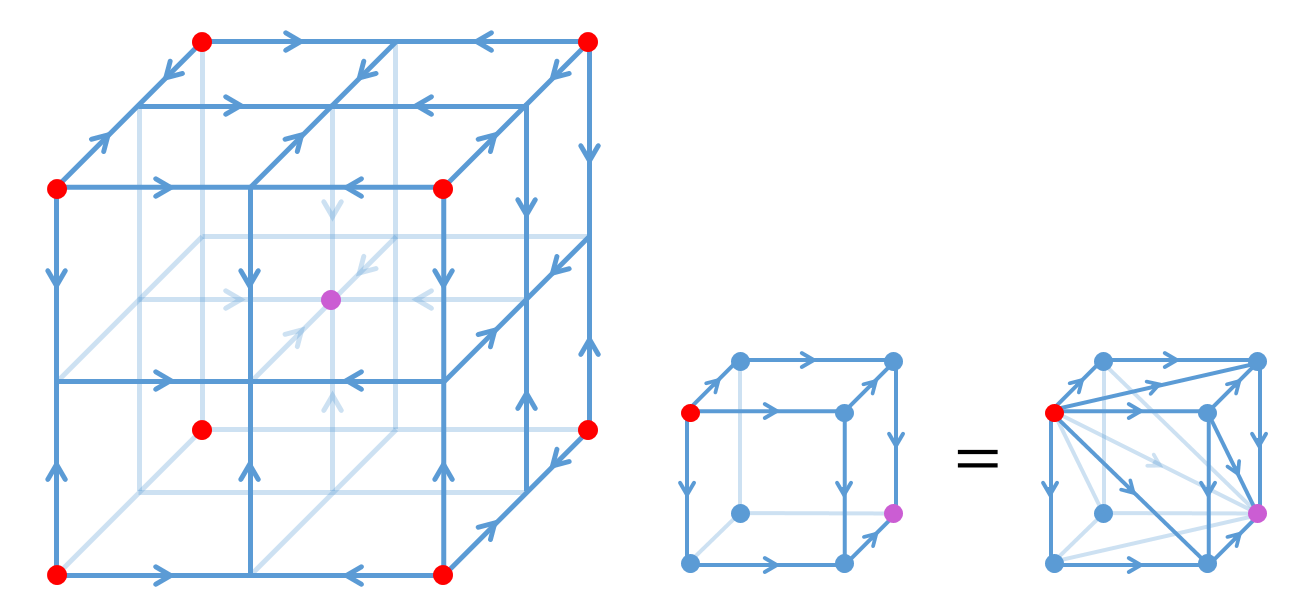}
	\caption{Introduce arrows to the edges of the tetrahedra. The center vertex of the $2\times2\times2$ cube is colored in purple now. Roughly speaking, the arrows point out from the red dots, and point to the purple dot. To avoid clutter, we don't show all the arrows on the edges.
}
	\label{arrows}
\end{figure}

\begin{figure}
	\centering
	\includegraphics[width=0.25\linewidth]{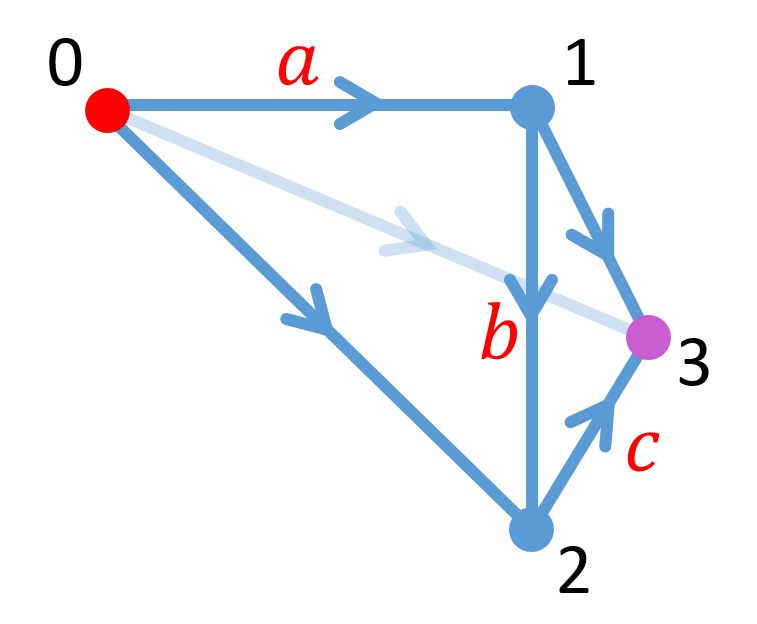}
	\caption{One tetrahedron with arrows introduced.
}
	\label{tarrows}
\end{figure}

In $u^{ijkl}(a,b,c)$, the order of indexes and the order of arguments are important. Now we introduce rules for reading off the tensor from the tetrahedron. The first step is to introduce arrows on the edges of the tetrahedra as shown in figure \ref{arrows}. Let us focus on one tetrahedron as shown in figure \ref{tarrows}. We label its vertices by $0,1,2,3$ such that the arrows point from the small numbers to the large numbers. This label is unique. The red dot is always 0, and the purple dot is always 3. Then we label the $01,12,23$ edges by $a,b,c$ respectively. They are chosen as the 3 independent labels of edges. And we label the $012,023,013,123$ faces by $i,j,k,l$ respectively. The tensor corresponding to this tetrahedron is written as $u^{ijkl}(a,b,c)$.

Each cube is divided into 6 tetrahedra. They are related with each other by 3D rotations and 3D reflections. So we assume that the tensors corresponding to them are the same $u$. And the tensor network type wave function is given by
\be
\langle \Omega (u)| = \sum_{\{a\}}\sum_{i,j,k,l\dots}\dots u^{ijkl}(a_1,a_2,a_3)\dots \la \{a\}| ,  \label{eq:omegau}
\ee
where $\sum_{\{a\}}$ means the summation over all allowed boundary configurations, and $\sum_{i,j,k,l,\dots}$ is the summation over the labels of faces. Each tetrahedron on the boundary will contribute a $u$ tensor, and $\dots u^{ijkl}(a_1,a_2,a_3)\dots$ is the product of these $u$ tensors. Now it becomes obvious that face labels $i,j,k,l,\dots$ are the indexes to be contracted.

\begin{figure}
	\centering
	\includegraphics[width=0.3\linewidth]{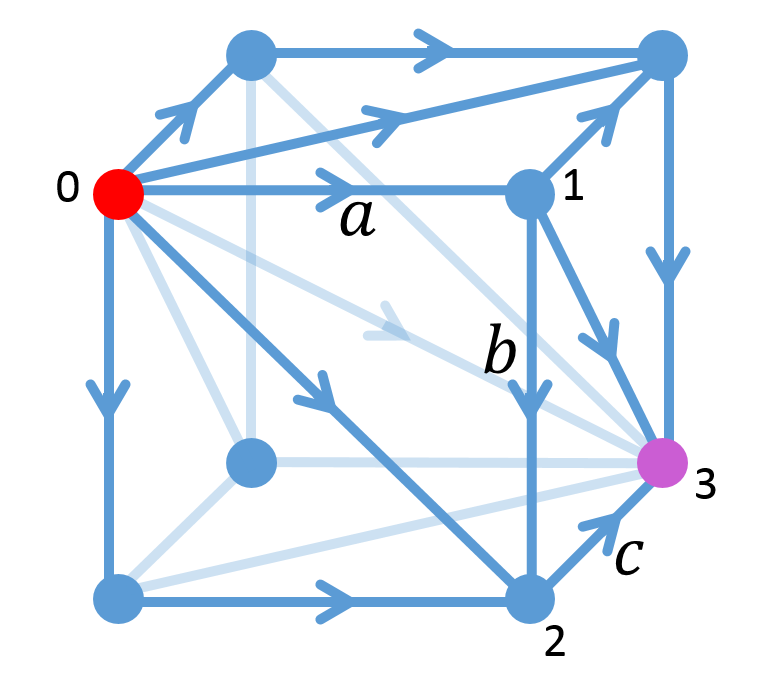}
	\caption{The 0123 tetrahedron in the cube. The labels $a,b,c$ lie on the previous blue edges of the cube.
}
	\label{tincube}
\end{figure}

We can recover the previous $\la \Omega |$ in equation (\ref{Omega1}, \ref{Omega2}) by choosing the $u^{ijkl}(a,b,c)$ to be
\be
u^{ijkl}(a,b,c)=\left\{
 \begin{aligned}
 e^{\frac{\beta(a+2b+c)}{8} }&,& i=j=k=l=1,\\
 0&,&\textrm{Otherwise},
 \end{aligned}
 \right.
\ee
where $a,b,c$ take values $\{1,-1\}$ corresponding to objects $\{1,-1 \}$. Let's briefly explain how to obtain the above $u$. In (\ref{eq:ising2}), blue edges in figure \ref{bluered} labeled by $a$ will contribute a factor $e^{\beta a}$ to $\la \Omega|\Psi\ra$. Note that $a,b,c$ in figure \ref{tincube} correspond to the blue edges in figure \ref{bluered}. The edge labeled by $a$ is shared by 8 tetrahedra, and our choice of orientation is such that $a$ is the first argument (according to the rule in figure \ref{tarrows}) in all of them. Similarly, the edge labeled by $b$ is shared by 4 tetrahedra, and $b$ is the second argument in all of them. Finally, the edge labeled by $c$ is shared by 8 tetrahedra, and $c$ is the third argument in all of them.

For simple illustration how to follow the flow of the boundary state (\ref{eq:omegau}) under repeated use of the above RG map,  in the following we consider the case that the dimension of face labels is 1 i.e. $N=1$.
And $u^{ijkl}(a,b,c)$ reduces to $u(a,b,c)$.

\begin{figure}
	\centering
	\includegraphics[width=0.6\linewidth]{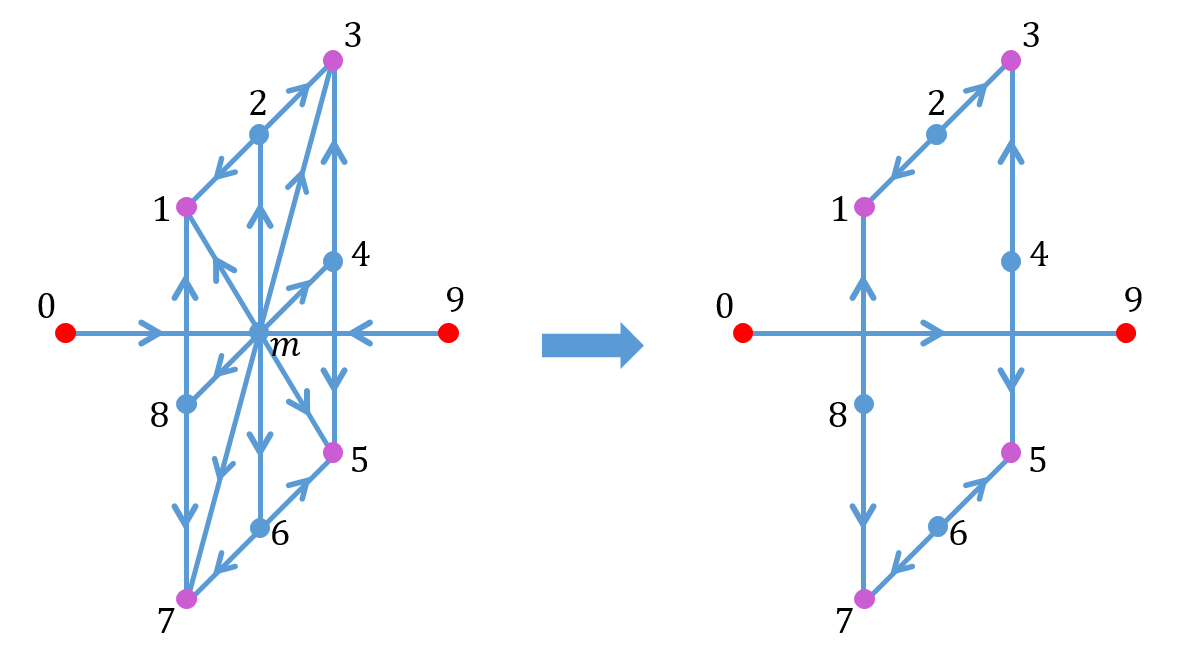}
	\caption{The first step, first illustrated in figure \ref{step1}, now with arrows introduced. Here we relabel the vertices for convenience. These labels are just for indicating edges and faces. The purple dots are centers of $2\times2\times2$ cubes.
}
	\label{step1a}
\end{figure}

As said before, one RG step consists of 3 steps. Let us recall the first step, now as shown in figure \ref{step1a} with arrows introduced.
The left hand side of figure \ref{step1a} depicts the lattice before the RG step, and the right hand side the result of the RG step. The left and the right  of figure \ref{step1a} are respectively on the two boundaries of a 4D body which consists of eight 4-simplices. Now the 4D body becomes an operator $\mathcal{R}_1$ which consists of eight 4-cocycles $\alpha$ corresponding to these 4-simplices. And it maps the state defined on the left hand side to the state defined on the right hand side.
The operator $\mathcal{R}_1$ imposes that edges forming a triangle have to product to the identity. For example, in the 4-simplex $09m21$ there is a face $09m$, and we should have $a_{0m}a_{9m}=a_{09}$ where $a_{0m}$ is the object on the edge $0m$.

To satisfy the fusion rules, on the right hand side of figure \ref{step1a} there are 9 independent labels which are chosen as $a_{01},a_{02},a_{03},a_{04},a_{05},a_{06},a_{07},
a_{08},a_{09}$. Then the labels on other edges are fixed by them according to the fusion rules. For example $a_{21}=a_{02}a_{01}$. And the state defined on the right hand side takes the form
\be
\sum_{a_{01},a_{02},\dots,a_{09}}\mathcal{M}^R_1(a_{01},a_{02},
\dots,a_{09}) \la a_{01},a_{02},\dots,a_{09} |.   \label{eq:RGedwf}
\ee

On the left of figure \ref{step1a}, there is one extra vertex $m$. Compared to the right, there is an extra independent edge degree of freedom $a_{0m}$. Given $a_{0m},a_{01},a_{02},a_{03},a_{04},a_{05},a_{06},a_{07},
a_{08},a_{09}$, all the labels on the left hand side can be fixed. For example, $a_{9m}=a_{09}a_{0m}$ when we consider the face $09m$. The state defined on the left hand side takes the form
\be
\sum_{a_{0m},a_{01},\dots,a_{09}}\mathcal{M}^L_1(a_{0m},a_{01},
\dots,a_{09}) \la a_{0m},a_{01},\dots,a_{09} |.
\ee
Each tetrahedron will contribute a $u$ tensor. And we can read off $\mathcal{M}^L_1$ as
\bea
\nonumber
\mathcal{M}^L_1&=&
u(a_{0m},a_{m2},a_{21})u(a_{0m},a_{m2},a_{23})
u(a_{0m},a_{m4},a_{43})\\
\nonumber
&\;&\times
u(a_{0m},a_{m4},a_{45}) u(a_{0m},a_{m6},a_{65})u(a_{0m},a_{m6},a_{67})\\
\nonumber
&\;&\times
u(a_{0m},a_{m8},a_{87})u(a_{0m},a_{m8},a_{81}) u(a_{9m},a_{m2},a_{21})\\
\nonumber
&\;&\times
u(a_{9m},a_{m2},a_{23})
u(a_{9m},a_{m4},a_{43})u(a_{9m},a_{m4},a_{45})\\
\nonumber
&\;&\times u(a_{9m},a_{m6},a_{65})u(a_{9m},a_{m6},a_{67})
u(a_{9m},a_{m8},a_{87})\\
&\;&\times u(a_{9m},a_{m8},a_{81}).
\eea
Here all the $a$'s can be expressed by $a_{0m}$, $a_{01}$,$a_{02}$,$\dots$,$a_{09}$. So the arguments of $\mathcal{M}^L_1$ are exactly $a_{0m}$,$a_{01}$,$a_{02}$,$\dots$,$a_{09}$.  $\mathcal{M}^R_1$ and $\mathcal{M}^L_1$ are related under RG by
\be
\mathcal{M}^R_1(a_{01},a_{02},
\dots,a_{09})=\sum_{a_{0m}}\mathcal{M}^L_1(a_{0m},a_{01},a_{02},
\dots,a_{09}).
\ee
When the labels $a_{01}$,$a_{02}$,$\dots$,$a_{09}$ of $\mathcal{M}^R_1$ are given, the labels $a_{01}$,$a_{02}$,$\dots$,$a_{09}$ in $\mathcal{M}^L_1$ are fixed. The edge $a_{0m}$ is an {\it interior edge} now buried between the left boundary condition and the RG operator $\mathcal{R}_1$, which should thus be summed over.
 In the 3+1 D toric code model, the RG operator is unity where fusion constraints on each face are satisfied.

In the 3D case, we follow the RG flow by performing an SVD to convert the boundary condition into a local form after every RG step as depicted in figure \ref{fig:3dblocking}.  We pursue a similar procedure here. But instead of breaking the blocked wavefunction (\ref{eq:RGedwf}) into two pieces, we need to break it into 8 pieces, one attached to each remaining tetrahedron on the right hand side of figure \ref{step1a}.    Assume that after the {\it break-up},  each tetrahedron on the right hand side of figure \ref{step1a} is attached to a tensor $v(a,b,c)$ .  We denote the state constructed from $v(a,b,c)$ as
\be
\sum_{a_{01},a_{02},\dots,a_{09}}\widetilde{\mathcal{M}}^R_1(a_{01},a_{02},
\dots,a_{09}) \la a_{01},a_{02},\dots,a_{09}|,
\ee
where $\widetilde{\mathcal{M}}^R_1$ is given by
\bea
\nonumber
\widetilde{\mathcal{M}}^R_1&=&v(a_{09},a_{92},a_{21})
v(a_{09},a_{92},a_{23})v(a_{09},a_{94},a_{43})\\
\nonumber
&\;&\times
v(a_{09},a_{94},a_{45})
v(a_{09},a_{96},a_{65})v(a_{09},a_{96},a_{67})\\
&\;&\times
v(a_{09},a_{98},a_{87})v(a_{09},a_{98},a_{81}).
\eea
Here all the $a$'s can be expressed by $a_{01},a_{02},\dots,a_{09}$. So the arguments of $\widetilde{\mathcal{M}}^R_1$ are exactly $a_{01},a_{02},\dots,a_{09}$.
Insisting that the face-degrees of freedom remain cut-downed to 1-dimensional (which is imposed  by hand when we pick the tensor $v(a,b,c)$ to carry no extra indices attached to the faces),  it is generically not possible to reproduce $\mathcal{M}^R_1$ precisely. We will try to optimise $v(a,b,c)$ such that
\be
\epsilon_1 \equiv \sum_{a_{01},\dots,a_{09}} |\mathcal{M}^R_1(a_{01},
\dots,a_{09})-\widetilde{\mathcal{M}}^R_1(a_{01},
\dots,a_{09})|^2
\ee
is minimal. The optimisation is done numerically here. Once $v(a,b,c)$ is found, we normalize it to
\be
v(a,b,c)\rightarrow \frac{v(a,b,c)}{\sqrt{\frac{1}{8}\sum_{a,b,c}|v(a,b,c)|^2}}.
\ee

\begin{figure}
	\centering
	\includegraphics[width=0.6\linewidth]{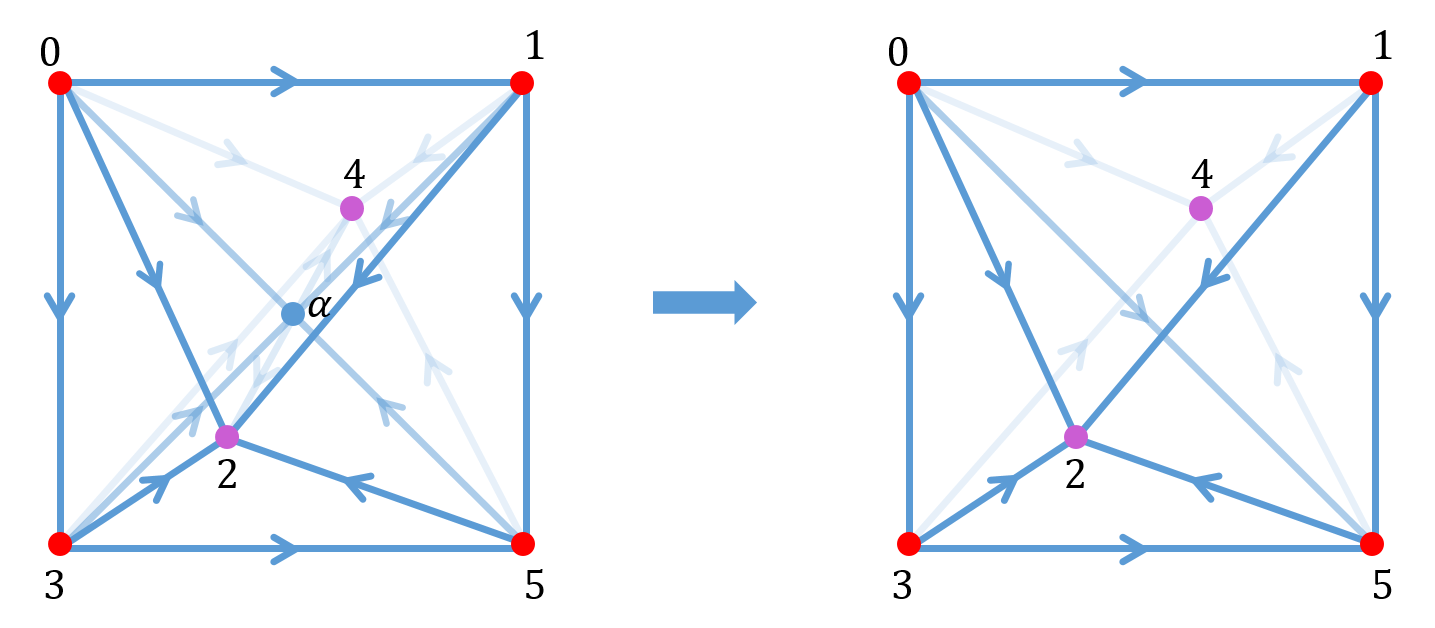}
	\caption{The second step with arrows introduced. Here we relabel the vertices for convenience. The purple dots are centers of $2\times2\times2$ cubes.
}
	\label{step2a}
\end{figure}

Now let us consider the second step. The procedure is similar. With arrows introduced the second step is shown in figure \ref{step2a}. On the right hand side there are 5 independent labels which are chosen as $a_{01},a_{02},a_{03},a_{04},a_{05}$. And the state defined on the right hand side takes the form
\be
\sum_{a_{01},a_{02},\dots,a_{05}}\mathcal{M}^R_2(a_{01},a_{02},
\dots,a_{05}) \la a_{01},a_{02},\dots,a_{05}|.
\ee
On the left hand side there are 6 independent labels which are chosen as $a_{0\a},a_{01},a_{02},a_{03},a_{04},a_{05}$. And the state defined on the left hand side takes the form
\be
\sum_{a_{0\a},a_{01},\dots,a_{05}}\mathcal{M}^L_2(a_{0\a},a_{01},
\dots,a_{05}) \la a_{0\a},a_{01},\dots,a_{05}|.
\ee
Previously we introduce $v(a,b,c)$ and approximate $\mathcal{M}^R_1$ by $\widetilde{\mathcal{M}}^R_1$. Now each tetrahedron on the left hand side will contribute $v(a,b,c)$. And we can read off $\mathcal{M}^L_2$ as
\bea
\nonumber
\mathcal{M}^L_2 &=& v(a_{01},a_{1\a},a_{\a 2})
v(a_{03},a_{3\a},a_{\a 2})
v(a_{15},a_{5\a},a_{\a 2})\\
\nonumber
&\;&\times v(a_{35},a_{5\a},a_{\a 2}) v(a_{01},a_{1\a},a_{\a 4})
v(a_{03},a_{3\a},a_{\a 4})\\
&\;&\times
v(a_{15},a_{5\a},a_{\a 4})v(a_{35},a_{5\a},a_{\a 4}).
\eea
Here all the $a$'s can be expressed by $a_{0\a}$,$a_{01}$,$a_{02}$,$\dots$,$a_{05}$. So the arguments of $\mathcal{M}^L_2$ are exactly $a_{0\a}$,$a_{01}$,$a_{02}$,$\dots$,$a_{05}$. And we can obtain $\mathcal{M}^R_2$ from $\mathcal{M}^L_2$ by the formula
\be
\mathcal{M}^R_2(a_{01},a_{02},
\dots,a_{05})=\sum_{a_{0\a}}\mathcal{M}^L_2(a_{0\a},a_{01},a_{02},
\dots,a_{05}).
\ee
Now we have to repeat the procedure above to break up  $\mathcal{M}^R_2$.
We introduce another new function $t(a,b,c)$ to the tetrahedra on the right hand side of figure \ref{step2a} from which we define the state
\be
\sum_{a_{01},a_{02},\dots,\a_{05}}\widetilde{\mathcal{M}}^R_2(a_{01},a_{02},
\dots,a_{05}) \la a_{01},a_{02},\dots,a_{05}|,
\ee
where $\widetilde{\mathcal{M}}^R_2$ is given by
\bea
\nonumber
\widetilde{\mathcal{M}}^R_2 &=& t(a_{01},a_{15},a_{52})
t(a_{03},a_{35},a_{52})t(a_{01},a_{15},a_{54})\\
&\;&\times
t(a_{03},a_{35},a_{54}).
\eea
Here all the $a$'s can be expressed by $a_{01},a_{02},\dots,a_{05}$. So the arguments of $\widetilde{\mathcal{M}}^R_2$ are exactly $a_{01},a_{02},\dots,a_{05}$. Now we would like to find the best $t(a,b,c)$ to approximate $\mathcal{M}^R_2$ by $\widetilde{\mathcal{M}}^R_2$ such that
\be
\epsilon_2 \equiv \sum_{a_{01},\dots,a_{05}} |\mathcal{M}^R_2(a_{01},
\dots,a_{05})-\widetilde{\mathcal{M}}^R_2(a_{01},
\dots,a_{05})|^2
\ee
is minimal. This is again achieved by numerical method. Once $t(a,b,c)$ is found, we normalize it to
\be
t(a,b,c)\rightarrow \frac{t(a,b,c)}{\sqrt{\frac{1}{8}\sum_{a,b,c}|t(a,b,c)|^2}}.
\ee

\begin{figure}
	\centering
	\includegraphics[width=0.8\linewidth]{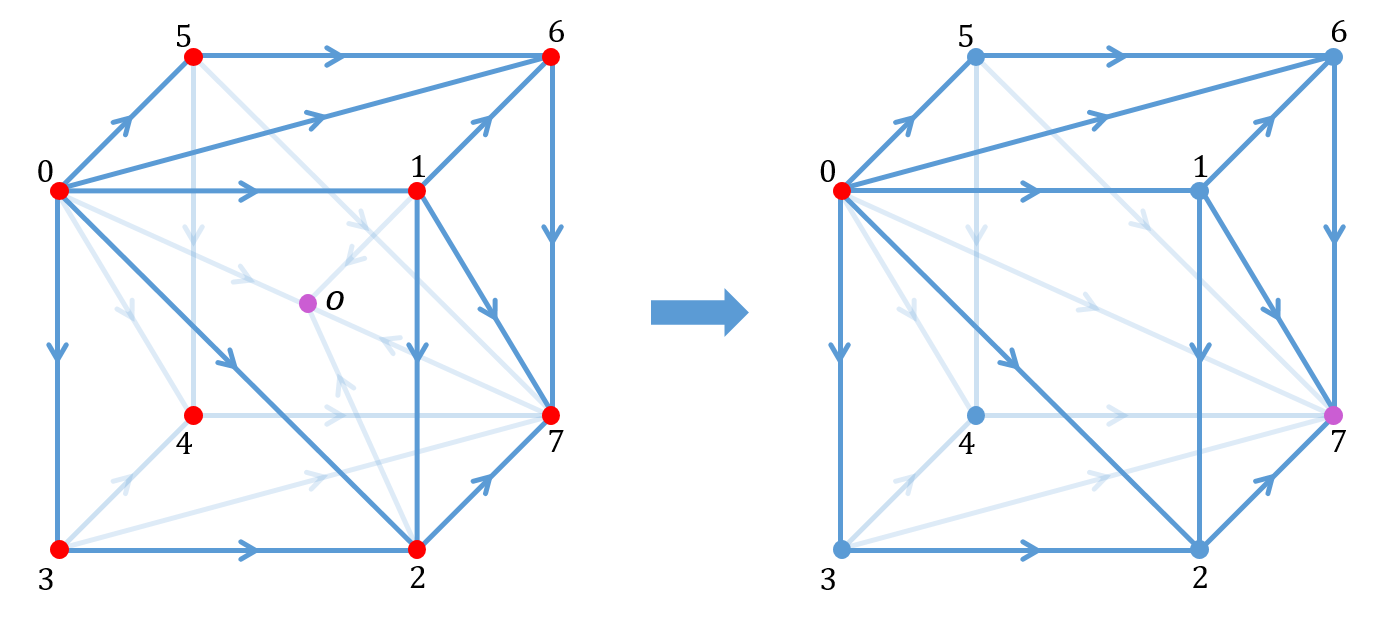}
	\caption{The third step with arrows introduced. Here we relabel the vertices for convenience. Vertex $o$ is the center of the original $2\times2\times2$ cube, and vertex $7$ becomes the center of the coarse grained $2\times2\times2$ cube.
}
	\label{step3a}
\end{figure}

We would then proceed to the third step. With arrows introduced the third step is shown in figure \ref{step3a}. On the right hand side there are 7 independent labels which are chosen as $a_{01}$,$a_{02}$,$a_{03}$,$a_{04}$,$a_{05}$,$a_{06}$,$a_{07}$. And the state defined on the right hand side takes the form
\be
\sum_{a_{01},a_{02},\dots,a_{07}}\mathcal{M}^R_3(a_{01},a_{02},
\dots,a_{07}) \la a_{01},a_{02},\dots,a_{07}|.
\ee
On the left hand side there are 8 independent labels which are chosen as $a_{0o},a_{01},a_{02},\dots,a_{07}$. And the state defined on the left hand side takes the form
\be
\sum_{a_{0o},a_{01},\dots,a_{07}}\mathcal{M}^L_3(a_{0o},a_{01},
\dots,a_{07}) \la a_{0o},a_{01},\dots,a_{07} |.
\ee
Previously we introduce $t(a,b,c)$ and approximate $\mathcal{M}^R_2$ by $\widetilde{\mathcal{M}}^R_2$. Now each tetrahedron on the left hand side of figure \ref{step3a} will contribute $t(a,b,c)$. And we can read off $\mathcal{M}^L_3$ as
\bea
\nonumber
\mathcal{M}^L_3 &=&
t(a_{01},a_{12},a_{2o})
t(a_{03},a_{32},a_{2o})
t(a_{01},a_{16},a_{6o})\times \\
\nonumber
&\;&
t(a_{05},a_{56},a_{6o})t(a_{05},a_{54},a_{4o})
t(a_{03},a_{34},a_{4o})\times \\
\nonumber
&\;&
t(a_{12},a_{27},a_{7o})
t(a_{16},a_{67},a_{7o})t(a_{56},a_{67},a_{7o})\times \\
&\;&
t(a_{54},a_{47},a_{7o})
t(a_{32},a_{27},a_{7o})
t(a_{34},a_{47},a_{7o})
\eea
Here all the $a$'s can be expressed by $a_{0o},a_{01},a_{02},\dots,a_{07}$. So the arguments of $\mathcal{M}^L_3$ are exactly $a_{0o},a_{01},a_{02},\dots,a_{07}$. And we can obtain $\mathcal{M}^R_3$ from $\mathcal{M}^L_3$ by the formula
\be
\mathcal{M}^R_3(a_{01},a_{02},
\dots,a_{07})=\sum_{a_{0o}}\mathcal{M}^L_3(a_{0o},a_{01},a_{02},
\dots,a_{07}).
\ee
Now we introduce a new function $u^{(1)}(a,b,c)$ to the tetrahedra on the right hand side of figure \ref{step3a}. And we can obtain the state
\be
\sum_{a_{01},a_{02},\dots,\a_{07}}\widetilde{\mathcal{M}}^R_3(a_{01},a_{02},
\dots,a_{07}) \la a_{01},a_{02},\dots,a_{07} |,
\ee
where $\widetilde{\mathcal{M}}^R_3$ is given by
\bea
\nonumber
\widetilde{\mathcal{M}}^R_3 &=& u^{(1)}(a_{01},a_{12},a_{27})
u^{(1)}(a_{01},a_{16},a_{67})\\
\nonumber
&\;& \times
u^{(1)}(a_{05},a_{56},a_{67})
u^{(1)}(a_{05},a_{54},a_{47})\\
&\;& \times
u^{(1)}(a_{03},a_{34},a_{47})
u^{(1)}(a_{03},a_{32},a_{27}).
\eea
Here all the $a$'s can be expressed by $a_{01},a_{02},\dots,a_{07}$. So the arguments of $\widetilde{\mathcal{M}}^R_3$ are exactly $a_{01},a_{02},\dots,a_{07}$. Now we would like to find the best $u^{(1)}(a,b,c)$ to approximate $\mathcal{M}^R_3$ by $\widetilde{\mathcal{M}}^R_3$ such that
\be
\epsilon_3 \equiv \sum_{a_{01},\dots,a_{07}} |\mathcal{M}^R_3(a_{01},
\dots,a_{07})-\widetilde{\mathcal{M}}^R_3(a_{01},
\dots,a_{07})|^2
\ee
is minimal. This is achieved by numerical method. Once $u^{(1)}(a,b,c)$ is found, we normalize it to
\be
u^{(1)}(a,b,c)\rightarrow \frac{u^{(1)}(a,b,c)}{\sqrt{\frac{1}{8}\sum_{a,b,c}|u^{(1)}(a,b,c)|^2}}.
\ee

We can note that after one RG step, the wave function becomes
\be
\la \Omega(u^{(1)})|= \sum_{\{a\}} \dots u^{(1)}(a_1,a_2,a_3)\dots \la \{a\} |,
\ee
which takes the same form as the origin wave function in (\ref{eq:omegau}). The only difference is replacing $u(a,b,c)$ by $u^{(1)}(a,b,c)$.
And we can iterate this RG step. After $n$ RG steps, we would get $u^{(n)}(a,b,c)$, and the wave function is
\be
\la \Omega(u^{(n)})|= \sum_{\{a\}} \dots u^{(n)}(a_1,a_2,a_3)\dots \la \{a\}|.
\ee
We expect that when $n$ is large enough, $u^{(n)}(a,b,c)$ will arrive at some fixed point and $\la \Omega(u^{(n)})|$ becomes one of the fixed point wave functions.

Recall that when we choose $u(a,b,c)$ as
\be
u_\b(a,b,c)\equiv e^{\frac{\beta(a+2b+c)}{8}},
\ee
the overlap $\la \Omega|\Psi\ra_{\mathcal{M}} $ is the partition function of 3D Ising model. And it is interesting to investigate where $u_\b(a,b,c)$ will flow to. There is a parameter $\beta$ in $u_\b(a,b,c)$. We find that when changing $\beta$, the $u_\b(a,b,c)$ will flow to different fixed points. And it is easy to check that $u_0(a,b,c)$ and $u_\infty(a,b,c)$ are two fixed points.
Our numerical result shows that when $0\leq\beta\leq0.27$, $u_\b(a,b,c)$ will flow to the fixed point $u_0(a,b,c)$. However when $\b= 0.28$, our numerical method breaks down since the errors $\epsilon_1,\epsilon_2,\epsilon_3$ wouldn't converge to 0 when $n$ becomes larger. And this suggests that something special happens in the range $\beta\in(0.27,0.28)$ --  which is a signature of getting close to the critical point of the 3D Ising model. In literature (See for example for a collection of numerical results in \cite{Hasenbusch_2010}), the critical point of the 3D Ising model is $\b_c=0.22165463(8)$. Our result is surprisingly good with a choice of face bond dimension truncated down to $N=1$. Increasing the face dimension $N$ should improve our accuracy. This will be demonstrated in a forth-coming paper\cite{kaixin}.

 \section{Tensor Network reconstruction of AdS/CFT? -- Some Numerics } \label{sec:holo}
We have proposed construction of RG operators analytically from topological data of $D+1$ dimensional topological lattice model, and demonstrated explicitly in $D+1=2,3,4$ dimensions in a family of models. Since these are operators responsible for blocking, repeated concatenation of them  $\lim_{N\to\infty} U(\mathcal{C})^N$ produces a MERA like network, which is thus an analytic holographic network, which, with the right boundary condition that is itself the fixed point or one that could flow to the fixed point, describes exact CFT's. We have given examples which map to well known 2D integrable models, or the 3D Ising model.  

These strange correlators bear a close resemblance to the p-adic tensor network that reproduces many aspects of the p-adic AdS/CFT dictionary \cite{Hung:2019zsk, Chen:2021qah, Chen:2021ipv}.The resemblance lies in the fact that the bulk of the tensor network is characterized by an associative algebra (the fusion algebra of the CFT operators) and this is generalized to (higher) fusion categories in the general construction in this paper. The p-adic tensor network that reproduces the p-adic CFT partition function also comes in the form of a strange correlator, such that at the fixed point, the boundary state is an eigenstate of the bulk holographic network.
 It is thus tempting that the current holographic network may bear resemblance to the AdS/CFT correspondence.

\subsection{Bulk-Boundary Correlators}
For simplicity, we work with the 2D Ising CFT that can be obtained from the $SU(2)_3$. The CFT has a small central charge and is not expected to have a semi-classical gravitational dual. However we would like to see how such a broad-brush would work.
We computed numerically two point functions where one operator is inserted at the UV layer given by figure \ref{fig:Isingbc}, and the other at the $n$-th layer of the holographic network. i.e.
\begin{align}
&\langle \mathcal{O}(n=0,x=y=0) \mathcal{O}(n,x,y) \rangle  \nonumber \\
&= \langle \Omega(T_{\Lambda_0})| \sigma^z(x=y=0) U^{n}(\mathcal{C}) \sigma^z(x,y) U^{n+1}(\mathcal{C}) \cdots |\Psi\ra
   \nonumber \\
&=  \langle \Omega(T_{\Lambda_0})| \sigma^z(x=y=0) U^{n}(\mathcal{C}) \sigma^z(x,y) |\Psi_{\Lambda_n}\rangle.
\end{align}

\begin{figure}
	\centering
	\includegraphics[width=1\linewidth]{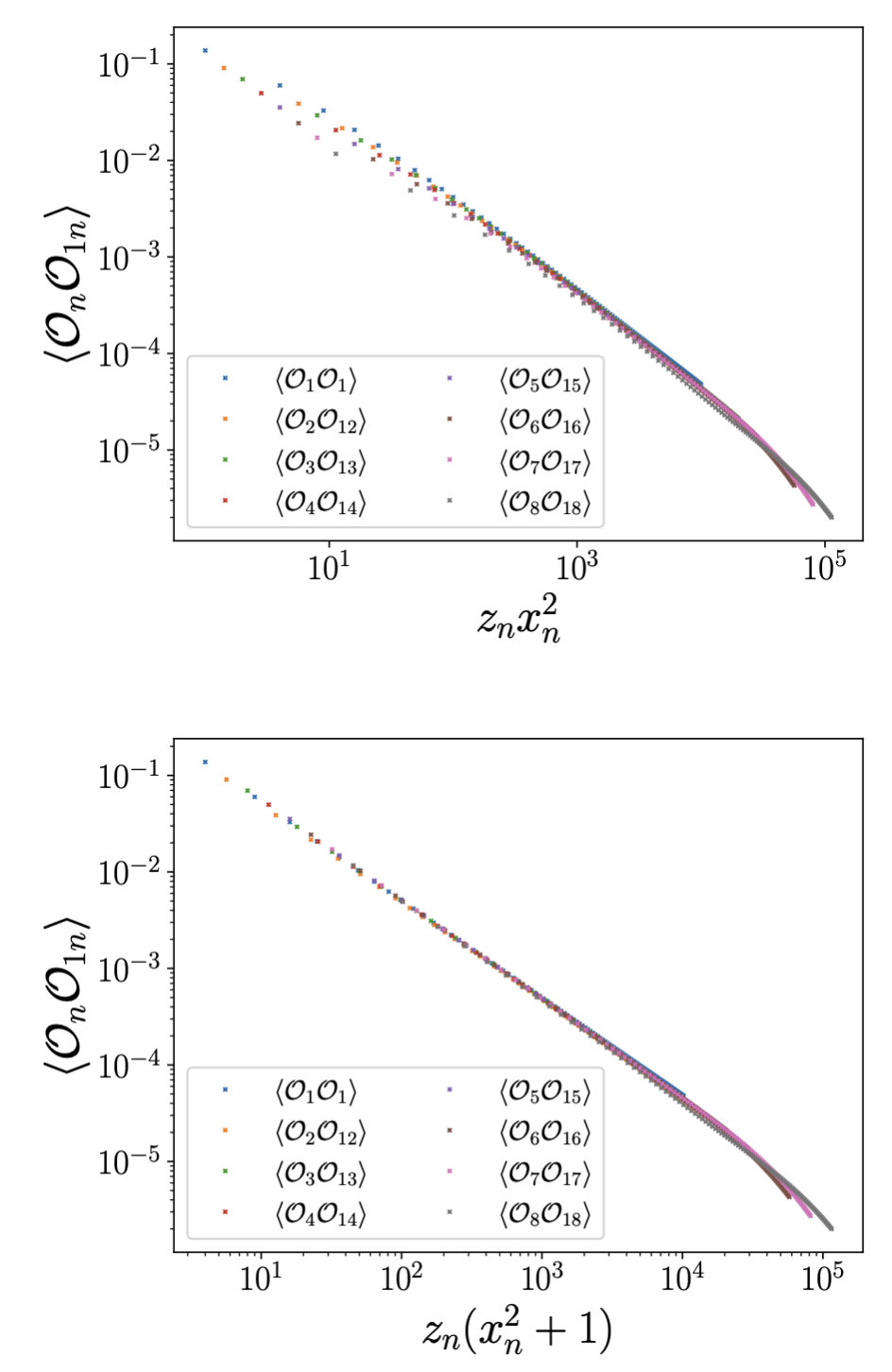}
	\caption{$\mathcal{O}_n$ denotes operator insertion at the $n$-th layer. $\mathcal{O}_{1n}$ corresponds to operation insertion at layer 1 (which is subsequently blocked with nearby tensors to the $n$-th layer -- for computational convenience). $\la\mathcal{O}_n \mathcal{O}_{1n}\ra $ is thus $\la\mathcal{O}_n \mathcal{O}_1\ra$.  We make a log-log plot against $z_n x_n^2$ at the top, and $z_n (1+x_n^2)$ in the bottom. Here, $z_n = (\sqrt{2})^{n-1} $ and $x_1 = z_n x_n $. The AdS/CFT dictionary would suggest that $\la\mathcal{O}_n \mathcal{O}_1\ra =  (z_n(x_n^2+1))^{-\Delta}$, where $\Delta$ is the conformal dimension of the CFT primary operator $\mathcal{O}_1$.
	All the plots of different $n$ collapse very nicely to the same line in the  bottom plot.}
	\label{fig:holographicplot}
\end{figure}

The result is illustrated in figure \ref{fig:holographicplot}. In the bottom plot, the plots at different $n$ collapse to the same plot in a more convincing way than the top plot, promisingly, suggesting that the ``bulk-boundary propagator'' in this holographic network is indeed given by $[z/(z^2 + x^2)]^{\Delta}$. This is, admittedly, not yet conclusive before we can explore more models and more propagators of different conformal dimensions.

\subsection{Reconstruction Wedge}
\begin{figure}
	\centering
	\includegraphics[width=1\linewidth]{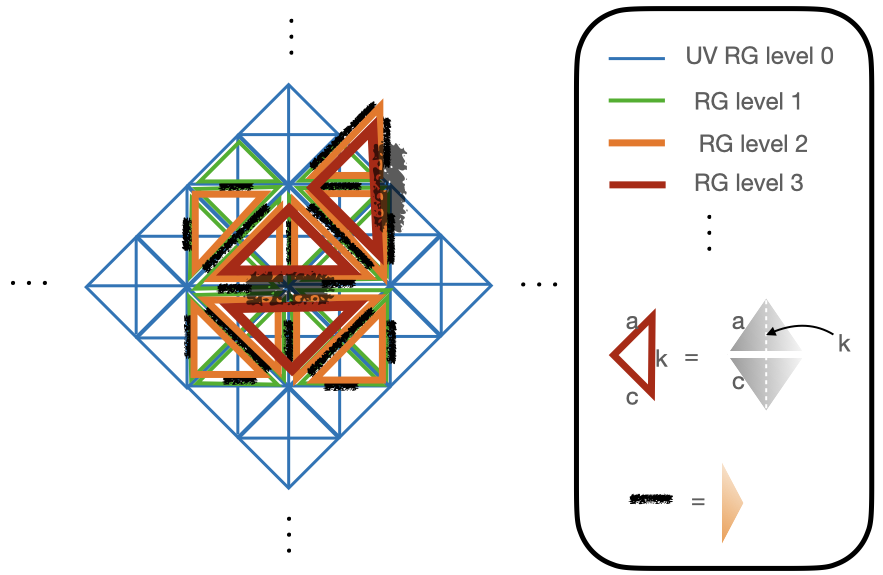}
	\caption{The picture depicts layers of tetrahedra on top of each other, implementing the RG process discussed in figure \ref{fig:Mong}, where two triangles are merged into one recursively. These layers of triangles are projections of the tetrahedron that is color - coded depending on the layer. The larger triangles contain smaller triangles from UV layers that eventually blocked into them.   The entanglement wedge of any edge belonging to any triangle at a given layer is given by all the smaller triangles contained inside this triangle. }
	\label{fig:entanglement_wedge}
\end{figure}

It is not hard to obtain the entanglement wedge, or reconstruction wedge of any edge in the interior.  In the tree RG network in 2D depicted in figure \ref{RG2d}, the entanglement wedge  at the UV boundary is self-explanatory.
In the case of a 3D bulk, the holographic network is depicted in figure \ref{fig:entanglement_wedge}. It is a projection of the 3D structure into two dimensions. To reconstruct action of edges in the interior layer from the UV layer, the UV degrees of freedom involved correspond to edges of triangles in the UV layer that appear to be contained inside the triangle in the interior layer concerned. The explicit transformation that converts an operator in the bulk into an operator from the UV boundary requires construction of the inverse of the holographic networks. This can be readily done using the orthogonality condition given by
\be
\sum_y d_y \left [ \begin{tabular}{ccc}  $a$& $b$& $x$ \\ $c$ & $d$ & $y$ \end{tabular} \right ] \left [ \begin{tabular}{ccc} $a$& $b$& $x'$ \\ $c$ & $d$ & $y$ \end{tabular} \right ] = \frac{\delta_{x x'} N^x_{ab}N^x_{cd}}{d_x}
\ee.

\section{Conclusions and Outlook} \label{sec:con}
Renormalisation group flow is the key connecting topological data in a TQFT characterising topological symmetries, the ``skeleton'' of a theory, to its ``flesh'' -- namely explicit path-integrals of a continuous field theory.  In this paper, we considered an exact holographic network that descends from coarse graining a topological wave-function. These holographic networks correspond to RG operators that preserve a given set of ``categorical symmetries'' associated to the TQFT in $D+1$ dimensions. Each fixed point of the RG operator determines either a TQFT or a CFT in $D$ dimensions that preserve the categorical symmetry concerned (or spontaneousely breaking part of it).  For bulk TQFTs in the family of Dijkgraaf-Witten models in arbitrary dimensions or Turaev type in 3 dimensions,  fixed points describing TQFT's are constructed using the (co-) product of (higher-) Frobenius algebra of the input fusion category $\mathcal{C}$.  

At $D=2$ we constructed exact strange correlators that reproduce RCFT path-integral. We explain at least from an RG standpoint why strange correlators could reproduce CFT partition functions -- the direct product state chosen for $\langle \Omega|$ is a seed boundary condition that could flow to the eigenstate of the RG operator constructed from the TQFT. The RG operator is producing an RG flow that explicitly protects the topological symmetries characterised by the TQFT. Our construction also provides an explicit relation between a discrete tensor network and the continuous path-integral of the field theory and directly relates the Turaev-Viro state sum to CFT conformal blocks in a way analogous to the construction based on Reshetikhin-Turav formulation of TQFT in \cite{Fuchs:2002cm}. 
 
 We note that \cite{Fuchs:2021zrx} discusses the reconstruction of RCFT partition functions via Levin-Wen string net models. It would be useful to explore the precise connection between the current discussion and the construction there.
 
We note that each eigenstate of the RG operator corresponding to CFTs corresponds to phase transitions between topological fixed points, which has been pointed out in \cite{Chatterjee:2022tyg,Chatterjee:2022kxb} .  This allows us to search numerically for CFTs by interpolating between topological fixed points. We illustrated these ideas in $ D=2, 3$. At $D=2$, we find that critical couplings of $SU(2)_k$ models can be recovered to at least one significant figure with very small bond-dimension compared to known theoretical values. We note that our numerical procedure  can in fact also efficiently generate the Frobenius algebra as topological fixed points, recovering known analytic results.  We also show that the $3D$ Ising model can be taken as a strange correlator with the $4D$ toric code, and the critical temperature can be obtained by looking for phase transition points between two topological fixed points of the corresponding RG operator corresponding to the smooth and rough boundaries associated to two higher Frobenius algebra.
We note that in the literature, it is known that different Frobenius algebras in the input category produce isomorphic/equivalent condensation \cite{Hu:2017lrs}, which is often taken to be the {\it same} spatial boundary condition. In our construction, it should be evident that there are phase transitions (such as the 2D Ising model) between algebras considered to be physically the same in  \cite{Hu:2017lrs}. It suggests that isomorphic/equivalent condensation algebras should be considered as physically distinct in one lower dimension.

We briefly look into the possibility of recovering some features of the AdS/CFT dictionary, such as the bulk-boundary propagator. The numerical result is, encouragingly, compatible with the AdS/CFT dictionary but not yet conclusive. 

The approach we adopted and developed to construct CFT partition functions via RG operators following from TQFT in one higher dimensions can be considered as an explicit realization of the Wick-rotation and holographic dualities discussed in \cite{Freed:2018cec, Lootens:2021tet, Ji:2019jhk, Kong:2020cie, Albert:2021vts, Chatterjee:2022kxb, Chatterjee:2022tyg, Liu:2022cxc, Kong:2017etd, Kong:2019byq, Kong:2019cuu, Kong:2020iek, Kong:2020jne, Kong:2021equ, Xu:2022rtj, Aasen:2016dop, Frank1,Lootens:2019ghu, Gaiotto:2020iye, Bhardwaj:2020ymp, Apruzzi:2021nmk, Moradi:2022lqp}.

Our explicit realization should open the door to constructing more novel CFTs in general dimensions.
At present, we are trying to identify the nature of the tricritical point that should be a 2D CFT, and we would like to construct more examples, both in 2D and 3D.  To our knowledge, our algorithm for searching for 2+1D topological fixed points of the 4D RG operator is essentially a first tensor network renormalization method applied to a 3D tensor network. We have since improved the algorithm to allow for larger bond dimensions. And these results would appear in a forth-coming paper \cite{kaixin}.
We should also be able to construct partition functions of (1+1 D) spin CFT by considering phase transitions between super Frobenius algebra previousely used to characterize fermionic spatial boundary conditions following from fermionic anyon condensations \cite{ Aasen:2017ubm, Gaiotto:2015zta, Wan:2016php, Lou:2020gfq}. This will be left for future work.

We explored whether the holographic duality exhibited in the current context is related to the AdS/CFT correspondence in its usual sense. The result is not incompatible, but certainly requires further scrutiny, which is currently underway.

\begin{small}
\textit{Acknowledgements.---} LYH acknowledges the support of NSFC (Grant
No. 11922502, 11875111) and the Shanghai Municipal Science and Technology Major Project
(Shanghai Grant No.2019SHZDZX01), and Perimeter Institute for hospitality as a part of the Emmy Noether Fellowship programme.
Part of this work was instigated in KITP during the program qgravity20.  LC acknowledges support of NSFC (Grant No. 12047515, 12305080) and the start up funding of
South China University of Technology.
We thank Gong Cheng and Zheng-Cheng Gu for collaboration on related problem.
We thank Bartek Czech, Babak Haghighat, Binxin Lao, Jiaqi Lou, Han Ma, Xiao-Liang Qi, Frank Verstraete, Gabriel Wong and Qifeng Wu for useful discussions. We thank Liang Kong in particular for discussions and detailed comments to our draft.
\end{small}

\appendix

\section{Fixed point wavefunctions of 2D Dijkgraaf-Witten RG operator}\label{2Dappendix}

In DW theory of group $G$, the basis of state at each site is given by $|g\rangle$, $g \in G$. We can alternatively use irreducible representations of the group to construct a basis.
They are related by
\be
|\mu, a,b\rangle = \sqrt{ \frac{|\mu|}{|G|}} \mathcal{N}_\mu \sum_g \rho^\mu_{ab}(g) |g\rangle,
\ee
where $\mu$ labels an irreducible representation, and $|\mu|$ is the dimension of the representation.
Therefore we can take $\otimes_i^N \rho^\mu(g)_{ab}$ as a complete basis for wavefunctions of $N$ sites.
Consider the action of the RG operator described in figure \ref{RG2d} and that the group cohomology is trivial. A triangle is thus equal to $\delta_{g_1 g_2, h}$. We can thus consider pairs of sites. Summation of a basis wave-function with the triangle with given output $h = g_1 g_2$ gives
\begin{align}
&\sum_{g_1 \in G} \rho^\mu_{a_1b_1}(g_1) \rho^\nu_{a_2b_2}(g_1^{-1} h) = \frac{|G|}{|\mu|} \delta_{\mu\nu} \delta_{a_1 c} \delta_{b_1 a_2} \rho^\mu_{c b_2}(h) \nonumber \\
 &= \frac{1}{|\mu|} \delta_{\mu\nu} \delta_{b_1 a_2} \sum_{g_1g_2 = h} \sum_m  \rho^\mu_{a_1 m}(g_1) \rho^\mu_{m b_2} (g_2). \label{eq:neighbour_RG}
\end{align}
This suggests that for whatever wave-function we started with, after the action of a triangle (a single 3-point vertex of the RG operator) on neighbouring sites, the wave-function would be projected into the direction corresponding to the same representation on neighbouring sites contracted with each other. Therefore upon multiple layers of the RG operator acting on any given boundary state, it would be projected into the form of (\ref{eq:mps_gprep}) up to an overall normalization. The most general fixed point would be a direct sum of these fixed point solutions.

\begin{figure}
	\centering
	\includegraphics[width=0.8\linewidth]{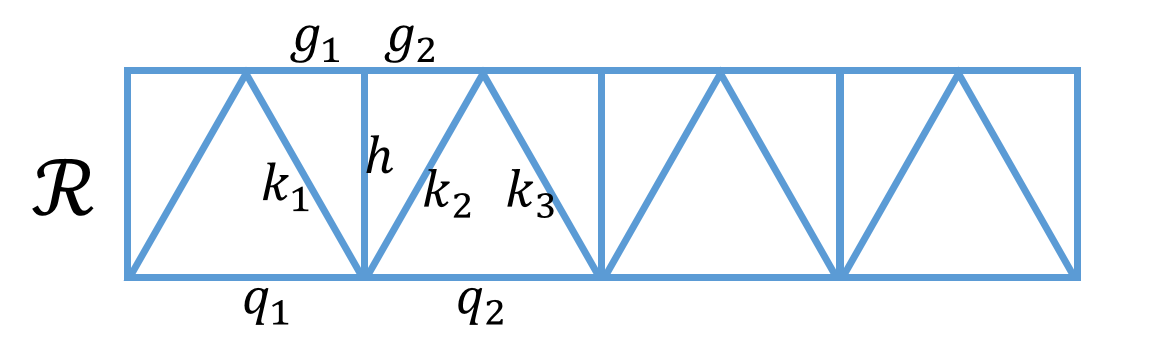}
	\caption{A different choice of the RG operator $\mathcal{R}$ compared to $U(G,\alpha)$ in figure \ref{RG2d}. Here an extra pair of triangle playing the role of ``disentangler'' is inserted between triangles that are responsible for blocking. }
	\label{fig:MERA}
\end{figure}

We note that the choice of RG operator is not unique. For example, one can construct an RG operator that would carry a form that is more closely resembling the MERA tensor network. It corresponds to a triangulation that is depicted in figure \ref{fig:MERA}. The most general fixed point for arbitrary $G$ but trivial $\alpha$ however remains the same as in (\ref{eq:mps_gprep}), following a consideration analogous to  (\ref{eq:neighbour_RG}).

\section{Frobenius algebra in $n$ D Dijkgraaf-Witten Models}\label{DWF}

Let's consider n-dimension DW theories characterized by group G as an example. The n-dimension body is triangulated into n-simplexes, where each edge is labeled by a group element $g_i\in G,$
and the n-simplex is assigned a value $\alpha(g_1,g_2,\dots,g_n) \in H^n(G, U(1))$, where $H^n(G, U(1))$ denotes the n-cohomology, and $g_1,g_2,\dots,g_n$ are the labels on the edges $i_0i_1,i_1i_2,\dots,i_{n-1}i_n$ if the vertices of the n-simplex are $i_0,i_1,i_2,\dots,i_n$ with $i_0<i_1<i_2<\dots<i_n$. Since $\alpha(g_1,g_2,\dots,g_n) \in H^n(G, U(1))$, it satisfies the cocycle condition $d\alpha=1$, where
\be\label{da}
d\alpha(g_0,g_1,\dots,g_n)\equiv \prod_{i=0}^{n+1} \alpha(\dots,g_{i-2},g_{i-1}g_i,g_{i+1},\dots)^{(-1)^i}.
\ee
This condition guarantees that the partition function of the closed n-dimension manifold is invariant under different triangulations.

When the n-dimension body has a boundary, to retain its topological nature, we should construct a partition function which is invariant under different triangulations of both the bulk and the boundary. The boundary is triangulated into (n-1)-simplexes, where each edge is labeled by a group element $h_i\in H\subseteq G,$
and the (n-1)-simplex is assigned a value $\beta(h_1,h_2,\dots,h_{n-1})$, where  $h_1,h_2,\dots,h_{n-1}$ are the labels on the edges $i_0i_1,i_1i_2,\dots,i_{n-2}i_{n-1}$ if the vertices of the (n-1)-simplex are $i_0,i_1,i_2,\dots,i_{n-1}$ with $i_0<i_1<i_2<\dots<i_{n-1}$. The partition function takes the form
\be\label{partitionf}
Z=\sum_{\{g_i\}}\sum_{\{h_i\}}\prod_{\D_n}\alpha_{\D_n}
\prod_{\D_{n-1}}\beta_{\D_{n-1}}
,
\ee
where the summations are over all allowed configurations, each n-simplex $\D_n$ in the bulk contributes $\alpha_{\D_n}$, and each (n-1)-simplex $\D_{n-1}$ on the boundary contributes $\beta_{\D_{n-1}}$. To make this partition function invariant under different triangulations of both the bulk and the boundary, we should have
\begin{align} \label{eq:higher_frob}
& \a d\b \equiv \nonumber \\
&\alpha(h_0,h_1,\dots,h_{n-1})\prod_{i=0}^{n} \b(\dots,h_{i-2},h_{i-1}h_i,h_{i+1},\dots)^{(-1)^i} \nonumber \\
& =1.
\end{align}

We note that both the fixed point solutions for the 4D DW bulk belongs to this class, where $\beta$ is used to construct the boundary state $\la \Omega |$.
In the 2D case, the MPS representation of fixed point boundary states $\la \Omega|$ we have obtained generally contain auxiliary indices -- geometrically we are still attaching a generalized $\beta$ to each edge at the boundary with auxiliary indices attached to the vertices at the end of the edge such that two edges sharing the same vertex would have their shared auxiliary index contracted.
We note however in this generalized sense they are still solution to equation \ref{eq:higher_frob}.

\bibliography{ref}

\end{document}